\documentclass[journal]{IEEEtran}
\usepackage[english]{babel}
\usepackage{algorithm,algorithmicx}
\usepackage{tcolorbox}
\usepackage{tikz}
\newcommand*\circled[1]{\tikz[baseline=(char.base)]{
		\node[shape=circle,draw,inner sep=2pt] (char) {#1};}}
\usepackage{framed}
\usepackage{algpseudocode}
\usepackage{amsthm,nccmath}
\usepackage{amsmath,bm}
\usepackage{bigints}
\usepackage{amssymb}
\usepackage{enumitem}
\usepackage{graphicx}
\usepackage{graphics}
\usepackage{tabularx}
\usepackage{multirow}
\usepackage[font=small,labelfont=bf]{caption}
\usepackage[font=small]{subcaption}
\usepackage{float}
\usepackage{epstopdf}
\usepackage{footnote}
\makesavenoteenv{tabular}
\makesavenoteenv{table}
\usepackage{bbm}
\usepackage{amsthm}
\usepackage{stmaryrd}
\usepackage{amsmath,bm}
\usepackage{amssymb}
\usepackage{mathtools, cuted}
\usepackage{kantlipsum,setspace}
\usepackage[normalem]{ulem}

\usepackage{dblfloatfix}
\theoremstyle{plain}
\newtheorem{theorem}{Theorem}[section]
\newtheorem{lemma}[theorem]{Lemma}
\newtheorem*{lemma*}{Lemma}

\theoremstyle{definition}

\newtheorem*{defn*}{Definition}

\theoremstyle{remark}
\newtheorem*{rem}{Remark}

\newcommand\copyrighttext{%
	\footnotesize \textcopyright~2019~IEEE. Personal use of this material is permitted. Permission from IEEE must be obtained for all other uses, in any current or future media, including reprinting/republishing this material for advertising or promotional purposes, creating new collective works, for resale or redistribution to servers or lists, or reuse of any copyrighted component of this work in other works.~DOI:~10.1109/TSP.2019.2946017.}

\usepackage{tikz}
\usetikzlibrary{calc,trees,positioning,arrows,chains,shapes.geometric,%
	decorations.pathreplacing,decorations.pathmorphing,shapes,%
	matrix,shapes.symbols}

\tikzstyle{startstop} = [rectangle, draw, rounded corners, align=center, minimum width=3cm, minimum height=1cm,text centered]
\tikzstyle{decision} = [diamond, draw, fill=blue!20, 
text width=4.5em, text badly centered, node distance=3cm, inner sep=0pt]
\tikzstyle{block} = [rectangle, draw, fill=blue!10, align=center, rounded corners, minimum width=3cm, minimum height=1cm]
\tikzstyle{blockcast} = [rectangle, draw, fill=red!25, align=center, rounded corners, minimum width=3cm, minimum height=0.45cm]
\tikzstyle{line} = [draw, -latex']
\tikzstyle{cloud} = [draw, ellipse,fill=red!20, node distance=3cm,
minimum height=2em]

\title{Decentralized Gaussian Filters for Cooperative Self-localization and Multi-target Tracking}
\newcommand\copyAlex{%
	\begin{tikzpicture}[remember picture,overlay]
	\node[anchor=south,yshift=4pt] at (current page.south) {\fbox{\parbox{\dimexpr\textwidth-\fboxsep-\fboxrule\relax}{\copyrighttext}}};
	\end{tikzpicture}%
}

\author{
	\IEEEauthorblockN{Pranay Sharma,~\IEEEmembership{Student Member,~IEEE}, Augustin-Alexandru Saucan,~\IEEEmembership{Member,~IEEE}, \\ 
	Donald J. Bucci Jr.,~\IEEEmembership{Member,~IEEE} and Pramod K. Varshney,~\IEEEmembership{Life Fellow,~IEEE}}\\
	\thanks{P. Sharma, A.-A. Saucan and P. K. Varshney are with the Department of Electrical Engineering and Computer Science, Syracuse University, Syracuse, New York-13244 (email: {psharm04, asaucan, varshney}@syr.edu). D. J. Bucci is with Lockheed Martin Advanced Technology Labs, Cherry Hill, New Jersey-08002 (e-mail: donald.j.bucci.jr@lmco.com). P. Sharma and A.-A. Saucan are considered equal contributors.}
}

\begin{document}
\maketitle
\copyAlex
\begin{abstract}
	Scalable and decentralized algorithms for Cooperative Self-localization (CS) of agents, and Multi-Target Tracking (MTT) are important in many applications. In this work, we address the problem of Simultaneous Cooperative Self-localization and Multi-Target Tracking (SCS-MTT) under target data association uncertainty, i.e., the associations between measurements and target tracks are unknown. Existing CS and tracking algorithms either make the assumption of no data association uncertainty or employ a hard-decision rule for measurement-to-target associations. We propose a novel decentralized SCS-MTT method for an unknown and time-varying number of targets under association uncertainty. Marginal posterior densities for agents and targets are obtained by an efficient belief propagation (BP) based scheme while data association is handled by marginalizing over all target-to-measurement association probabilities. Decentralized single Gaussian and Gaussian mixture implementations are  provided based on average consensus schemes, which require communication only with one-hop neighbors. An additional novelty is  a decentralized Gibbs mechanism for efficient evaluation of the product of Gaussian mixtures. Numerical experiments show the improved CS and MTT performance compared to the conventional approach of separate localization and target tracking. 
\end{abstract}


\IEEEpeerreviewmaketitle

\section{Introduction}
\IEEEPARstart{N}{etworks} consisting of mobile interconnected agents with different sensing capabilities are  
commonly found in surveillance \cite{aghajan09_book}, target tracking \cite{bar11_book}, intelligent transportation systems \cite{soatti17_itsc, soatti18_tits}, environmental monitoring \cite{corke10} and robotics \cite{bullo09_book}  applications. In GPS-denied environments and for agents with limited  
power, cooperative self-localization (CS) schemes that rely on inter-agent measurements become necessary. The objective of multi-target tracking (MTT) is the estimation of the trajectories of an unknown and time-varying number of targets. At any time instant, the sensors of an agent produce two kinds of measurements: inter-agent measurements - by observing other agents in proximity, and target measurements - by observing the targets that are within the measurement range of the agent. Due to the collaborative nature of CS, the inter-agent measurements are unambiguous, i.e., the identity of the neighboring agent is known for each inter-agent measurement. On the other hand, targets are non-cooperative and the measurement-to-target associations are not known. Clutter and missed detections also affect the target measurement set. 


In~\cite{spawn09}, CS is achieved via the SPAWN (Sum-Product Algorithm over a Wireless Network) method which relies on Belief Propagation (BP) \cite{SPA01,Yedidia2005jul} for an efficient evaluation of marginal agent posterior densities. The factorization of a joint posterior density is leveraged by BP to efficiently compute marginals. Techniques that address MTT under association uncertainty can be classified as hard (finding the most likely association map) ~\cite{bibl:alex_fusion2018}, \cite{vermaak05_taes} and soft or marginal-based (computing the target state marginal distribution over all measurement-to-target associations)~\cite{bibl:bar_shalom_PDAF2009}. MTT with multiple static agents is addressed in \cite{nannuru15_icassp, delande11_icassp, nannuru16_taes, meyer17_tsp} and \cite{saucan17_tsp}. 

In~\cite{meyer16_tsipn}, an iterative BP message-passing method is proposed for simultaneous cooperative self-localization and target tracking. That is, the target measurements are used for CS in addition to the inter-agent measurements. State inference for both the agents and targets benefits from the exchange of probabilistic information between the CS and tracking tasks. However, the number of targets is assumed fixed and known in \cite{meyer16_tsipn}. In addition, perfect association between measurements and targets is assumed known at each agent. These two assumptions are relaxed in \cite{meyer17_tsp}, which employs the BP message passing approach of \cite{williams14_taes, williams10_issnip} to compute marginal measurement-to-target association probabilities followed by marginal target densities. However, the algorithm is centralized and without sensor self-localization. For a general overview of BP-based methods for MTT, we refer the reader to \cite{meyer18_proc_ieee}. Methods in both \cite{meyer16_tsipn} and \cite{meyer17_tsp} rely on particle representations of agent and target probability densities and the BP messages. Particle filters (PF) \cite{bibl:DoucetBook} are methods for sequential estimation of the state vector in highly non-linear and/or non-Gaussian state systems. However, the computational and communication requirements of PF-based methods can be quite high.

A  simultaneous CS-MTT (SCS-MTT) method for intelligent transportation systems was proposed in \cite{brambilla18_ssp}, where a MAP rule is employed to select the measurement-to-target associations with the highest marginal probabilities. Additionally, a decentralized single Gaussian implementation is given. In \cite{frohle18_wcnc}, a centralized BP method for agent localization is proposed, which only uses target measurements. The number of targets is assumed known. This is extended in \cite{meyer18_icc}, where the number of targets is unknown. The agents can exchange their location information as well as the target measurements to assist each other. A BP-based method for CS is proposed in \cite{meyer18_fusion} where association uncertainty is considered for the inter-agent measurements. BP-based methods for SCS-MTT under measurement and/or dynamic model uncertainties were proposed in \cite{meyer16_globecom} and \cite{soldi18_fusion}. In \cite{pranay18_asilomar}, we proposed a centralized PF implementation of a SCS-MTT filter for an unknown and time-varying number of targets and in the presence of association uncertainty for target measurements.



\subsection{Our Contributions} \label{subsec_contri}

We propose an efficient, decentralized BP message passing based algorithm for simultaneous cooperative self-localization (of mobile agents) and multi-target tracking (SCS-MTT), under measurement-to-target association uncertainty, extending the work in \cite{meyer17_tsp} and~\cite{meyer16_tsipn}. As in \cite{meyer17_tsp}, the data association problem is solved using an iterative BP-based approach \cite{williams14_taes}. Unlike \cite{meyer18_icc}, target measurements are not shared across agents.

The factor graph of the joint posterior over agent and target states has cycles and several message orderings are possible. The novelty of our contribution also lies in the ordering of messages that ensures a reduced amount of data exchange over the network. Additional novelties are our decentralized Gaussian-based (DG) and decentralized Gaussian-Mixture based (DGM) implementations of the algorithm in DG-SCS-MTT and DGM-SCS-MTT filters respectively. The filters achieve network-wide consensus over the target beliefs, i.e., over the means, covariance matrices and component weights of the Gaussian Mixture (GM). For most kinematic tracking applications, the communication loads of DG-SCS-MTT and DGM-SCS-MTT are significantly smaller than the PF implementations.


Computing the target belief by marginalizing over all the possible associations leads to a GM even when the target prior is a Gaussian density. Hence, the DG-SCS-MTT filter employs a moment matching approach to approximate the resulting GM target belief with a single Gaussian.
In case of GMs, the decentralized computation of target beliefs involves a product of GM likelihood messages (stored at different agents) and a GM prior. The number of components in the complete GM product is exponential in the number of agents. Thus, we propose a novel  decentralized Gibbs mechanism, extending the centralized Gibbs approach proposed in \cite{nonparametricbp10}, to sample only the components of the GM product with the highest weights, and thus approximate the entire product. In parallel, the agents sample local Gaussian components followed by a synchronization step where a consensus is reached among the agents regarding the parameters of the resulting product component. 
Our numerical results show that the performance of the decentralized algorithm that employs these techniques is similar to its centralized counterpart. Numerical experiments exhibit improved performance of both DG-SCS-MTT and DGM-SCS-MTT filters
when compared to a separate SPAWN \cite{spawn09} (for localization) and MTT~\cite{meyer17_tsp} approach.

The paper is organized as follows. The system model and notation is discussed in Section \ref{sec_sys_mod}, followed by the proposed SCS-MTT filter in Section \ref{sec_prop_method}. The decentralized Gaussian-mixture and single Gaussian implementations are given in Sections \ref{sec_gmm} and \ref{sec_dist_gauss} respectively. We present the simulation results in Section \ref{sec_sim}, followed by conclusion in Section \ref{sec_conc}.

\section{System Model and notation} \label{sec_sys_mod}
The notations, assumptions, and the resulting system model are presented in the following sections. The system model is essentially a combination of the system models in \cite{meyer17_tsp, meyer16_tsipn}. Therefore, a lot of notation is also borrowed from \cite{meyer17_tsp, meyer16_tsipn}.
\subsection{Notation}
\subsubsection{Agent and target states}
For $a,b \in \mathbb{N}$, we denote with $[a:b]$ the set of positive integers $\{ a, a+1, \cdots, b \}$. We denote the set of agents by $\mathcal{A} \triangleq [ 1:S]$, and the set of Potential Targets (PTs) by $\mathcal{T} \triangleq [ 1:K ]$, where $K$ is the maximum possible number of PTs present. The state of agent $s$ at time $n$ is denoted by $\mathbf{y}_{n,s} \in \mathbb{R}^{d_a}$. 

PT $k \in \mathcal{T}$ is described at time $n$, by state $\mathbf{x}_{n,k} \in \mathbb{R}^{d_t}$ alongside a binary variable, $r_{n,k}$, that indicates its existence at time $n$ ($r_{n,k} = 1$ for presence, $0$ for absence). The time-varying number of targets is accounted for via the variables $\{ r_{n,k} \}$ while target existence can be inferred from the probability of existence $\text{Pr}(r_{n,k}=1)$. We further define the joint state vector of all the PTs at time $n$, $\mathbf{x}_{n} \triangleq \left[ \mathbf{x}_{n,1}^{T} , \cdots, \mathbf{x}_{n,K}^{T} \right]^T$, and the across-time vector, $\mathbf{x} \triangleq \left[ \mathbf{x}_{0}^{T} , \cdots, \mathbf{x}_{n}^{T} \right]^T$. In an analogous manner, we introduce the joint vectors at time $n$, $\mathbf{r}_{n}$ and $\mathbf{y}_{n}$, and across-time vectors $\mathbf{r}$ and $\mathbf{y}$. Let $\tilde{\mathbf{x}}_{n,k} = [\mathbf{x}_{n,k}^T, r_{n,k}]^T$ be the \textit{augmented} state vector for PT $k$ at time $n$. We also define $\tilde{\mathbf{x}}_n = [ \mathbf{x}_n^{T}, \mathbf{r}_n^T ]^T$ and $\tilde{\mathbf{x}} = [ \mathbf{x}^{T}, \mathbf{r}^T ]^T$. In addition, we introduce the notation $\int (\cdot) d \tilde{\mathbf{x}}_{n,k} \triangleq \sum_{r_{n,k} \in \{0,1\} } \int (\cdot) d\mathbf{x}_{n,k}   $. If $f(\tilde{\mathbf{x}}_{n,k}) \equiv f(\mathbf{x}_{n,k}, r_{n,k})$ is the augmented state probability density for PT $k$, then the probability of existence at time $n$ is $P_{n,k}^e \triangleq \text{Pr}(r_{n,k}=1) = \int f(\mathbf{x}_{n,k}, 1) d\mathbf{x}_{n,k}$.  

\subsubsection{Inter-agent Measurements} \label{sec_ag_meas}
For agent $s$, let $\mathcal{A}_{n,s} \subseteq \mathcal{A}$ denote the set of its neighboring agents, i.e., agents that are within its inter-agent measurement range, at time $n$. Let $\mathbf{w}_{s,\ell;n} \in \mathbb{R}^{d_w}$ be the measurement that agent $s$ makes  with respect to the neighboring agent $\ell \in \mathcal{A}_{n,s}$, at time $n$. The inter-agent measurement likelihood is  denoted as $ f(\mathbf{w}_{s,\ell;n} \vert \mathbf{y}_{s,n}, \mathbf{y}_{\ell,n}) $. The stacked vector of the inter-agent measurements at agent $s$ at time $n$ is denoted by $\mathbf{w}_{s;n}$. Let $\mathbf{w}_n \triangleq \left[ \mathbf{w}_{1;n}^T , \cdots, \mathbf{w}_{S;n}^{T} \right]^T$ and $\mathbf{w} \triangleq \left[ \mathbf{w}_1^T ,\cdots, \mathbf{w}_n^T \right]^T$.

\subsubsection{Target Measurements}  \label{sec_tg_meas}
Agent $s$ observes a subset $\mathcal{T}_{n,s} \subset \mathcal{T}$ of PTs that are within its target-measurement range. We also define the set of agents observing PT $k$ at time $n$ as $ \mathcal{A}_{n,k} = \{s\in \mathcal{A} \colon k \in \mathcal{T}_{n,s}  \}$. Since targets are non-cooperative, the collection of target measurements suffers from missed detections, clutter and association uncertainty. Let $M_n^s$ be the number of target measurements gathered by agent $s$ at time $n$. Let $\mathbf{z}_n^{s} \triangleq [  (\mathbf{z}_{n,1}^{s})^T , \cdots,   (\mathbf{z}_{n,M_n^{s}}^{s})^T ]^T$ be an arbitrarily ordered collection of these measurements, with $\mathbf{z}_{n,m}^{s} \in \mathcal{R}^{d_z}$ $\forall \ m \in \mathcal{M}_{n,s} \triangleq [ 1:M_n^s ]$. Furthermore, let $\mathbf{z}_n \triangleq [  (\mathbf{z}_n^{1})^T, \cdots, (\mathbf{z}_n^{S})^T ]^T$, $\mathbf{z} \triangleq [  (\mathbf{z}_1)^T, \cdots, (\mathbf{z}_n)^T ]^T$, $\mathbf{m}_n  \triangleq [ M_n^{1} \cdots M_n^{s} ]^T $ and $\mathbf{m} \triangleq [  \mathbf{m}_1^T, \cdots, \mathbf{m}_n^T ]^T$. The likelihood of measurement $\mathbf{z}_{n,m}^s$ made by agent $s$, if it corresponds to PT $k$ is $f(\mathbf{z}_{n,m}^s \vert \mathbf{y}_{n,s}, \mathbf{x}_{n,k}) $. A PT $\mathbf{x}_{n,k}$ is detected by agent $s$ with probability $P_D^s(\mathbf{x}_{n,k})$. Finally, $\mathbf{z}_{n}^s$ also contains clutter measurements, independently sampled from a Poisson point process. The rate of clutter points is $\lambda_{n}^s$ and their probability distribution is $f_{n,s}^{FA}(\mathbf{z})$, for measurement $\mathbf{z}$. 

\subsection{Assumptions} \label{sec:assumptions}
Our assumptions in this work stem from~\cite{meyer17_tsp, meyer16_tsipn} and are provided in the following: 
\\ \hspace*{0.5cm}(A1) Agent and target states are \textit{a priori} independent and evolve independently in time according to Markov processes. 
\\ \hspace*{0.5cm}(A2) The communication graph $\mathcal{G}_n$ that spans the decentralized network of agents is connected at all times and the communication links between the agents are bidirectional.
\\ \hspace*{0.5cm}(A3) Given the current agent states $\mathbf{y}_n$ and  augmented target states $\tilde{\mathbf{x}}_n$, the measurements $\mathbf{w}_n$ and  $\mathbf{z}_n$ are conditionally independent of past $(\mathbf{w}_{1:n-1}, \mathbf{z}_{1:n-1})$ and future $(\mathbf{w}_{n+1:\infty}, \mathbf{z}_{n+1:\infty})$ measurements, i.e., $f( \mathbf{w}_n, \mathbf{z}_n \vert \mathbf{w}_{1:n-1}, \mathbf{z}_{1:n-1}, \mathbf{w}_{n+1:\infty}, \mathbf{z}_{n+1:\infty}, \mathbf{y}_n, \tilde{\mathbf{x}}_n ) = f( \mathbf{w}_n, \mathbf{z}_n \vert \mathbf{y}_n, \tilde{\mathbf{x}}_n )$.
\\ \hspace*{0.5cm}(A4) Current agent and target states $\mathbf{y}_n$, $\tilde{\mathbf{x}}_n$, are conditionally independent of all the past measurements $\mathbf{w}_{0:n-1}, \mathbf{z}_{0:n-1}$ given the previous states $\mathbf{y}_{n-1}$ and $\tilde{\mathbf{x}}_{n-1}$. 
\\ \hspace*{0.5cm}(A5) Given $\mathbf{y}_n$, the inter-agent measurements $\mathbf{w}_{s,\ell;n}$ and $\mathbf{w}_{s',\ell';n}$ are conditionally independent if $(s,\ell)\neq (s',\ell')$.
\\ \hspace*{0.5cm}(A6) Given $\mathbf{y}_n$ and $\tilde{\mathbf{x}}_n$, the target  and agent measurements $\mathbf{z}_{n}$  and $\mathbf{w}_{n}$ are conditionally independent and furthermore $f( \mathbf{w}_n, \mathbf{z}_n\vert \mathbf{y}_n, \tilde{\mathbf{x}}_n)= f( \mathbf{w}_n\vert \mathbf{y}_n)f( \mathbf{z}_n\vert \mathbf{y}_n, \tilde{\mathbf{x}}_n)$.   
\\ \hspace*{0.5cm}(A7) At any time $n$, an existing target can generate at most one measurement at any agent, and any target measurement at an agent is generated by at most one existing target \cite{bar11_book, mahler07_book}. The detection process is independent for different targets and across different agents.
\\ \hspace*{0.5cm}(A8) Target measurements $\mathbf{z}_{n}$ suffer from origin uncertainty, i.e., the associations between the individual measurements of $\mathbf{z}_{n}$ and the PT $\tilde{\mathbf{x}}_{n}$ are unknown. Some measurements are due to clutter and some PTs are not detected. 
\\ \hspace*{0.5cm}(A9) Inter-agent measurements do not suffer from origin uncertainty. Agent $s$ knows that measurement $\mathbf{w}_{s,\ell;n}$ originates from agent $\ell \in \mathcal{A}_{n,s}$ and the inter-agent measurement links are bidirectional, i.e., $\ell \in \mathcal{A}_{n,s} \Leftrightarrow s \in \mathcal{A}_{n,\ell}$ for $s, \ell \in \mathcal{S}$.   
\\ \hspace*{0.5cm}(A10) Each agent knows its own prior and dynamic model and the priors and dynamic models of all PTs. All agents have synchronized internal clocks.

The SPAWN approach \cite{spawn09} addresses the problem of self-localization without MTT. In \cite{meyer16_tsipn}, a perfect knowledge of the target-to-measurement associations is assumed. Also, the number of targets is known and time-invariant.
In~\cite{meyer17_tsp}, these assumptions are removed, but the agents have perfect knowledge of their positions, i.e., fixed sensors case. In this work, we extend~\cite{meyer16_tsipn} by relaxing the assumption of known origins of target measurements, and accommodate an unknown, time-varying number of targets as in~\cite{meyer17_tsp}.
\begin{table}[t!]
	\centering
	\begin{tabular}{|c|c|c|}
		\hline
		$r_{n-1,k}$ & $r_{n,k}$ & $f \left( \mathbf{x}_{n,k}, r_{n,k} \vert \mathbf{x}_{n-1,k}, r_{n-1,k} \right)$ \\ \hline
		0 & 1 & $P_{n,k}^B f_b \left( \mathbf{x}_{n,k} \right)$ \\ \hline
		0 & 0 & $\left( 1 - P_{n,k}^B \right) f_D \left( \mathbf{x}_{n,k} \right)$ \\ \hline
		1 & 0 & $\left( 1 - P_{n,k}^S \right) f_D \left( \mathbf{x}_{n,k} \right)$ \\ \hline
		1 & 1 & $P_{n,k}^S(\mathbf{x}_{n,k}) f \left( \mathbf{x}_{n,k} \vert \mathbf{x}_{n-1,k} \right)$ \\ \hline
	\end{tabular}
	\caption{ State transition kernels for different values of target existence indicators. The function $f_D(\cdot)$ is a dummy pdf~\cite{meyer17_tsp}.}
	\label{tabel_transition_pdf}
\end{table}
\subsection{System Model}
Under assumption (A1), we denote the  agent transition densities with $f \left( \mathbf{y}_{n,s} \vert \mathbf{y}_{n-1,s} \right)$ $\forall$ $s$. For PT $k$, the transition kernel $  f(\tilde{\mathbf{x}}_{n,k} \vert \tilde{\mathbf{x}}_{n-1,k})  \equiv f( \mathbf{x}_{n,k}, r_{n,k} \vert \mathbf{x}_{n-1,k}, r_{n-1,k})$ accounts for target birth, death and evolution (in case of survival) as listed in Table \ref{tabel_transition_pdf}. The dynamic kernel is a function of the indicator variable $r_{n,k}$. Here, $P_{n,k}^B$ is the birth probability, $f_b \left( \mathbf{x}_{n,k} \right)$ is the birth pdf, $P_{n,k}^S(\cdot)$ is the survival probability, and $f \left( \mathbf{x}_{n,k} \vert \mathbf{x}_{n-1,k} \right)$ is the state transition pdf. Under assumption (A1), the joint pdf of $[\mathbf{y}^T, \mathbf{x}^T, \mathbf{r}^T]^T$ given by
\begin{align}
f ( \mathbf{y}, \underbrace{\mathbf{x}, \mathbf{r}}_{\tilde{\mathbf{x}}} ) &= \prod\nolimits_{s=1}^S f \left( \mathbf{y}_{0,s} \right) \prod\nolimits_{n'=1}^{n} f \left( \mathbf{y}_{n',s} \vert \mathbf{y}_{n'-1,s} \right) \nonumber \\
& \times \prod\nolimits_{k=1}^K f \left( \tilde{\mathbf{x}}_{0,k} \right) \prod\nolimits_{n'=1}^{n} f \left( \tilde{\mathbf{x}}_{n',k} \vert \tilde{\mathbf{x}}_{n'-1,k} \right). \label{eq_state_evol}
\end{align}
To solve the data association problem of assumption (A8), i.e., finding the associations between measurements and PTs, we use the redundant formulation of association variables proposed in \cite{williams14_taes}. \textit{Target oriented association variables} define the PT-measurement associations at sensor $s$ at time $n$:
\begin{equation}
\label{eq_tg_assoc_var}
a_{n,k}^{s} \triangleq
\begin{cases}
m \in \mathcal{M}_{n,s} & \text{PT } k \text{ generated } \mathbf{z}_{n,m}^s \text{ at time } n, \\
0 & \text{PT } k \text{ is not detected at time }n. 
\end{cases}
\end{equation}
The \textit{measurement-oriented association variables} are
\begin{equation}
\label{eq_ag_assoc_var}
b_{n,m}^{s} \triangleq
\begin{cases}
k \in \mathcal{K} & \text{PT } k \text{ generated } \mathbf{z}_{n,m}^s \text{ at time } n, \\
0 &  \mathbf{z}_{n,m}^s \text{ is a clutter measurement}.
\end{cases}
\end{equation}
Further, we define stacked vectors of association variables: $\mathbf{a}_n^{s} \triangleq [ a_{n,1}^{s},\cdots ,a_{n,K}^{s} ]^T$, $\mathbf{a}_n \triangleq [ ( \mathbf{a}_n^{1} )^T,\cdots , ( \mathbf{a}_n^{S} )^T ]^T, \mathbf{a} \triangleq [ \mathbf{a}_1^T,\cdots , \mathbf{a}_n^T ]^T$, and $\mathbf{b}_n^{s} \triangleq [ b_{n,1}^{s},\cdots ,b_{n,M_n^{s}}^{s} ]^T, \mathbf{b}_n \triangleq [ ( \mathbf{b}_n^{1} )^T,\cdots , ( \mathbf{b}_n^{S} )^T ]^T, \mathbf{b} \triangleq [ \mathbf{b}_1^T,\cdots ,\mathbf{b}_n^T ]^T$. Note that $\mathbf{a}_n^s$ and $\mathbf{b}_n^s$ are redundant, meaning one can be derived from the other. We define the  indicator function $\Psi ( a_{n,k}^{s}, b_{n,m}^{s} )$
\begin{align}
\label{eq_assoc_Psi}
\Psi \left( a_{n,k}^{s}, b_{n,m}^{s} \right) \triangleq \begin{cases}
0 & \text{if } a_{n,k}^{s}=m \text{ and } b_{n,m}^{s} \neq k \\
& \text{or } a_{n,k}^{s} \neq m \text{ and } b_{n,m}^{s} = k \\ 
1 & \text{otherwise}
\end{cases}
\end{align}
where $\{ \Psi ( a_{n,k}^{s}, b_{n,m}^{s} ) \}_{k,m}$ collectively enforce the association variables $\mathbf{a}_{n}^{s}$ and $\mathbf{b}_{n}^{s} $ to be consistent \cite{williams14_taes}. Under the assumptions (A3-A9), the joint measurement likelihood becomes
\begin{align}
\label{eq_joint_lklhd}
& f \left( \mathbf{z}, \mathbf{w} \vert \mathbf{y}, \tilde{\mathbf{x}}, \mathbf{a}, \mathbf{m} \right) = f \left( \mathbf{w} \vert \mathbf{y} \right) f \left( \mathbf{z} \vert \mathbf{y}, \tilde{\mathbf{x}}, \mathbf{a}, \mathbf{m} \right) = \\
& \prod_{n'} \prod_{s} f \left( \mathbf{z}^{s}_{n'} \big| \mathbf{y}_{n',s}, \tilde{\mathbf{x}}_{n'}, \mathbf{a}^{s}_{n'}, M^{s}_{n'} \right) \prod_{\mathclap{\ell \in \mathcal{A}_{n',s}}} f \left( \mathbf{w}_{s,\ell;n'} \big| \mathbf{y}_{n',s}, \mathbf{y}_{n',\ell} \right). \nonumber
\end{align}
Since under assumption (A7) each measurement is caused by a target or clutter, the target measurement likelihood further factorizes as
\begin{align}
    & f(\mathbf{z}_{n}^{s} \vert \mathbf{y}_{n,s}, \tilde{\mathbf{x}}_{n}, \mathbf{a}_{n}^{s}, M_{n}^{s} ) =  \label{eq_tgt_meas_lklhd} \\
    & \underbrace{\prod\nolimits_{m : b_{n,m}^s = 0} f_{n,s}^{FA} \left( \mathbf{z}_{n,m}^{s} \right)}_{\text{clutter measurements}} \ \times \ \underbrace{\prod\nolimits_{\substack{m : b_{n,m}^s = k \\ \text{and } r_{n,k} = 1}} f \left( \mathbf{z}_{n,m}^{s} \big| \mathbf{x}_{n,k}, \mathbf{y}_{n,s} \right)}_{\text{Measurements from existing targets}} \nonumber
\end{align}
which we can rewrite as follows
\begin{multline}
\hspace{-3mm} f(\mathbf{z}_{n}^{s} \vert \mathbf{y}_{n,s}, \tilde{\mathbf{x}}_{n}, \mathbf{a}_{n}^{s}, M_{n}^{s} ) \propto \prod\nolimits_{k\in \mathcal{K}} g_{k}( \tilde{\mathbf{x}}_{n,k}, \mathbf{y}_{n,s}, a_{n,k}^{s}; \mathbf{z}_{n}^{s}) \hspace{-2mm} \label{eq:prod_g}
\end{multline}
where the normalization factor depends only on $f_{n,s}^{FA}(\cdot)$, hence only on the measurements $\mathbf{z}_n^{s}$. For $m \in \ \mathcal{M}_n^{s}$
\begin{equation}
\label{eq_g}
g_{k} \left( {\mathbf{x}}_{n,k}, r_{n,k} = 1, \mathbf{y}_{n,s}, a_{n,k}^s = m; \mathbf{z}_{n}^{s} \right)
= \mfrac{f \left( \mathbf{z}_{n,m}^{s} \vert \mathbf{x}_{n,k}, \mathbf{y}_{n,s} \right)}{f_{n,s}^{{FA}} \left( \mathbf{z}_{n,m}^{s} \right)} 
\end{equation}
while $g_{k} ( {\mathbf{x}}_{n,k}, r_{n,k} = 1, \mathbf{y}_{n,s}, a_{n,k}^s = 0; \mathbf{z}_{n}^{s} ) = 1$. For absent targets ($r_{n,k} = 0$), $g_{k} ( {\mathbf{x}}_{n,k}, r_{n,k} = 0, \mathbf{y}_{n,s}, a_{n,k}^s = m; \mathbf{z}_{n}^{s} ) = 1, \ \forall \ m =0,\dots, M_{n}^{s} $.

The association variables $\mathbf{a}$, $\mathbf{b}$ and the number of measurements $\mathbf{m}$ are assumed conditionally independent across time and across agents, given the states of agents and targets. Thus, the joint distribution of association variables and the number of measurements, factorizes as 
\begin{align}
& p ( \mathbf{a}, \mathbf{b}, \mathbf{m} \vert \mathbf{y}, \tilde{\mathbf{x}}) = \prod_{n'=1}^{n} \prod_{s=1}^S p ( a_{n'}^{s}, b_{n'}^{s}, M_{n'}^{s} \vert \mathbf{y}_{n',s}, \tilde{\mathbf{x}}_{n'}) \label{eq_prior_assoc} \\
& \propto \prod_{n'=1}^{n} \prod_{s=1}^S \prod_{k=1}^K h_k ( \tilde{\mathbf{x}}_{n',k},  a_{n',k}^{s}, \mathbf{y}_{n',s} ) \prod_{m=1}^{M_{n'}^{s}} \Psi \left( a_{n',k}^{s}, b_{n',m}^{s} \right) \nonumber 
\end{align}
where the normalization constant depends only on the clutter rate $\lambda_n^s$ and the number of measurements $\mathbf{m}$ \cite{maskell06_fusion}. The term $h_k(\cdot)$ is defined as
\begin{equation}
\label{eq_h}
h_k ( \mathbf{x}_{n,k}, 1, a_{n,k}^{s}, \mathbf{y}_{n,s} )= \begin{cases}
\frac{P_D^{s} \left( \mathbf{x}_{n,k}\right)}{\lambda^{s}_n}, & \text{ if } a_{n,k}^{s} \in \mathcal{M}_n^{s} \\
1 - P_D^{s} \left( \mathbf{x}_{n,k}\right), & \text{ if } a_{n,k}^{s}=0
\end{cases}
\end{equation}
and $h_k ( \mathbf{x}_{n,k}, r_{n,k} = 0, a_{n,k}^{s} , \mathbf{y}_{n,s}) = 1(a_{n,k}^{s})$ where $1(a) = 1$ if $a=0$ and $1(a) = 0$ otherwise. $\Psi \left( \cdot, \cdot \right)$ is defined in (\ref{eq_assoc_Psi}).

\section{The SCS-MTT filter} 
\label{sec_prop_method}
We perform agent and target state inference using the marginal posterior densities. These are obtained from the joint posterior density using the following factorization, which is derived by extending analogous results in \cite{meyer17_tsp} and \cite{meyer16_tsipn}.
\begin{lemma} 
\label{lemma_joint_factor}
The joint posterior density of all the agent and PT states, given inter-agent and target measurements, up to time $ n$, admits the factorization
\begin{align}
& f (\mathbf{y}, \tilde{\mathbf{x}}, \mathbf{a}, \mathbf{b} \vert \mathbf{z}, \mathbf{w} ) \propto \left[  \prod_{k=1}^K f \left( \tilde{\mathbf{x}}_{0,k} \right) \prod_{ n'=1}^{ n} f \left( \tilde{\mathbf{x}}_{ n',k} \vert \tilde{\mathbf{x}}_{ n'-1,k} \right) \right] \nonumber \\
& \prod_{s=1}^S \hspace{-1mm} \Bigg\{ \hspace{-1mm}  f \left( \mathbf{y}_{0,s} \right)\:\: \mathclap{\prod_{ n'=1}^{ n}} \:\:\:  \Bigg[ f \left( \mathbf{y}_{ n',s} \vert \mathbf{y}_{ \: \mathclap{ n'} -1,s} \right) \Big( \:\:\:\:\:\: \mathclap{\prod_{ \ell \in \mathcal{A}_{n',s}}} \:\:\: f \left( \mathbf{w}_{s,\ell; n'} \vert \mathbf{y}_{\: \mathclap{n'}\:,s}, \mathbf{y}_{\: \mathclap{ n'} \:,\ell} \right) \Big)  \nonumber \\
&   \prod_{k=1}^K \bigg( v_k^s ( \tilde{\mathbf{x}}_{ n',k}, a_{ n',k}^{s}, \mathbf{y}_{ n',s}; \mathbf{z}_{ n'}^{s} )   \prod_{m=1}^{M_{ n'}^{s}} \Psi \left( a_{n',k}^{s}, b_{ n',m}^{s} \right) \bigg) \Bigg] \Bigg\} \label{eq_joint_post_2}
\end{align}
\end{lemma}
where, $v_k^s ( \tilde{\mathbf{x}}_{ n',k}, a_{ n',k}^{s}, \mathbf{y}_{ n',s}; \mathbf{z}_{ n'}^{s} ) $
\begin{align*}	 
& \triangleq \begin{cases}
\frac{P_D^s(\mathbf{x}_{ n',k}) f \left( \mathbf{z}_{ n',m}^{s} \vert \mathbf{x}_{ n',k}, \mathbf{y}_{ n',s} \right)}{\lambda_{ n'}^s f_{ n',s}^{{FA}} \left( \mathbf{z}_{ n',m}^{s} \right)}, & \begin{matrix}
            \text{ if } a_{ n',k}^s = m \neq 0 \\
            \text{ and } r_{ n',k}=1
            \end{matrix} \\
1 - P_D^s(\mathbf{x}_{ n',k}), & \text{ if } a_{ n',k}^s = 0,\; r_{ n',k}=1 \\
1, & \text{ if } a_{ n',k}^s = 0, \; r_{ n',k}=0 \\
0, & \text{ otherwise}.
\end{cases}
\end{align*}

\begin{figure*}[t!]
	\centering
	\includegraphics[width=0.85\textwidth]{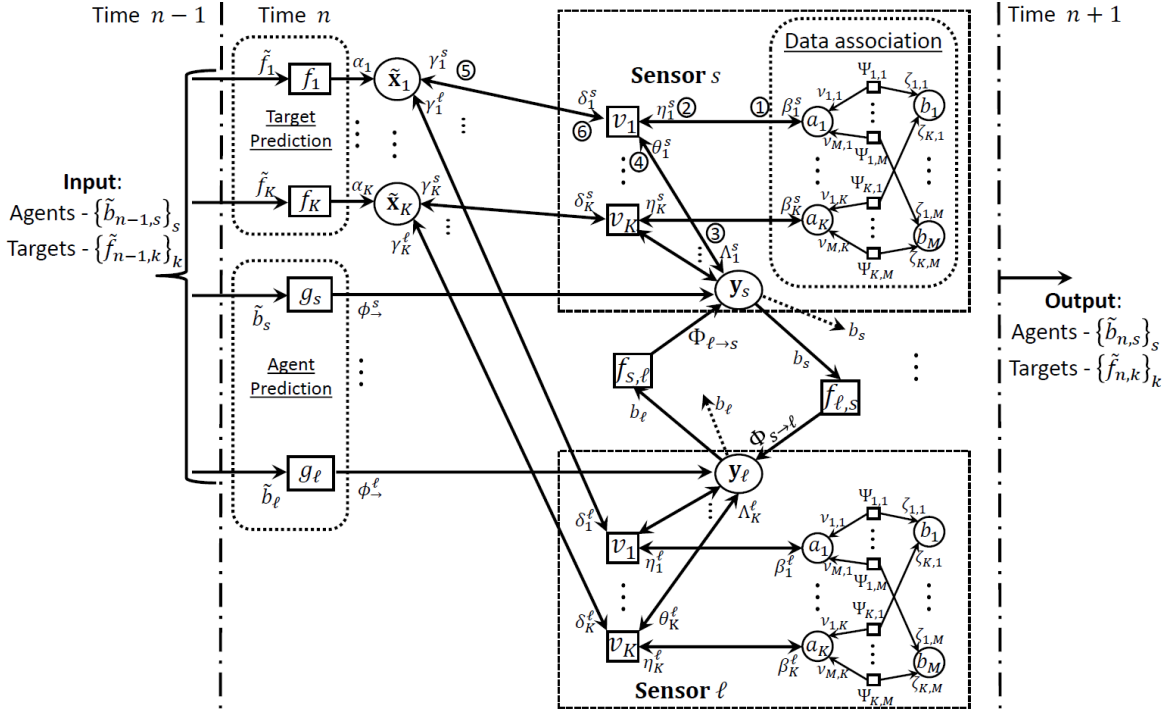}  
	\caption{Factor graph representing the factorization of (\ref{eq_joint_post_2}), for one time step. The factor nodes are shown as rectangles, while the variable nodes are shown as ovals. Time index $n$ has been omitted from notations and messages passed between nodes are represented as annotations on each link. Following are the factor nodes: for PT $k$, $f_{k} \triangleq f ( \tilde{\mathbf{x}}_{n,k} \vert \tilde{\mathbf{x}}_{n-1,k} )$; for agent $s$, $g_{s} \triangleq f ( \mathbf{y}_{n,s} \vert \mathbf{y}_{n-1,s} )$, $v_k \triangleq v ( \mathbf{x}_{n,k}, r_{n,k}, a_{n,k}^{s}, \mathbf{y}_{n,s} ; \mathbf{z}_n^{s} )$; for agent $s$ measuring another agent $\ell$, $f_{s,\ell} \triangleq f ( \mathbf{w}_{s,\ell;n} \vert \mathbf{y}_{n,s}, \mathbf{y}_{n,\ell} )$, $\Psi_{k,m} \triangleq \Psi ( a_{n,k}^{s}, b_{n,m}^{s} )$. The numbered circles $1-6$ in the block corresponding to sensor $s$ demonstrate the order in which messages are computed. The beliefs broadcast by the agents at the beginning of each outer loop are shown by arrows $b_s$ and $b_{\ell}$ coming out of agent state nodes $\mathbf{y}_{s}$ and $\mathbf{y}_{\ell}$ respectively.}
	\label{fig_factor_graph}
\end{figure*}

\begin{proof}
    Note that the number of target measurements $M_n^{s}$ becomes fixed when conditioning on $\mathbf{z}_n^{s}$. Applying Bayes' rule
    \begin{align}
    & f \left( \mathbf{y}, \mathbf{x}, \mathbf{r}, \mathbf{a}, \mathbf{b} \vert \mathbf{z}, \mathbf{w} \right) = f \left( \mathbf{y}, \mathbf{x}, \mathbf{r}, \mathbf{a}, \mathbf{b}, \mathbf{m} \vert \mathbf{z}, \mathbf{w} \right) \label{eq:factor_big} \\
    & \propto \underbrace{f \left( \mathbf{y}, \mathbf{x}, \mathbf{r} \right)}_{(i)} \cdot \underbrace{p \left( \mathbf{a}, \mathbf{b}, \mathbf{m} \vert \mathbf{y}, \mathbf{x}, \mathbf{r} \right)}_{(ii)} \cdot \underbrace{f \left( \mathbf{z}, \mathbf{w} \vert \mathbf{y}, \mathbf{x}, \mathbf{r}, \mathbf{a}, \mathbf{b}, \mathbf{m} \right)}_{(iii)} \nonumber
    \end{align}
    where $(i)$ represents the joint distribution (\ref{eq_state_evol}) of the agent and target states  up to time $n$; $(ii)$ represents the data association and detection of the targets given the agent and augmented target states (\ref{eq_prior_assoc})-(\ref{eq_h}); and $(iii)$ represents the joint measurement likelihood, given the states of all the agents and targets, and their data association relationships (\ref{eq_joint_lklhd})-(\ref{eq_g}). Substituting the expressions for $(i)-(iii)$ into (\ref{eq:factor_big}), and defining $v ( \tilde{\mathbf{x}}_{n,k}, a_{n,k}^{s}, \mathbf{y}_{n,s}; \mathbf{z}_{n}^{s} ) \triangleq h_k ( \tilde{\mathbf{x}}_{n,k}, a_{n,k}^{s}, \mathbf{y}_{n,s} ) \times g_k ( \tilde{\mathbf{x}}_{n,k}, a_{n,k}^{s}, \mathbf{y}_{n,s}; \mathbf{z}_{n}^{s} )$, we obtain (\ref{eq_joint_post_2}).
\end{proof}
The marginals associated with (\ref{eq_joint_post_2}) can be efficiently computed via BP algorithms that exploit the structure embedded in its factorization. The factor graph corresponding to (\ref{eq_joint_post_2}) for a fixed time step $n$ is shown in Figure~\ref{fig_factor_graph}. This factor graph represents a combination of the factor graph containing the agent and target states from \cite[Figure 2]{meyer16_tsipn} and the factor graph corresponding to target measurement uncertainty of~\cite{meyer17_tsp}. 

\subsection{The SCS-MTT filter: the BP message passing scheme} \label{sec_bp_filter}
In this section, we describe our proposed message passing algorithm for inferring the marginal densities of targets $b(\tilde{\mathbf{x}}_{n,k})$ and agents $b(\mathbf{y}_{n,s})$ at time $n$ corresponding to the joint density  of (\ref{eq_joint_post_2}). For an introduction to BP, the reader is directed to~\cite{Yedidia2005jul}. Since the factor graph of Figure~\ref{fig_factor_graph} has cycles, multiple message ordering schemes exist. Similar to \cite{meyer17_tsp}, we assume that: (i) messages are not sent backward in time, and (ii) marginal association probabilities are evaluated via BP at each agent. 


At each beginning of time step $n$, using the agent belief ${b}_{} (\mathbf{y}_{n-1,s})$ from the previous time step, agent $s$ computes the \textit{prediction message} $\phi_{\to n} ( \mathbf{y}_{n,s} )$ given by
\begin{equation}
\label{eq_msg_phi}
\phi_{\to n}( \mathbf{y}_{n,s} ) = {\textstyle \int} f ( \mathbf{y}_{n,s} \vert \mathbf{y}_{n-1,s} ) {b}_{} (\mathbf{y}_{n-1,s})  d \mathbf{y}_{n-1,s}.
\end{equation}
 Additionally, each agent also computes locally, the predicted messages $\alpha_{\to n} ( \mathbf{x}_{n,k}, r_{n,k} )$ for all PTs $k \in \mathcal{T}$, using the target state beliefs at the previous time step $\tilde{b}_{} ( \tilde{\mathbf{x}}_{n-1,k} )$
\begin{equation}
\alpha_{\to n} ( \tilde{\mathbf{x}}_{n,k} ) = {\textstyle \int} f_{} ( \tilde{\mathbf{x}}_{n,k} \vert \tilde{\mathbf{x}}_{n-1,k}) {b}_{}( \tilde{\mathbf{x}}_{n-1,k} ) \ d\tilde{\mathbf{x}}_{n-1,k}
\label{eq_msg_alpha}
\end{equation}
where the transition density $f_{}(\tilde{\mathbf{x}}_{n,k} \vert \tilde{\mathbf{x}}_{n-1,k})$ (Table \ref{tabel_transition_pdf}) incorporates target birth and death in addition to its kinematic model. Note that (\ref{eq_msg_phi})-(\ref{eq_msg_alpha}) correspond to the Chapman--Kolmogorov equations in the prediction step of the recursive Bayesian filters and incorporate the agent and target dynamic models.

Before the start of the message passing scheme, the agent beliefs at the current time-step $b(\mathbf{y}_{n,s})$ are initialized with the predicted beliefs $\phi_{\to n} ( \mathbf{y}_{n,s} )$, $\forall$ $s\in \mathcal{S}$. Synchronously and in parallel, the agents run the iterative message passing scheme, referred to as the outer BP loop in Algorithm \ref{alg_BP}. Each agent $s$ executes the loop $P$ times. Subsequently, we present the BP outer-loop messages in the order in which they are evaluated in Algorithm \ref{alg_BP} while also indicating the corresponding nodes and messages in the factor graph of Figure~\ref{fig_factor_graph}.

At the beginning of an outer loop, each agent broadcasts its belief $b(\mathbf{y}_{n,s})$ to its neighboring agents $\ell \in \mathcal{A}_{n,s}$ (see the arrows coming out of $\mathbf{y}_s, \mathbf{y}_{\ell}$ in Figure~\ref{fig_factor_graph}). The time subscript will be dropped for the rest of this section since all subsequent messages only involve variables at time $n$.  We shall use the term ``target'' generically, and ``PT'' when referring to a specific potential target $k$. We present the expressions of different BP messages involving agent $s$. 

The current beliefs $b(\mathbf{y}_{\ell})$ of neighboring agents $\ell \in \mathcal{A}_{n,s}$ are broadcast, and received at agent $s$. Next, agent $s$ computes the \textit{likelihood} messages $\Phi_{\ell \rightarrow s}$ (line \ref{alg_line_Phi}, Algorithm \ref{alg_BP}) using
\begin{align}
\label{eq_msg_Phi}
\Phi_{\ell \rightarrow s} ( \mathbf{y}_s ) = {\textstyle \int} f ( \mathbf{w}_{s,\ell} \vert \mathbf{y}_s, \mathbf{y}_{\ell} ) b( \mathbf{y}_{\ell}) d \mathbf{y}_{\ell}. 
\end{align}
By marginalizing over the state of agent $\ell$, the message $\Phi_{\ell \rightarrow s}$ represents the likelihood of agent $s$ for the measurement $\mathbf{w}_{s,\ell}$ taken by agent $s$ with respect to agent $\ell$. This is followed by locally computing the single-target association weights $\beta_k^s ( a_{k}^{s} =m)$ between the local measurements $\mathbf{z}_n^s$ (at agent $s$) and the target set $\mathcal{K}$. For all the PTs $k \in \mathcal{K}$ and $m \in \left\{ 0,\cdots, M^s \right\}$, these weights $\beta_k^s ( a_{k}^{s} =m)$ (line \ref{alg_line_beta} in Algorithm \ref{alg_BP} and \circled{1} in Figure \ref{fig_factor_graph}) are given as	 
\begin{align}
&\beta_k^s(a_k^s = m) = {\textstyle \int} v_k^s ( \tilde{\mathbf{x}}_{k}, a_{k}^{s}, \mathbf{y}_{s}; \mathbf{z}^{s} ) \delta_k^{s} ( \tilde{\mathbf{x}}_{k})  \theta_{k}^s ( \mathbf{y}_s ) d \tilde{\mathbf{x}}_k d\mathbf{y}_{s} =  \label{eq_msg_beta} \nonumber \\	
& \hspace{-2mm} \begin{cases}
\mfrac{\int P_D^{s}(\mathbf{x}_{k})f( \mathbf{z}_m^{s} \vert \mathbf{y}_s, \mathbf{x}_k) \delta_k^{s} ( \mathbf{x}_{k}, 1 ) \theta^{s}_{k} ( \mathbf{y}_{s} ) d\mathbf{y}_{s} d\mathbf{x}_{k} }{\lambda^{s} f_{s}^{\text{FA}} \left( \mathbf{z}_m^{s} \right)} & \hspace{-2mm} \text{if } m \neq 0 \\
1 - {\textstyle \int} P_D^{s}(\mathbf{x}_k)  \delta_k^{s} ( \mathbf{x}_{k}, 1 ) d\mathbf{x}_{k}, & \hspace{-2mm}  \text{if } m = 0
\end{cases}
\end{align}
for $k \in \mathcal{T}_s$, and $\beta_k^s (m) = 0$ for $k \notin \mathcal{T}_s$ and $\forall \ m$. At the first iteration of the outer BP-loop, we initialize the messages as $\delta_k^{s} ( \tilde{\mathbf{x}}_{k})= \alpha_{\to n} ( \tilde{\mathbf{x}}_{k} ) $ and $\theta_{k}^s ( \mathbf{y}_s ) =\phi_{\to n} ( \mathbf{y}_s )$. In other words, the association weights in the first outer loop iteration are estimated by marginalizing with respect to the predicted agent and target densities.

Next, these weights $\left\{ \beta_k^s (m) \right\}$ are used to evaluate the messages $\left\{ \eta_k^s (m) \right\}$ (line \ref{alg_line_eta} in Algorithm \ref{alg_BP} and \circled{2} in Figure \ref{fig_factor_graph}). This is achieved by a second, inner BP loop which involves message exchanges between the local association variables $\mathbf{a}^s_{n}$ and $\mathbf{b}^s_{n}$ of agent $s$ \cite{williams14_taes}. Similar to other track-oriented marginal filters such as the JPDAF~\cite{bibl:bar_shalom_PDAF2009}, the SCS-MTT filter evaluates the single--target association weights $\beta_k^s$ followed by an efficient BP evaluation of the marginal association probabilities. Additionally in SCS-MTT, the uncertainty in the position of agent $s$ is accounted for in $\beta_k^s$ by marginalizing over the message $\theta_{k}^s$.

\begin{algorithm}[t!]
	\caption{SCS-MTT outer BP-loop - in parallel $\forall$ $s\in \mathcal{S}$}
	\label{alg_BP}
	\begin{algorithmic}[1]
		\State{\textbf{Input}: Predicted beliefs $\phi_{\to n} ( \mathbf{y}_{s} )$, $\left\{ \alpha_{\to n} (\tilde{\mathbf{x}}_k) \right\}_{k \in \mathcal{T}}$}
		\State{\textbf{Initialize}: $b(\mathbf{y}_{s}) \leftarrow \phi_{\to n}(\mathbf{y}_{s})$, $\theta_k^s(\mathbf{y}_{s}) \leftarrow b(\mathbf{y}_{s}) $ and $\delta_k^{s}(\tilde{\mathbf{x}}_k) \leftarrow \alpha_{\to n}(\tilde{\mathbf{x}}_k) $, $\forall \ k \in \mathcal{T}_s$}
		\For{$p \leftarrow 1$ to $P$ } \hfill (Outer BP iterations)
		\State{Broadcast $b(\mathbf{y_{s}})$ and receive $b_{}(\mathbf{y}_{\ell})$ $\forall \ \ell \in \mathcal{A}_s$}
		\For{all $\ell \in \mathcal{A}_{s}$}
		\State{\label{alg_line_Phi}} Compute $\Phi_{\ell \rightarrow s} \left( \mathbf{y}_{s} \right)$ via (\ref{eq_msg_Phi})
		\EndFor
		\For{all $k \in \mathcal{T}_{s}$}
		\State{\label{alg_line_beta}		
			Compute $\beta_k^s\left( a_{k}^{s} \right)$ via (\ref{eq_msg_beta})}
		\State{\label{alg_line_eta} Compute $\eta_k^s \left( a_{k}^{s} \right)$ as in~\cite{williams14_taes} \hfill (Data Association)}
		\State{\label{alg_line_Lambda} Compute $\Lambda_{k}^s(\mathbf{y}_s)$ via (\ref{eq_msg_Lambda})}
		\EndFor
		\State{\label{alg_line_agt_belief} Update agent belief $b_s( \mathbf{y}_{s} )$ via (\ref{eq_ag_belief})}
		\For{all $k \in \mathcal{T}_{s}$}		
		\State{\label{alg_line_theta} Compute $\theta_k^s ( \mathbf{y}_{s} )$ via (\ref{eq_msg_theta})}		
		\EndFor				
		\For{all $k \in \mathcal{T}$}
		\State{\label{alg_line_gamma} Compute $\gamma_k^{s} ( \mathbf{x}_{k}, r_{k} )$ via (\ref{eq_msg_gamma}) if $k \in \mathcal{T}_s$}
		\State{Set $\gamma_k^{s} ( \mathbf{x}_{k}, r_{k} )=1$ if $k \notin \mathcal{T}_s$} 
		\State{\label{alg_line_trgt_belief} Network consensus to update $b( \mathbf{x}_{k}, r_{k} )$ via (\ref{eq_tg_belief}) }
		\State{\label{alg_line_delta} Compute $\delta_k^{s} (\tilde{\mathbf{x}}_k)$ via (\ref{eq_msg_delta}) if $k \in \mathcal{T}_s$ and $p\neq P$}
		\EndFor								
		\EndFor
		\State{\textbf{Return} $ b(\mathbf{y}_s)$ and $ b(\tilde{\mathbf{x}}_k)$ $\forall $ $k \in \mathcal{T}$.}
	\end{algorithmic}
\end{algorithm}
The $\eta_k^s$ messages are subsequently used to evaluate the likelihood messages $\Lambda_k^s ( \mathbf{y}_{s} )$ (line \ref{alg_line_Lambda} in Algorithm \ref{alg_BP} and \circled{3} in Figure~\ref{fig_factor_graph}), sent from the factor node $v_k^s$ of each  PT $k \in \mathcal{T}_{s}$, to the agent state node $\mathbf{y}_{s}$. 
\begin{align}		
& \Lambda_k^s ( \mathbf{y}_{s} ) = \sum\nolimits_{m = 0}^{M^{s}} {\textstyle \int} v_k^s (\tilde{\mathbf{x}}_k, m, \mathbf{y}_s; \mathbf{z}_m^s ) \eta_k^s ( m ) \delta_k^{s} ( \tilde{\mathbf{x}}_{k}) d\tilde{\mathbf{x}}_{k} \nonumber \\
&= \sum\nolimits_{ m = 1}^{M^{s}} \mfrac{\eta_k^s ( m )}{\lambda^{s}  f_{s}^{\text{FA}} ( \mathbf{z}_m^{s} )} {\textstyle \int} P_D^s (\mathbf{x}_k) \delta_k^{s} ( \mathbf{x}_{k}, 1 ) f ( \mathbf{z}_m^{s} \vert \mathbf{x}_{k}, \mathbf{y}_{s} ) d\mathbf{x}_{k} \nonumber   \\
& \qquad + \ \eta_k^s (0) \cdot \big[ 1 - {\textstyle \int}  P_D^s(\mathbf{x}_{k}) \delta_k^{s} ( \mathbf{x}_{k}, 1 ) d\mathbf{x}_{k} \big]. \label{eq_msg_Lambda}
\end{align}
The message $\Lambda_k^s ( \mathbf{y}_{s} )$ can be seen as a likelihood function for the target measurements made by sensor $s$ which also incorporates the target position uncertainty, via $\delta_k^{s} ( \tilde{\mathbf{x}}_{k})$, and association uncertainty, via $\eta_k^s (a_k^s)$. The updated belief for agent $s$ can now be evaluated in a Bayesian manner, that is, by multiplying the predicted message $\alpha_{\to n}(\mathbf{y}_s)$ (i.e., prior) with the inter-agent likelihood messages $\Phi_{\ell \rightarrow s}$ $\forall$ $\ell \in \mathcal{A}_{s}$ and agent-to-target likelihood messages $\Lambda_{k}^s$ $\forall$ $k \in \mathcal{T}_{s}$. More specifically the updated agent belief (line \ref{alg_line_agt_belief} in Algorithm \ref{alg_BP}) is given as
\begin{equation}
\label{eq_ag_belief}
b ( \mathbf{y}_s ) \propto \phi_{\to n} ( \mathbf{y}_s ) \prod\nolimits_{\ell \in \mathcal{A}_{s}} \Phi_{\ell \rightarrow s} ( \mathbf{y}_s ) \prod\nolimits_{k \in \mathcal{T}_{s}} \Lambda_k^s ( \mathbf{y}_{s} )
\end{equation}
and normalized as $\int b ( \mathbf{y}_s ) d\mathbf{y}_s = 1$ in order to represent an approximation to the agent posterior probability density. Note that the product of agent-to-target likelihood messages $\Lambda_{k}^s$ $\forall$ $k \in \mathcal{T}_{s}$ in (\ref{eq_ag_belief}) represents the probabilistic transfer of information from target tracking to agent localization. In contrast, for separate localization and MTT algorithms, there are no agent-to-target likelihood messages $\Lambda_{k}^s$ in the agent belief as probabilistic information is only passed down from the agents to the targets. Thus, the messages $\Lambda_{k}^s$ $\forall$ $k \in \mathcal{T}_{s}$ lead SCS-MTT methods to improved agent localization as compared to separate localization and MTT methods.

Next, using the updated agent belief computed in (\ref{eq_ag_belief}), the message $\theta_{k}^s \left( \mathbf{y}_{s} \right)$ (line \ref{alg_line_theta} in Algorithm \ref{alg_BP} and \circled{4} in Figure \ref{fig_factor_graph}), sent from $\mathbf{y}_{s}$ to factor node $v_k^s$, $\forall \ k \in \mathcal{T}_{s}$ is computed as
\begin{align}
\label{eq_msg_theta}
\theta_{k}^s ( \mathbf{y}_s ) \propto \phi_{\to n} ( \mathbf{y}_s ) \prod_{\ell \in \mathcal{A}_{s}} \Phi_{\ell \rightarrow s} ( \mathbf{y}_s ) \prod_{k' \in \mathcal{T}_{s} \setminus \left\{ k \right\}} \Lambda_{k'}^s ( \mathbf{y}_{s} )
\end{align}
and normalized, i.e., $\int \theta_{k}^s ( \mathbf{y}_s ) d\mathbf{y}_s =1 $. The message $\theta_{k}^s$ represents the belief in the localization of agent $s$ without the benefit of  PT $k$ (i.e., $\theta_{k}^s( \mathbf{y}_s ) \propto b ( \mathbf{y}_s ) / \Lambda_k^s( \mathbf{y}_s )$) and is referred to as the \textit{extrinsic information}~\cite{meyer16_tsipn} on agent $s$, seen by  PT $k$.

Next, for each PT $k \in \mathcal{K}$, the likelihood message $\gamma_k^{s}( \mathbf{x}_{k}, r_{k} )$ (line \ref{alg_line_gamma} in Algorithm \ref{alg_BP} and \circled{5} in Figure \ref{fig_factor_graph}) from factor node $v_k^s$ to the variable node $\tilde{\mathbf{x}}_k$ is computed as 
\begin{align}
\label{eq_msg_gamma}
& \gamma_k^{s} ( \mathbf{x}_{k}, r_{k} ) = \sum\nolimits_{m=0}^{M^{s}} {\textstyle \int} v_k^s ( \mathbf{x}_k, r_k, \mathbf{y}_s ; \mathbf{z}_m^s ) \eta_k^s ( m ) \theta_k^s ( \mathbf{y}_{s} ) d\mathbf{y}_{s} \nonumber \\
&= \begin{cases}
\sum\nolimits_{m = 1}^{M^{s}} \frac{\eta_k^s ( m ) P_D^s(\mathbf{x}_k)}{\lambda^{s} f_s^{\text{FA}} ( \mathbf{z}_m^{s} )} \int  f ( \mathbf{z}_m^{s} \vert \mathbf{x}_{k}, \mathbf{y}_{s} ) \theta_k^s ( \mathbf{y}_s ) d\mathbf{y}_s  \\
\qquad + \ \eta_k^s ( 0 ) \left( 1 - P_D^s(\mathbf{x}_k) \right), \hfill \text{ for } r_k = 1 \\
\eta_k^s (0), \hfill \text{ for } r_k = 0
\end{cases}
\end{align}
if $k \in \mathcal{T}_s$ and $\gamma_k^{s} ( \mathbf{x}_{k}, r_{k} ) = 1$ otherwise. The message $\gamma_k^{s}$ represents a likelihood function for  PT $k$ with respect to the measurements made by agent $s$. It accounts for the uncertainty in the position of agent $s$, via $\theta_k^s$, and the uncertainty in the association of the measurements to  PT $k$, via $\eta_k^s ( a_k^s )$. Given the messages $\gamma_k^{s}$ from all the agents $s \in \mathcal{S}$, we update the target beliefs (line \ref{alg_line_trgt_belief} in Algorithm \ref{alg_BP}) in a decentralized way as
\begin{equation}
\label{eq_tg_belief}
b ( \mathbf{x}_{k}, r_k ) \propto \alpha_{\to n} ( \mathbf{x}_{k}, r_k) \prod\nolimits_{s \in \mathcal{S}} \gamma_k^{s} ( \mathbf{x}_{k}, r_k ).
\end{equation}
Note that (\ref{eq_tg_belief}) involves network consensus, i.e., agent $s$ obtains $b ( \tilde{\mathbf{x}}_{k})$ even if PT $k$ is not observed by agent $s$. Network consensus is implementation dependent, i.e., it depends on the  representation of the messages as discrete particle  sets or Gaussian mixtures. In Section~\ref{sec_tg_belief_gm} and Section~\ref{sec_dist_gauss}, we provide algorithms for GM and single Gaussian implementations. Furthermore, the target belief is normalized as $\sum_{r_k \in \{0,1\}} \int  b ( \mathbf{x}_{k}, r_k )d\mathbf{x}_k = 1$. Note that (\ref{eq_tg_belief}) is reminiscent of the Bayesian multi-sensor update of a target with prior density $\alpha_{\to n}(\tilde{\mathbf{x}}_{k})$ and sensor likelihood functions $\gamma_k^{s}(\tilde{\mathbf{x}}_{k})$. 

Finally, for $k\in \mathcal{T}_s$, we compute the $\delta_k^{s} ( \mathbf{x}_{k}, r_{k} )$ messages (\circled{6} in Figure~\ref{fig_factor_graph}, sent from $\tilde{\mathbf{x}}_k$ to the factor node $v_k^s$) as
\begin{equation}
\label{eq_msg_delta}
\delta_k^{s} ( \mathbf{x}_{k}, r_k) \propto \alpha_{\to n}( \mathbf{x}_{k}, r_k ) \prod\nolimits_{s' \in \mathcal{S} \setminus \left\{ s \right\}} \gamma_k^{s'} ( \mathbf{x}_{k}, r_k ).
\end{equation}
The message $\delta_k^{s}$ can be seen as the \textit{extrinsic information} on the state of  PT $k$ as seen by agent $s$ ($\delta_k^{s}( \mathbf{x}_{k}, r_k) \propto b ( \mathbf{x}_{k}, r_k ) / \gamma_k^{s}( \mathbf{x}_{k}, r_k)$). Note that (\ref{eq_msg_delta}) can be efficiently evaluated (or approximated) from the belief $b ( \tilde{\mathbf{x}}_k )$, hence avoiding additional network-consensus processes, as presented in Section~\ref{sec_tg_belief_gm} and Section~\ref{sec_dist_gauss} for the case of Gaussian mixture and single Gaussian implementations. Furthermore, the message $\delta_k^{s} ( \mathbf{x}_{k}, r_k)$ is only computed if $p\neq P$. At the end of the outer iterations, i.e., when $p = P$, the agent $b(\mathbf{y}_{s})$ and target $b(\mathbf{x}_{k}, r_k)$ beliefs represent estimates of their marginal probability densities for the $n$-th time step and are used as inputs for the next time step. 

Note the similarities between Algorithm \ref{alg_BP} and that of \cite{meyer16_tsipn}, with the exception that Algorithm \ref{alg_BP} also considers association uncertainty for target measurements which requires the computation of single-target association weights $\beta_k^s$ and the execution of the inner-BP loop, as done in~\cite{meyer17_tsp}. The inner-BP loop, as shown in~\cite{williams14_taes}, converges to a unique fixed point. In contrast, the convergence of the overall message passing scheme (outer and inner BP loops) is not guaranteed due to the presence of loops in the factor graph of Figure~\ref{fig_factor_graph}. This can lead to overconfident beliefs, as also shown in~\cite{meyer16_tsipn}, which in practice are countered by performing the outer-BP loop only once per time-step (i.e., $P=1$). The proposed message passing scheme with $P=1$ is shown in Section~\ref{sec_sim} and in~\cite{meyer16_tsipn} to accurately localize agents and targets. 

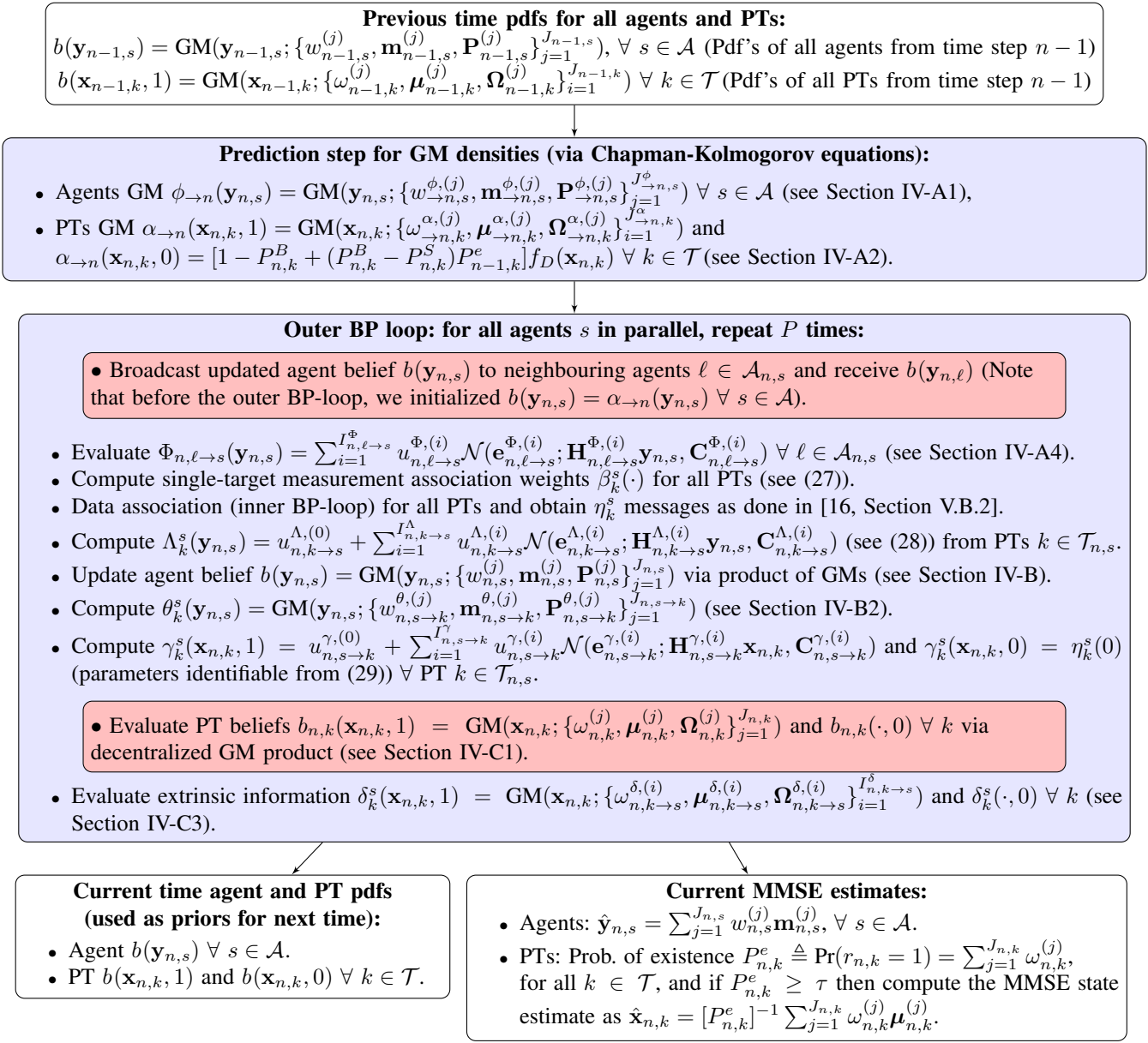
\begin{figure*}[!t]
	\centering
	\begin{tikzpicture}[auto]
	\node [startstop] (init) {\textbf{Previous time pdfs for all agents and PTs:}\\$b_{}(\mathbf{y}_{n-1,s})= \text{GM} ( \mathbf{y}_{n-1,s}; \{ {w}_{n-1,s}^{(j)}, \mathbf{m}_{n-1,s}^{(j)}, \mathbf{P}_{n-1,s}^{(j)} \}_{j=1}^{J_{n-1,s}})$, $\forall$ $s\in \mathcal{A}_{}$ (Pdf's of all agents from time step $n-1$) \\ ${b}_{}(\mathbf{x}_{n-1,k},1) = \text{GM} (  \mathbf{x}_{n-1,k}; \{ \omega_{n-1,k}^{(j)}, \bm{\mu}_{n-1,k}^{(j)}, \bm{\Omega}_{n-1,k}^{(j)} \}_{i=1}^{J_{n-1,k}} )$ $\forall$ $k\in \mathcal{T}_{}$ (Pdf's of all PTs from time step $n-1$)	};
	\node [block, text width=17.5cm, below=0.5cm of init] (predict) {\textbf{Prediction step for GM densities (via Chapman-Kolmogorov equations):} \\ 
		\begin{itemize}
		\item Agents GM $\phi_{\rightarrow n}(\mathbf{y}_{n,s}) = \text{GM} ( \mathbf{y}_{n,s}; \{ {w}_{\to n,s}^{\phi, (j)}, \mathbf{m}_{\to n,s}^{\phi, (j)}, \mathbf{P}_{\to n,s}^{\phi, (j)} \}_{j=1}^{J^{\phi}_{\to n,s}} )$ $\forall$ $s \in \mathcal{A}_{}$ (see Section~\ref{sec_phi_GM}),  
		\item PTs GM $\alpha_{\to n}(\mathbf{x}_{n,k},1)= \text{GM} ( \mathbf{x}_{n,k}; \{ {\omega}_{\to n,k}^{\alpha,(j)}, \bm{\mu}_{\to n,k}^{\alpha,(j)}, \bm{\Omega}_{\to n,k}^{\alpha, (j)} \}_{i=1}^{J^{\alpha}_{\to n,k}}) $ and \\$\alpha_{\to n} (\mathbf{x}_{n,k},0) = [1 - P_{n,k}^B + (P_{n,k}^B - P_{n,k}^S) P^e_{n-1,k}] f_D(\mathbf{x}_{n,k})$  $\forall$ $k\in \mathcal{T}_{}$ (see Section~\ref{sec_alpha_GM}). 
		\end{itemize} };
	\node [block, text width=17cm, below=0.5cm of predict] (Outerloop1) {\textbf{Outer BP loop: for all agents $s$ in parallel, repeat $P$ times:}
		\begin{itemize}
		\vspace*{1.3cm}
		\item Evaluate $\Phi_{n,\ell \rightarrow s} ( \mathbf{y}_{n,s} ) = \sum\nolimits_{i=1}^{I^{\Phi}_{n,\ell \to s}} u_{n,\ell \to s}^{\Phi,(i)} \mathcal{N} ( \mathbf{e}_{n,\ell \to s}^{\Phi,(i)}; \mathbf{H}_{n,\ell \to s}^{\Phi,(i)}\mathbf{y}_{n,s}, \mathbf{C}_{n,\ell \to s}^{\Phi,(i)} )$ $\forall$ $\ell \in \mathcal{A}_{n,s}$ (see Section~\ref{sec_Phi_Lambda_GM}).
		\item Compute single-target measurement association weights $\beta_k^s(\cdot)$ for all PTs (see (\ref{eq_msg_beta_gmm})).
		\item Data association (inner BP-loop) for all PTs and obtain $\eta_{k}^s$ messages as done in \cite[Section V.B.2]{meyer17_tsp}. 
		\item Compute $\Lambda_k^s(\mathbf{y}_{n,s}) = u_{n,k\to s}^{\Lambda,(0)}+\sum_{i=1}^{I_{n,k\to s}^{\Lambda}}  u_{n,k\to s}^{\Lambda,(i)} \mathcal{N}(\mathbf{e}_{n,k\to s}^{\Lambda,(i)}; \mathbf{H}_{n,k\to s}^{\Lambda,(i)}\mathbf{y}_{n,s}, \mathbf{C}_{n,k\to s}^{\Lambda,(i)}) $ (see (\ref{eq_msg_Lambda_gmm})) from PTs $k \in \mathcal{T}_{n,s}$.
		\item Update agent belief $b_{}(\mathbf{y}_{n,s}) =  \text{GM} ( \mathbf{y}_{n,s}; \{ {w}_{n,s}^{(j)}, \mathbf{m}_{n,s}^{(j)}, \mathbf{P}_{n,s}^{(j)} \}_{ j=1}^{J_{n,s}} )$ via product of GMs (see Section~\ref{sec_ag_belief_gibbs}). 
		\item Compute $\theta_k^s(\mathbf{y}_{n,s}) = \text{GM} ( \mathbf{y}_{n,s}; \{ {w}_{n,s\to k}^{\theta,(j)}, \mathbf{m}_{n,s\to k}^{\theta,(j)}, \mathbf{P}_{n,s\to k}^{\theta,(j)} \}_{ j=1 }^{J_{n,s\to k}})$ (see Section~\ref{sec_theta_GMM}). 
		\item Compute $\gamma_k^s(\mathbf{x}_{n,k},1) = u_{n,s\to {\tiny k}}^{\gamma,(0)}+\sum_{i=1}^{I^{\gamma}_{n,s\to k}}  u_{n,s\to k}^{\gamma,(i)} \mathcal{N}(\mathbf{e}_{n,s\to k}^{\gamma,(i)}; \mathbf{H}_{n,s\to k}^{\gamma,(i)}\mathbf{x}_{n,k}, \mathbf{C}_{n,s\to k}^{\gamma,(i)}) $ and $\gamma_k^s(\mathbf{x}_{n,k},0)=\eta_k^s (0)$ (parameters identifiable from (\ref{eq_msg_gamma_gmm})) $\forall$ PT $k \in \mathcal{T}_{n,s}$.
		\vspace{1.35cm}
		\item Evaluate extrinsic information $\delta_k^s(\mathbf{x}_{n,k},1) = \text{GM} ( \mathbf{x}_{n,k}; \{ {\omega}_{n,k\to s}^{\delta,(i)}, \bm{\mu}_{n,k\to s}^{\delta,(i)}, \bm{\Omega}_{n,k\to s}^{\delta,(i)} \}_{ i=1 }^{I^{\delta}_{n,k\to s}})$ and $\delta_k^s(\cdot,0)$ $\forall$ $k$ (see Section~\ref{sec_delta_GMM}).
		\end{itemize}	
	};
	\node[blockcast, align=left, anchor=west,text width=15cm, below=1.1cm of predict] (Broadcast2) { $\bullet$ Broadcast updated agent belief $b_{}(\mathbf{y}_{n,s})$ to neighbouring agents $\ell \in \mathcal{A}_{n,s}$ and receive $b_{}(\mathbf{y}_{n,\ell})$ (Note that before the outer BP-loop, we initialized $b_{}(\mathbf{y}_{n,s}) = \alpha_{\to n}(\mathbf{y}_{n,s})$ $\forall$ $s\in \mathcal{A}$).};
	\node[blockcast, align=left, anchor=west,text width=15cm, below=-2.2cm of Outerloop1] (Ptbelief) { $\bullet$ Evaluate PT beliefs $b_{n,k}(\mathbf{x}_{n,k},1) = \text{GM} ( \mathbf{x}_{n,k}; \{ {\omega}_{n,k}^{(j)}, \bm{\mu}_{n,k}^{(j)}, \bm{\Omega}_{n,k}^{(j)} \}_{ j=1 }^{J_{n,k}})$ and $b_{n,k}(\cdot,0)$ $\forall$ $k$ via decentralized GM product (see Section~\ref{sec_dist_gmm_hogwild}).};
	\node[startstop, text width=6.5cm, below left=0.5cm and -6.7cm of Outerloop1] (stop) {\textbf{Current time agent and PT pdfs \\(used as priors for next time):} 	
		\begin{itemize}
		\item Agent ${b}_{}(\mathbf{y}_{n,s})$ $\forall$ $s\in \mathcal{A}_{}$. 
		\item PT ${b}_{}(\mathbf{x}_{n,k},1)$ and ${b}_{}(\mathbf{x}_{n,k},0)$ $\forall$ $k\in \mathcal{T}$.
		\end{itemize} };
	\node[startstop, text width=10cm,below right=0.5cm and -10.3cm of Outerloop1] (Estimates) {\textbf{Current MMSE estimates:} 
		\begin{itemize}
		\item Agents: $\hat{\mathbf{y}}_{n,s} = \sum_{j=1}^{J_{n,s}} {w}_{n,s}^{(j)}\mathbf{m}_{n,s}^{(j)}$, $ \forall$ $s\in \mathcal{A}_{}$.
		\item PTs: Prob. of existence $ P^e_{n,k} \triangleq \text{Pr}(r_{n,k}=1) = \sum_{j=1}^{J_{n,k}} {\omega}_{n,k}^{(j)}$, \\
		for all $k\in \mathcal{T}$, and if $P^e_{n,k} \geq \tau$ then compute the MMSE state estimate as $\hat{\mathbf{x}}_{n,k} = [P^e_{n,k}]^{-1} \sum_{j=1}^{J_{n,k}}{\omega}_{n,k}^{(j)} \bm{\mu}_{n,k}^{(j)}$.
		\end{itemize}};
	\path [line] (init) -- (predict);
	\path [line] (predict) -- (Outerloop1);
	\path [line] (Outerloop1) -- (Estimates);	
	\path [line] (Outerloop1) -- (stop);		
	\end{tikzpicture}
	\caption{GM processing flowchart of the DGM-SCS-MTT filter at time $n$. Decentralized and local computations are colored in red and blue respectively. The various messages are given in generic GM form while the expressions of their parameters are given in Section~\ref{sec:gm_messages}.}\label{fig_flowchart}
\end{figure*}

\subsection{Agent and target inference} \label{sec:agtg_infer}

An MMSE estimate of the state of agent $s$ is obtained via $\hat{\mathbf{y}}_{n,s}= \int \mathbf{y}_s b(\mathbf{y}_s) d\mathbf{y}_s $, where $b(\mathbf{y}_s)$ is the agent marginal density estimated via Algorithm~\ref{alg_BP}. Based on the estimated marginal density $b(\mathbf{x}_k, r_k)$, PT $k$ is declared a valid target if the estimated probability of existence $P_{n,k}^e \triangleq \text{Pr}(r_{n,k}=1) =\int b(\mathbf{x}_k, 1) d\mathbf{x}_k $ is greater than a specified threshold $P_{n,k}^e \geq \tau$ (in this work $\tau=0.5$). Subsequently an MMSE state estimate is given as $\hat{\mathbf{x}}_{n,k}= \frac{1}{P_{n,k}^e}\int  \mathbf{x}_k b(\mathbf{x}_k, 1) d\mathbf{x}_k$.

\section{Decentralized Gaussian Mixture SCS-MTT filter} \label{sec_gmm}
In this section, we present the Gaussian Mixture (GM) implementation of the messages of Section~\ref{sec_bp_filter}. We denote a Gaussian pdf over $\mathbf{x}\in \mathbb{R}^d$, with mean $\mathbf{m}$ and covariance matrix $\mathbf{P}$ as $\mathcal{N} ( \mathbf{x}; \mathbf{m}, \mathbf{P} )$. A GM density function $\sum_{j=1}^J w^{(j)} \mathcal{N} ( \mathbf{x}; \mathbf{m}^{(j)}, \mathbf{P}^{(j)} )$ is compactly denoted as $\text{GM} \big( \mathbf{x}; \{ (w^{(j)}, \mathbf{m}^{(j)}, \mathbf{P}^{(j)} ) \}_{j=1}^J \big)$. Except for likelihood messages, GM messages are normalized $\sum_{j=1}^J w^{(j)} = 1$ for agents while for targets $\sum_{j=1}^J w^{(j)} \leq 1$. Throughout this section, we employ the following GM assumptions:  
\begin{itemize}
    \item[G1] The agent dynamic model is Gaussian with transition kernel $f(\mathbf{y}_{n,s} \vert \mathbf{y}_{n-1,s}) = \mathcal{N}(\mathbf{y}_{n,s}; \mathbf{A}_{ n,s} \mathbf{y}_{n-1,s}, \mathbf{Q}_{n,s})$. 
    \item[G2] The target dynamic model {(Table \ref{tabel_transition_pdf})} involves a constant probability of survival $P_{n,k}^S(\mathbf{x}) = P_{n,k}^S$, a dynamic model $f(\mathbf{x}_{n,k}, 1 \vert \mathbf{x}_{n-1,k}, 1) = P_{n,k}^S \mathcal{N} (\mathbf{x}_{n,k}; \mathbf{B}_{n,k} \mathbf{x}_{n-1,k}, \bm{\Sigma}_{n,k})$. We also assume a GM birth density $f_b(\mathbf{x}_{n,k}) = \text{GM} ( \mathbf{x}_{n,k}; {\{ (\omega_{n,k}^{B,(j)}, \bm{\mu}_{n,k}^{B,(j)}, \bm{\Omega}_{n,k}^{B,(j)} )\}_{j=1}^{J^B_{n,k}} } )$ with probability of birth $P_{n,k}^B =\sum_{j=1}^{J^B_{n,k}} { \omega_{n,k}^{B,(j)}} $.
    \item[G3] The inter-agent measurement model of agent $s$ measuring agent $\ell$ is linear with Gaussian noise: $f(\mathbf{w}_{s,\ell} \vert  \mathbf{y}_s, \mathbf{y}_{\ell}) = \mathcal{N} \left(\mathbf{w}_{s,\ell}; \mathbf{D}_s\mathbf{y}_s + \mathbf{F}_{\ell} \mathbf{y}_{\ell}, \mathbf{W}_s \right)$.
    \item[G4] The target measurement model of agent $s$ is linear with Gaussian noise: $f(\mathbf{z} \vert  \mathbf{y}_s, \mathbf{x}_k) = \mathcal{N}(\mathbf{z}; \mathbf{G}_s \mathbf{y}_s + \mathbf{E}_s \mathbf{x}_k, \mathbf{R}_s)$, and a constant probability of detection $p_{n,s}^{D}(\mathbf{x}) = P_{n,s}^D$.
    \item[G5] The initial marginal densities of agents and PTs are assumed GM. 
\end{itemize}

The constant probability of survival and of detection is a common requirement in GM implementations of MTT filters (e.g., \cite{vo06_tsp,bibl:bn_vo_labeledRFS_TSP2014}). Note that the proposed GM-SCS-MTT filter can easily accommodate GM dynamic kernels for both agents (G1) and targets (G2), and GM likelihood functions for both inter-agent (G3) and target (G4) measurements. For compactness, we present the GM expressions for the BP messages of Algorithm~\ref{alg_BP} under assumptions G1-G5, which, as we will show further, lead to the following generic GM expressions, for the agent and PT beliefs
\begin{align}
	b(\mathbf{y}_{n,s}) &=  \sum\nolimits_{j=1}^{J_{n,s}} w_{n,s}^{(j)} \mathcal{N}(\mathbf{y}_{n,s}; \mathbf{m}^{(j)}_{n,s}, \mathbf{P}^{(j)}_{n,s}), \label{eq:ag_gmm_gen} \\
	b(\mathbf{x}_{n,k},1) &= \sum\nolimits_{j=1}^{J_{n,k}} \omega_{n,k}^{(j)} \mathcal{N}(\mathbf{x}_{n,k}; \bm{\mu}^{(j)}_{n,k}, \bm{\Omega}^{(j)}_{n,k}), \label{eq:tg_gmm_gen}
\end{align}
and for the extrinsic information messages
\begin{align}
\theta_{n,k}^s(\mathbf{y}_{n,s}) &= \; \; \mathclap{\sum_{j=1}^{J^{\theta}_{n,s\to k}}} \ \ \ w_{n,s\to k}^{\theta, (j)} \mathcal{N}(\mathbf{y}_{n,s}; \mathbf{m}^{\theta,(j)}_{n,s\to k}, \mathbf{P}^{\theta,(j)}_{n,s\to k}), \label{eq:ag_ex_gmm_gen} \\
\delta_{n, k}^s(\mathbf{x}_{n,k},1) &= \ \ \mathclap{\sum_{j=1}^{J^{\delta}_{n,k\to s}}} \ \ \ \omega_{n,k\to s}^{\delta,(j)} \mathcal{N}(\mathbf{x}_{n,k}; \bm{\mu}^{\delta,(j)}_{n,k\to s}, \bm{\Omega}^{\delta,(j)}_{n,k\to s}). \label{eq:tg_ex_gmm_gen}
\end{align}
\begin{rem}
In practice, as well as in Section \ref{sec_sim}, the nonlinear measurement models are often linearized (for example, using the extended Kalman filter \cite[Ch. 2.1]{bibl:beyondKalman}).
\end{rem}
Such generic forms for all GM messages are shown in the flowchart of Figure \ref{fig_flowchart} while detailed expressions for the GM parameters are presented in the following. The properties of Gaussian functions~\cite[Ch. 3.8]{bibl:beyondKalman} and G1-G4 allow the derivation of closed form GM expressions for the GM-SCS-MTT messages. In Section \ref{sec:gm_messages}, the GM parameters of the prediction and likelihood messages are given. The computation of the GM beliefs (\ref{eq:ag_gmm_gen})-(\ref{eq:tg_gmm_gen}) and the extrinsic information (\ref{eq:ag_ex_gmm_gen})-(\ref{eq:tg_ex_gmm_gen}) requires the product of several GM terms. Exact computation of these is computationally prohibitive and incurs a high communication cost. Therefore in Section~\ref{sec_ag_belief_gibbs}, we propose a centralized and efficient algorithm to select high-weight Gaussian components from the GM product based on Gibbs sampling~\cite{nonparametricbp10}. In Section~\ref{sec_tg_belief_gm} a decentralized Gibbs algorithm is proposed for efficiently evaluating the target beliefs. The special case of this algorithm for a single Gaussian implementation is discussed in Section~\ref{sec_dist_gauss}.

\subsection{GM prediction and likelihood messages} \label{sec:gm_messages}

\subsubsection{Agent Prediction Messages} \label{sec_phi_GM}

We start with the belief ${b}(\mathbf{y}_{n-1,s})$ of agent $s$ computed at the previous time $n-1$ and with parameters similar to (\ref{eq:ag_gmm_gen}). Assuming G1 and substituting the GM representation of ${b}(\mathbf{y}_{n-1,s})$ in (\ref{eq_msg_phi}), we obtain the agent predicted message $\phi_{\to n}( \mathbf{y}_{n,s}) = \text{GM} \big( \mathbf{y}_{n,s}; \{ ({w}_{\to n, s}^{\phi,(j)}, \mathbf{m}_{\to n, s}^{\phi,(j)}, \mathbf{P}_{\to n,s}^{\phi,(j)} )\}_{j=1}^{J^{\phi}_{\to n, s}} \big)$ with $J^{\phi}_{\to n, s}= J^{}_{n-1, s}$ Gaussian components with parameters given in Table~\ref{table_gmm_predict}. As seen in Figure~\ref{fig_flowchart} and discussed in Section~\ref{sec_bp_filter}, before the BP iterations begin, the current agent belief $b_{}(\mathbf{y}_{n,s})$ is initialized with $\phi_{\rightarrow n} ( \mathbf{y}_{n,s} )$. Also, we initialize $\theta_{n,k}^s ( \mathbf{y}_s )$ with $\phi_{\rightarrow n} ( \mathbf{y}_{n,s} )$.

\subsubsection{Target Prediction Messages} \label{sec_alpha_GM}
Similarly, assuming a GM belief such as (\ref{eq:tg_gmm_gen}) for PT $k$ at $n-1$, under assumption G2 and from (\ref{eq_msg_alpha}) we obtain the predicted GM message $\alpha_{\to n}( \mathbf{x}_{n,k}, 1) = \text{GM} \big( \mathbf{x}_{n,k}; \{ ({\omega}_{\to n,k}^{\alpha, (j)}, \bm{\mu}_{\to n,k}^{\alpha, (j)}, \bm{\Omega}_{\to n,k}^{\alpha, (j)}) \}_{j=1}^{J_{\to n,k}^{\alpha}} \big)$. The $J_{ \to n,k}^{\alpha} = J_{n-1, k} + J_{n,k}^B$ Gaussian components of $\alpha_{\to n}( \mathbf{x}_{n,k}, 1)$ are the union of surviving and birthed tracks,  $$\underbrace{\{ ({\omega}_{\to n,k}^{S,(j)}, \bm{\mu}_{\to n,k}^{S,(j)}, \bm{\Omega}_{\to n,k}^{S,(j)}) \}_{j}}_{J_{n-1,k} \text{ surviving components}} \bigcup \underbrace{\{ ({\omega}_{\to n,k}^{B,(j)}, \bm{\mu}_{\to n,k}^{B,(j)}, \bm{\Omega}_{\to n,k}^{B,(j)}) \}_{j}}_{J_{n,k}^B \text{ new birthed components}}$$ where the component parameters are given in Table~\ref{table_gmm_predict}. Similarly, $\alpha_{\to n} (\mathbf{x}_{n,k},0) = [1 - P_{n,k}^B + (P_{n,k}^B - P_{n,k}^S) P^e_{n-1,k}] f_D(\mathbf{x}_{n,k})$. Also, we initialize $\delta_{n,k}^{s} ( \tilde{\mathbf{x}}_{n,k})= \alpha_{\to n} ( \tilde{\mathbf{x}}_{n,k} ) $. Henceforth, we drop the time index $n$ since all the following messages correspond to the current time instant.

\subsubsection{Single-target association weights} \label{sec_beta_GM}
Using the generic GM representations for $\theta_k^s$ (\ref{eq:ag_ex_gmm_gen}) and $\delta_k^s$ (\ref{eq:tg_ex_gmm_gen}), the single-target association weight $\beta_k^s(m)$ in (\ref{eq_msg_beta}), for $m\in [1:M_{n}^s]$ becomes
\begin{equation}
\nonumber \label{eq_msg_beta_gmm}
\beta_k^s(m)\;\, \mathclap{=} \hspace{7mm} \mathclap{\sum\nolimits_{j=1}^{J^{\theta}_{s\to k}}} \hspace{5mm} \sum\nolimits_{i=1}^{J^{\delta}_{k \to s}} \mfrac{P_D^s {\omega}_{k\to s}^{\delta, (i)} {w}_{s\to k}^{\theta, (j)}}{\lambda^s f_{s}^{\text{FA}}(\mathbf{z}_m^s)} \mathcal{N}(\mathbf{z}_m^s; \mathbf{m}_{s, k}^{(i,j)}, \mathbf{P}_{s,k}^{(i,j)}),
\end{equation}
where
\begin{align}
\mathbf{m}_{s,k}^{(i,j)} &= \mathbf{E}_s \bm{\mu}_{k \to s}^{\delta, (i)} + \mathbf{G}_s \mathbf{m}_{s \to k}^{\theta, (j)} \nonumber\\
\mathbf{P}_{s,k}^{(i,j)} &= \mathbf{R}_s + \mathbf{E}_s \bm{\Omega}_{k\to s}^{\delta, (i)} \mathbf{E}_s^T + \mathbf{G}_s \mathbf{P}_{s\to k}^{\theta, (j)} \mathbf{G}_s^T. \nonumber 
\end{align}
For $m=0$, $\beta_k^s(0) = 1-P_D^s \sum_{i=1}^{J_{k\to s}^{\delta}} {\omega}_{k\to s}^{\delta,(i)}$. These weights are then used to compute the messages $\{ \eta_k^s (m) \}$ using the inner BP loop \cite{williams14_taes}. The $\eta_k^s$ messages are subsequently used in the computations of the following likelihood messages.
\begin{table*}[t!]
	
	\begin{subtable}[h]{0.99\textwidth}
		\centering
		\begin{tabular}{|c|c|c|c|}
			\hline
			\textbf{GM Message} &  \textbf{Weights} & \textbf{Means} & \textbf{Covariance Matrices} \\
			\hline
			
			$\phi_{\rightarrow n} (\mathbf{y}_{n,s})$ & $w_{\to n,s}^{\phi,(j)} =w_{n-1,s}^{(j)}$ & $\mathbf{m}_{\to n,s}^{\phi,(j)} = \mathbf{A}_{n,s} \mathbf{m}_{n-1,s}^{(j)}$ & $\mathbf{P}_{\to n,s}^{\phi,(j)} = \mathbf{Q}_{n,s} + \mathbf{A}_{n,s} \mathbf{P}_{n-1,s}^{(j)} \mathbf{A}_{n,s}^T$ \\
			\hline
			
			\multirow{2}{*}{$\alpha_{\to n} (\mathbf{x}_{n,k},1)$}  & $\omega_{\to n,k}^{S,(j)} = P_{n,k}^S\omega_{n-1,k}^{(j)}$ & $\bm{\mu}_{\to n,k}^{S, (j)} =\mathbf{B}_{n,k} \bm{\mu}_{n-1,k}^{(j)}$ & $\bm{\Omega}_{\to n,s}^{S,(j)} = \bm{\Sigma}_{n,k} + \mathbf{B}_{n,k} \bm{\Omega}_{n-1,k}^{(j)} \mathbf{B}_{n,k}^T$ \\
			\cline{2-4}
			
			& $\omega_{\to n,k}^{B,(j)} = \omega_{n,k}^{B,(j)}(1-P^e_{n-1,k})$ & $\bm{\mu}_{\to n,k}^{B,(j)} =  \bm{\mu}_{n,k}^{B,(j)}$ & $ \bm{\Omega}_{\to n,k}^{B,(j)} = \bm{\Omega}_{n,k}^{B,(j)}$ \\
			\hline		    
			
		\end{tabular}
		\vspace{-0.2cm}
		\caption{GM parameters of prediction messages for agents (\ref{eq_msg_phi}) and PTs (\ref{eq_msg_alpha}).}
		\label{table_gmm_predict}
	\end{subtable}

	\vspace{0.05 in}
	\begin{subtable}[h]{0.99\textwidth}
		\centering
		\tabcolsep=0.1cm		
		\begin{tabular}{|c|c|c|c|c|}
			\hline
			\textbf{Message} & \textbf{Weights} & \textbf{Residuals} & \textbf{Obs. matrix} & \textbf{Covariance Matrices} \\
			\hline
			$\Phi_{\ell \rightarrow s} ( \mathbf{y}_s )$ & $u_{\ell \to s}^{\Phi, (i)} = w_{\ell}^{(i)}$ & $\mathbf{e}_{\ell \to s}^{\Phi, (i)} = \mathbf{w}_{s, \ell} - \mathbf{F}_{\ell} \mathbf{m}_{\ell}^{(i)}$ &$\mathbf{H}_{\ell \to s}^{\Phi,(i)} = \mathbf{D}_s$ & $\mathbf{C}_{\ell \to s}^{\Phi,(i)} = \mathbf{W}_s + \mathbf{F}_{\ell} \mathbf{P}_{\ell}^{(i)} \mathbf{F}_{\ell}^T$ \\
			\hline
			$\Lambda_k^s \left( \mathbf{y}_{s} \right)$ &  \begin{tabular}{c} $u_{k \to s}^{\Lambda, (m,j)} = \mfrac{\eta_{k}^s ( m )  P_D^s {\omega}_{k\to s}^{\delta, (j)}}{\lambda^{s} f_{s}^{\text{FA}} ( \mathbf{z}_m^{s} )}$ \\ $u_{k \to s}^{\Lambda, (0)} =\eta_k^s(0)[1-P_D^s\sum_{j} \omega_{k\to s}^{\delta, (j)}]$ \end{tabular} & $\mathbf{e}_{k\to s}^{\Lambda, (m,j)} = \mathbf{z}_m^{s} - \mathbf{E}_s \bm{\mu}_{k\to s}^{\delta,(j)}$ & $\mathbf{H}_{k \to s}^{\Lambda,(m,j)} = \mathbf{G}_s$ & \begin{tabular}{c} $\mathbf{C}_{k \to s}^{\Lambda,(m,j)} = \mathbf{R}_s + \mathbf{E}_s \bm{\Omega}_{k\to s}^{\delta, (j)} \mathbf{E}_s^T$ \\ (independent of $m$) \end{tabular} \\
			\hline
			$\gamma_k^{s} ( \mathbf{x}_{k}, 1 )$ & \begin{tabular}{c} $u_{s\to k}^{\gamma, (m,j)} = \mfrac{\eta_k^s ( m ) P_D^s {w}_{s\to k}^{\theta, (j)} }{ \lambda^{s} f_s^{\text{FA}} ( \mathbf{z}_m^{s} )}$ \\$u_{s\to k}^{\gamma, (0)} = \eta_k^s(m)[1-P_D^{s}] $\end{tabular}  & $\mathbf{e}_{s\to k}^{\gamma, (m,j)} = \mathbf{z}_m^{s} - \mathbf{G}_s \mathbf{m}_{s\to k}^{\theta,(j)}$ &$\mathbf{H}_{s \to k}^{\gamma,(m,j)} = \mathbf{E}_s$ & \begin{tabular}{c} $\mathbf{C}_{s\to k}^{\gamma,(m,j)} = \mathbf{R}_s + \mathbf{G}_s \mathbf{P}_{s\to k}^{\theta,(j)} \mathbf{G}_s^T$ \\ (independent of $m$) \end{tabular} \\
			\hline
		\end{tabular}
		\vspace{-0.2cm}
		\caption{GM parameters of likelihood messages for agents (\ref{eq_msg_Phi}), (\ref{eq_msg_Lambda}) and PTs (\ref{eq_msg_gamma}).}
		\label{table_gmm_likelihood}
	\end{subtable}
	\vspace{-0.1cm}
	\caption{Gaussian mixture parameters for the message passing scheme of Algorithm \ref{alg_BP}.}
	\vspace{-0.5cm}
	\label{table_gmm}
\end{table*}

\subsubsection{Agent likelihood messages} \label{sec_Phi_Lambda_GM}
During an outer-BP loop, using the generic GM form for the belief of agent $\ell$ (\ref{eq:ag_gmm_gen}) and under G3, the likelihood message $\Phi_{\ell \rightarrow s} ( \mathbf{y}_s )$ in (\ref{eq_msg_Phi}) becomes
\begin{align}
\label{eq_msg_Phi_gmm}
\Phi_{\ell \rightarrow s} ( \mathbf{y}_s )  = \hspace{4mm} \mathclap{\sum\nolimits_{i=1}^{ I_{\ell \to s}^{\Phi}}} \hspace{6mm} {u}_{\ell \to s}^{\Phi, (i)} \mathcal{N} \big( \mathbf{e}_{\ell \to s}^{\Phi, (i)}; \mathbf{H}_{\ell \to s}^{\Phi, (i)}\mathbf{y}_s, \mathbf{C}_{\ell \to s}^{\Phi, (i)} \big)
\end{align}
where $I_{\ell \to s}^{\Phi} = J_{\ell}$, the weights ${u}_{\ell \to s}^{\Phi, (i)}$, residuals $\mathbf{e}_{\ell \to s}^{\Phi, (i)}$, observation matrices $\mathbf{H}_{\ell \to s}^{\Phi, (i)}$ and covariance matrices $\mathbf{C}_{\ell \to s}^{\Phi, (i)}$ are given in Table~\ref{table_gmm_likelihood}. Similarly, using G4 and the GM expression for $\delta_k^s$ of (\ref{eq:tg_ex_gmm_gen}), the message $\Lambda_k^s ( \mathbf{y}_{s} )$ in (\ref{eq_msg_Lambda}) becomes
\begin{align} 
&\Lambda_k^s ( \mathbf{y}_{s} )  = u_{k\to s}^{\Lambda,(0)} + \label{eq_msg_Lambda_gmm} \\  &\sum\nolimits_{m=1}^{M^{s}} \sum\nolimits_{j=1}^{J_{k, s}^{\delta}}  u_{k\to s}^{\Lambda,(m,j)} \mathcal{N}(\mathbf{e}_{k\to s}^{\Lambda,(m,j)}; \mathbf{H}_{k \to s}^{\Lambda, (m,j)} \mathbf{y}_{s}, \mathbf{C}_{k\to s}^{\Lambda, (m,j)})\nonumber 
\end{align}
with parameters given in Table~\ref{table_gmm_likelihood}.

\subsubsection{Target likelihood messages} \label{sec_gamma_GM}
From (\ref{eq_msg_gamma}) and assuming G4 and the GM form (\ref{eq:ag_ex_gmm_gen}) for $\theta_k^s$, the $\gamma_k^{s}$ message becomes
\begin{align}
\label{eq_msg_gamma_gmm}
    & \gamma_k^{s} ( \mathbf{x}_{k}, 1 ) = u_{s\to k}^{\gamma, (0)} +  \\
    & \sum\nolimits_{ m = 1}^{M^{s}} \sum\nolimits_{j=1}^{J_{s, k}^{\theta}} u_{s\to k}^{\gamma, (m,j)} \mathcal{N}(\mathbf{e}_{s\to k}^{\gamma, (m,j)};\mathbf{H}_{s\to k}^{\gamma, (m,j)} \mathbf{x}_{k}, \mathbf{C}_{s\to k}^{\gamma, (m,j)}), \nonumber
\end{align}
with parameters given in Table~\ref{table_gmm_likelihood}. 

In the flowchart of Figure~\ref{fig_flowchart}, for the GM likelihood messages $\Lambda_k^s$ (\ref{eq_msg_Lambda_gmm}) and $\gamma_k^{s}$ (\ref{eq_msg_gamma_gmm}), for compactness, we employ a notation using a single summation. The correspondence between the double and single summation parameters for $\Lambda_k^s$ is given by any one-to-one mapping from  $[1:M^s] \times [1:J_{k, s}^{\delta}]$ to $[1:I_{k \rightarrow s}^{\Lambda}]$, where $I_{k \rightarrow s}^{\Lambda} = M^s \cdot J_{k, s}^{\delta}$  
An analogous one-to-one mapping yields the correspondence of parameters for $\gamma_k^{s}$.

\subsection{Agent belief via centralized GM product}	\label{sec_ag_belief_gibbs}	
The belief (\ref{eq_ag_belief}) of agent $s$, under the assumptions of the previous section, is given by the product of locally-available GM likelihood messages and has the generic form
\begin{align}
b(\mathbf{y}_s) & = \sum\nolimits_{j=1}^{J_{\to n, s}} w_{\to n, s}^{(j)} \mathcal{N} \big( \mathbf{y}_s; \mathbf{m}_{\to n, s}^{(j)}, \mathbf{P}_{\to n, s}^{(j)} \big) 
\label{eq:gen_gauss_prod} \\
& \hspace{-1cm}\times \prod\nolimits_{l \in \mathcal{N}_s} \Big(u_{l\to s}^{(0)}+ \sum\nolimits_{i_l=1}^{I_{l \to s}} u_{l\to s}^{(i_l)} \mathcal{N} \big( {\mathbf{e}}_{l\to s}^{(i_l)}; \mathbf{H}_{l\to s}^{(i_l)} \mathbf{y}_s, \mathbf{C}_{l\to s}^{(i_l)} \big) \Big)   \nonumber
\end{align}
where $\mathcal{N}_s = \mathcal{A}_s \cup \mathcal{T}_s$ is the set of neighboring agents and the targets observed by agent $s$ at time $n$. The GM in the first line represents the predicted message $\phi_{\to n}$ (\ref{eq_ag_belief}), where the superscript $\phi$ is dropped for clarity. The $L\triangleq \vert \mathcal{N}_s \vert$ GM likelihood terms in the second line represent the various inter-agent and target measurement likelihood terms ($\Phi_{\ell \rightarrow s}$ and $\Lambda_k^s$ respectively). Since both $\Phi$ and $\Lambda$ share the same GM likelihood structure, the superscripts $\Phi$ and $\Lambda$ are dropped for clarity and solely the index $l$ identifies each likelihood term as a $\Phi_{l \rightarrow s}$ (if $l\in \mathcal{A}_s$) or a  $\Lambda_l^s$ (if $l\in \mathcal{T}_s$) message. The corresponding parameters for each likelihood term $u_{l\to s}^{(i_l)}, {\mathbf{e}}_{l\to s}^{(i_l)}, \mathbf{H}_{l\to s}^{(i_l)}, \mathbf{C}_{l\to s}^{(i_l)}$ are defined in Table \ref{table_gmm_likelihood}. As already stated above, we replace the double superscript $(m,j)$ in parameters of $\Lambda_l^s$ with the single superscript $(i)$. 
Comparing (\ref{eq_msg_Phi_gmm}) with (\ref{eq_msg_Lambda_gmm}), each $\Lambda_l^s$ message has a constant term $u_{l\to s}^{(0)}\neq 0$, whereas for $\Phi_{l \rightarrow s}$, $u_{l \to s}^{(0)} = 0$.

\noindent For $i_l\neq 0$ let 
\begin{align}
\tilde{\mathbf{e}}_{l\to s}^{(i_l)} &\triangleq \big[\mathbf{H}_{l\to s}^{(i_l)}\big]^T \big[ \mathbf{C}_{l\to s}^{(i_l)} \big]^{-1} \mathbf{e}_{l\to s}^{(i_l)},\nonumber\\
\tilde{\mathbf{C}}_{l\to s}^{(i_l)} &\triangleq \big[\mathbf{H}_{l\to s}^{(i_l)} \big]^T\big[ \mathbf{C}_{l\to s}^{(i_l)} \big]^{-1} \mathbf{H}_{l\to s}^{(i_l)}, \label{eq_ag_lkhd_param} \\
c_{l\to s}^{(i_l)} & \triangleq \log \left( \mfrac{u_{l\to s}^{(i_l)}}{\sqrt{\det(2\pi \mathbf{C}_{l\to s}^{(i_l)})}} \right) - \mfrac{1}{2}\big[\big( \mathbf{e}_{l\to s}^{(i_l)} \big)^{T} \big( \mathbf{C}_{l\to s}^{(i_l)} \big)^{-1} \mathbf{e}_{l\to s}^{(i_l)} \big]. \nonumber
\end{align}
For $i_l=0$, let $ \tilde{\mathbf{e}}^{(0)}_{l\to s} = \mathbf{0}_{d_a}$, $\tilde{\mathbf{C}}^{(0)}_{l\to s} = \mathbf{0}_{d_a\times d_a}$ and  $c^{(0)}_{l\to s} = \log(u_{l\to s}^{(0)})$. We also define the $L$-length vector $\bm{i} \triangleq \left[ i_1,\cdots, i_L \right]$, where $i_l \in [0:I_{l\to s}]$, and the product space $\bm{I}_L \triangleq \bigtimes_{l=1}^L [0:I_{l\to s}]$. Furthermore, let 
\begin{equation}
    \tilde{\mathbf{C}}^{(\bm{i})} \triangleq \sum\nolimits_{l=1}^L \tilde{\mathbf{C}}_{l\to s}^{(i_l)}, \tilde{\mathbf{e}}^{(\bm{i})} \triangleq \sum\nolimits_{l=1}^L \tilde{\mathbf{e}}_{l\to s}^{(i_l)}, c^{(\bm{i})} \triangleq \sum\nolimits_{l=1}^L c_{l\to s}^{(i_l)}. \label{eq_ag_GM_comp}
\end{equation}
Then by the property of the product of Gaussian functions~\cite[Ch. 3.8]{bibl:beyondKalman}, the result of (\ref{eq:gen_gauss_prod}) is the $\text{GM} ( \mathbf{y}_s; \{ w_{s}^{(j,\bm{i})}, \mathbf{m}_{s}^{(j,\bm{i})}, \mathbf{P}_{s}^{(j, \bm{i})} \}_{ j\in[1:J_{\to n,s}], \bm{i}\in \bm{I}_L} )$ where
\begin{align}
\mathbf{P}_{s}^{(j,\bm{i})} &= \Big[ \big(\mathbf{P}_{\to n, s}^{(j)}\big)^{-1} + \tilde{\mathbf{C}}^{(\bm{i})} \Big]^{-1} \label{eq_prod_gm_exact_P}\\
\mathbf{m}_{s}^{(j,\bm{i})} &= \mathbf{P}_s^{(j,\bm{i})} \Big[ \big( \mathbf{P}_{\to n, s}^{(j)} \big)^{-1}\mathbf{m}_{\to n, s}^{(j)} + \tilde{\mathbf{e}}^{(\bm{i})} \Big] \label{eq_prod_gm_exact_m} \\
w_s^{(j,\bm{i})} &=  w_{\to n,s}^{(j)} \exp \left( c^{(\bm{i})} -\mfrac{1}{2} \big(\mathbf{m}_{\to n, s}^{(j)}\big)^T \hspace{-1mm} \big( \mathbf{P}_{\to n, s}^{(j)}\big)^{-1} \hspace{-1mm} \mathbf{m}_{\to n, s}^{(j)} \right) \nonumber \\
& \hspace*{-1cm} \times\exp \left( \mfrac{1}{2} \big(\mathbf{m}_s^{(j, \bm{i})}\big)^T \big( \mathbf{P}_s^{(j, \bm{i})}\big)^{-1} \mathbf{m}_s^{(j, \bm{i})}  \right)  \sqrt{\mfrac{\det(\mathbf{P}_s^{(j,\bm{i})})}{\det(\mathbf{P}_{\to n, s}^{(j)})}}. \label{eq_prod_gm_exact_w}
\end{align}
Although the computation of (\ref{eq_prod_gm_exact_P})-(\ref{eq_prod_gm_exact_w}) involves parameters that are locally available at each agent (\ref{eq_ag_belief}), it has computational complexity $O(J_{\to n, s}\prod_{l=1}^L I_{l\to s})$, i.e., exponential in the number $L$ of likelihood terms. In the following, we present a Gibbs-sampling based method that efficiently constructs a truncated GM approximation of (\ref{eq:gen_gauss_prod}) where only the $T$ highest scoring mixture components are retained. 


\subsubsection{GM product via Gibbs sampling} The Gibbs sampling approach borrows from the method in~\cite{nonparametricbp10} which involves the product of GM probability densities whereas (\ref{eq:gen_gauss_prod}) involves the product of a GM density with $L$ GM likelihood terms. The Gibbs procedure for the GM product of (\ref{eq:gen_gauss_prod}) is given in Algorithm \ref{alg_prod_gmm_cent} and referred to as Centralized Gibbs, since all the required messages are locally available. 

 To address the challenge of the high number $\prod_{l=1}^{L}I_{l\to s}$ of components of the likelihood product in (\ref{eq:gen_gauss_prod}), we aim to select component labels $\bm{i}$ from the product space $\bm{I}_L$ of likelihood components that lead to Gaussian components (\ref{eq_prod_gm_exact_P})-(\ref{eq_prod_gm_exact_w}) with high weights $w_s^{(j, \bm{i})}$. Ideally, this can be achieved by sampling independently with probability 
\begin{align}
	\text{Pr}(\bm{i} = [i_1, \dots, i_L]^T) & = \sum\nolimits_{j=1}^{J_{\to n, s}} \text{Pr}(j, \bm{i} = [i_1, \dots, i_L]^T) \nonumber \\
	&\propto \sum\nolimits_{j=1}^{J_{\to n, s}} w_s^{(j, \bm{i})}. \label{eq_ideal_GM_samp}
\end{align}
According to (\ref{eq_ideal_GM_samp}), vectors $\bm{i}$ that lead to higher weights (\ref{eq_prod_gm_exact_w}) are selected with higher probability. However, sampling from (\ref{eq_ideal_GM_samp}) is difficult as it requires the computation of all $\{ w_s^{(j, \bm{i})} \}$ which is again $O({J_{\to n, s}}\prod_{l=1}^L I_{l\to s})$. The Gibbs sampler constructs a finite Markov chain with stationary distribution (\ref{eq_ideal_GM_samp}) by iteratively sampling from conditional densities that are easily constructed. 
\begin{algorithm}[t!]
	\caption{Centralized Gibbs GM product for belief $b(\mathbf{y}_s)$}
	\label{alg_prod_gmm_cent}
	\begin{algorithmic}[1]
		\State{\textbf{Input} parameters of (\ref{eq:gen_gauss_prod}).}
		\State{\label{alg_prod_gmm_cent_init_sample}Sample ${i_l}$ $\forall\, l$ s.t. $\text{Pr}(i_l) \propto \varrho(i_l)$ where $\varrho(0) = u_{l \to s}^{(0)}$ and}
		\Statex{for $i_l \neq 0$, $\varrho(i_l)$ in (\ref{eq_alg_cent_init_label}).}
		\State{\label{alg2_global}Compute $c^{(\bm{i})} = \sum_{l=1}^L c_{l\to s}^{{i}_l}$, $\tilde{\mathbf{C}}^{(\bm{i})} = \sum_{l=1}^L \tilde{\mathbf{C}}_{l\to s}^{({i}_l)}$ and}
		\Statex{$\tilde{\mathbf{e}}^{(\bm{i})}=\sum_{l=1}^L \tilde{\mathbf{e}}_{l\to s}^{({i}_l)}$.}
		\For{\label{alg2_prod_gmm_cent_loopL_begin}$l \leftarrow 1$ to $L$ }
		\State{\label{alg2_L1}Compute $c^{(\bm{i}_{\neg l})} = {c}_{}^{(\bm{i})} - c_{l\to s}^{({i}_l)}$, $\tilde{\mathbf{C}}^{(\bm{i}_{\neg l})} = \tilde{\mathbf{C}}^{(\bm{i})} - \tilde{\mathbf{C}}_{l\to s}^{({i}_l)}$}
		\Statex \hspace{\algorithmicindent}{and $\tilde{\mathbf{e}}^{(\bm{i}_{\neg l})} = \tilde{\mathbf{e}}^{(\bm{i})} - \tilde{\mathbf{e}}_{l\to s}^{({i}_l)}$.}	
		\For{\label{alg_prod_gmm_cent_q_start}$q \leftarrow 0$ to $I_{l\to s}$}
		\State{Let $\bm{i}_{*q} \triangleq [i_{1}, \dots, i_{l-1}, q, i_{l+1}, \dots, i_{L}]$.}
		\State{\label{alg2_pre}Set $c^{(\bm{i}_{* q})} = {c}^{(\bm{i}_{\neg l})} + c_{l\to s}^{(q)}$, $\tilde{\mathbf{C}}^{(\bm{i}_{* q})} = \tilde{\mathbf{C}}^{(\bm{i}_{\neg l})} + \tilde{\mathbf{C}}_{l\to s}^{(q)}$,}
		\Statex{\hspace{\algorithmicindent}\hspace{\algorithmicindent}{and $\tilde{\mathbf{e}}^{(\bm{i}_{* q})} = \tilde{\mathbf{e}}^{(\bm{i}_{\neg l})} + \tilde{\mathbf{e}}_{l\to s}^{(q)}$.}}    	    
		\For{\label{alg_cgibbs_inner_for1} $j \leftarrow 1$ to $J_{\to n,s}$}
		\State{\label{alg2_mpc}Compute $\mathbf{P}_s^{(j,\bm{i}_{* q})}, \mathbf{m}_s^{(j,\bm{i}_{* q})}, w_s^{(j,\bm{i}_{*q})}$ (\ref{eq_prod_gm_exact_P})-(\ref{eq_prod_gm_exact_w}).}
		\EndFor \label{alg_cgibbs_inner_for2}
		\State{\label{alg_prod_gmm_cent_weight_q}Compute $ \pi_l(q\vert \bm{i}_{\neg l}) \propto \sum_{j=1}^{J_{\to n,s}} w_s^{(j,\bm{i}_{*q})}$}.
		\EndFor{\label{alg_prod_gmm_cent_q_end}}
		\State{\label{alg_prod_gmm_cent_samp_new}Sample new label $ q' \sim \pi_l(q \vert \bm{i}_{\neg l})$ and set $\bm{i}\leftarrow \bm{i}_{*q'}$,}
		\State{\label{alg_prod_gmm_cent_new}$\tilde{\mathbf{C}}^{(\bm{i})} \leftarrow \tilde{\mathbf{C}}^{(\bm{i}_{*q'})}$, $\tilde{\mathbf{e}}^{(\bm{i})} \leftarrow \tilde{\mathbf{e}}^{(\bm{i}_{*q'})}$, $c^{(\bm{i})} \leftarrow c^{(\bm{i}_{*q'})}$}.
		\EndFor{\label{alg_prod_gmm_cent_loopL_end}}			
		\State{Repeat the steps \ref{alg2_prod_gmm_cent_loopL_begin}-\ref{alg_prod_gmm_cent_loopL_end} for $T$ iterations.}
		\State{\textbf{Return} $c^{(\bm{i})}$, $\tilde{\mathbf{C}}^{(\bm{i})}$, and $\tilde{\mathbf{e}}^{(\bm{i})}$ for the distinct samples $\bm{i}\in \bm{I}_L$.}
	\end{algorithmic}
\end{algorithm}
The proposed Gibbs method starts by sampling an initial label vector $\bm{i}$ (line \ref{alg_prod_gmm_cent_init_sample}, Algorithm \ref{alg_prod_gmm_cent}), with probabilities $\text{Pr}(i_l) \propto \varrho(i_l)$ where $\varrho(0) = u_{l \to s}^{(0)}$ and for $i_l \neq 0$
\begin{align}
    & \varrho(i_l) = u_{l \to s}^{(i_l)} {\textstyle \int} \phi_{\to n}(\mathbf{y}_{s}) \mathcal{N} ( {\mathbf{e}}_{l\to s}^{(i_l)}; \mathbf{H}_{l\to s}^{(i_l)} \mathbf{y}_s, \mathbf{C}_{l\to s}^{(i_l)} ) d\mathbf{y}_{s} \nonumber \\
    &= u_{l \to s}^{(i_l)} \sum\nolimits_{j=1}^{J_{\to n, s}} w_{\to n, s}^{(j)}\times \label{eq_alg_cent_init_label} \\
    & \:\:\:  \mathcal{N} \big( {\mathbf{e}}_{l\to s}^{(i_l)}; \mathbf{H}_{l\to s}^{(i_l)} \mathbf{m}_{\to n,s}^{(j)}, \mathbf{C}_{l\to s}^{(i_l)} + \mathbf{H}_{l\to s}^{(i_l)} \mathbf{P}_{\to n,s}^{(j)} \big[ \mathbf{H}_{l\to s}^{(i_l)} \big]^T \big). \nonumber
\end{align}
These initial weights are based on the intuition that if $\mathcal{N}_s = \{ l \}$, i.e., agent $s$ has only one neighbor, (\ref{eq_alg_cent_init_label}) would give the weight contributed to by the $i_l$-th component of the likelihood, in the resulting GM in (\ref{eq:gen_gauss_prod}). This is followed by sequentially sampling new labels for each of the $L$ likelihood messages (lines~\ref{alg2_L1}-\ref{alg_prod_gmm_cent_new}, Algorithm \ref{alg_prod_gmm_cent}), from the conditional distributions of (\ref{eq_ideal_GM_samp}), i.e.,
\begin{align}
	& \pi_l( i_l \vert \bm{i}_{\neg l}) \triangleq \text{Pr}( i_l \vert \bm{i}_{\neg l} ) = \mfrac{\text{Pr}(\bm{i})}{\text{Pr}(\bm{i}_{\neg l})} \nonumber \\
	&=  \mfrac{\sum_{j=1}^{J_{\to n, s}} \text{Pr}(j,\bm{i})}{\sum_{q=1}^{I_{l \to s}} \sum_{j=1}^{J_{\to n, s}} \text{Pr}(j,\bm{i}_{*q})} = \mfrac{\sum_{j=1}^{J_{\to n, s}} w_s^{(j,\bm{i})}}{\sum_{q=1}^{I_{l\to s}} \sum_{j=1}^{J_{\to n, s}} w_s^{(j,\bm{i}_{*q})}} \label{eq_Gibbs_cent_cond}
\end{align}
where $\bm{i}_{\neg l} \triangleq [i_1, \dots, i_{l-1}, i_{l+1}, \dots, i_L]^T$ and $\bm{i}_{*q} \triangleq [i_{1}, \dots, i_{l-1}, q, i_{l+1}, \dots, i_{L}]^T$.  Each cycle (lines \ref{alg2_L1}-\ref{alg_prod_gmm_cent_new}) of the Gibbs sampler involves sampling a new component $q \in [0:I_{l\to s}]$ for each of the likelihood messages $l\in[1:L]$. Holding fixed the labels for messages $[1:L] \setminus \{l \}$, the parameters (\ref{eq_ag_GM_comp}) are computed by first removing the contribution of the old label $i_l$ (line \ref{alg2_L1}) and adding the contribution of each $q \in [0:I_{l\to s}]$ (line \ref{alg2_pre}). Next, the resulting components are used to update the predicted agent message (line \ref{alg2_mpc}) and the conditional distribution (\ref{eq_Gibbs_cent_cond}) is obtained by marginalizing out the prediction message labels $j\in [1:J_{\to n,s}]$ (line \ref{alg_prod_gmm_cent_weight_q}). A new label $i_l$ is sampled (line \ref{alg_prod_gmm_cent_samp_new}) and the corresponding parameters (\ref{eq_ag_GM_comp}) are updated before continuing the Gibbs cycle for the next likelihood message. The entire sampling procedure is repeated $T$ times and the parameters $(c^{(\bm{i})}, \tilde{\mathbf{C}}^{(\bm{i})}, \tilde{\mathbf{e}}^{(\bm{i})})$ corresponding to all distinct vectors $\bm{i}$ (i.e., two vectors differ in at least one entry) are returned. The $T$ highest scoring components according to $c^{(\bm{i})}$ are used to construct the truncated agent belief via (\ref{eq_prod_gm_exact_P})-(\ref{eq_prod_gm_exact_w}). The convergence of Algorithm \ref{alg_prod_gmm_cent} and the uniqueness of the stationary distribution follow from the regularity of the transition matrix $P_{\bm{i}, \bm{i}'} = \pi(\bm{i} \vert \bm{i}')>0$ (as $w_s^{(j,\bm{i})}>0$ $\forall$ $(j,\bm{i})$ from (\ref{eq_prod_gm_exact_w})). The convergence rate is geometrically fast~\cite[Section 4.3.3]{bibl:gallager_stoch2013}, i.e., $\vert [P^n]_{\bm{i}, \bm{i}'} - \text{Pr}(\bm{i}) \vert  \leq (1-2\vartheta)^n$ $\forall$ ${\bm{i}, \bm{i}'}$ and where $\vartheta = \min_{\bm{i}, \bm{i}'} P_{\bm{i}, \bm{i}'}$ is the least likely 1-step transition probability. All resulting samples are used since every distinct sample $\bm{i}$ contributes to an improved approximation of (\ref{eq:gen_gauss_prod}). Hence, no burn-in period is required. Due to the pre-computations at lines \ref{alg2_L1}-\ref{alg2_pre} in Algorithm \ref{alg_prod_gmm_cent}, the evaluation of (\ref{eq_prod_gm_exact_P})-(\ref{eq_prod_gm_exact_w}) at lines \ref{alg_cgibbs_inner_for1}-\ref{alg_cgibbs_inner_for2} for all $l\in [1:L]$ is $O({J_{\to n, s}}d_{a}^3)$, leading to a time complexity for Algorithm \ref{alg_prod_gmm_cent} of $O(T{J_{\to n, s}}d_{a}^3\sum_{l}^L I_{l\to s})$.

\subsubsection{Computation of extrinsic information $\theta_k^s (\cdot)$} \label{sec_theta_GMM}
The $\theta_k^s$ message (\ref{eq_msg_theta}) represents the extrinsic information sent from agent $s$ to the PT $k$. If $l \in \mathcal{T}_s$ in the generic product of (\ref{eq:gen_gauss_prod}), then $\theta_l^s(\mathbf{y}_s) \propto b(\mathbf{y}_s)/\Lambda_l^s(\mathbf{y}_s)$ appears to be a ratio of GMs (which is not a GM in general). An alternate procedure based on (\ref{eq_msg_theta}) is described next. First note from line \ref{alg2_L1} of Algorithm \ref{alg_prod_gmm_cent} that the parameters $c^{(\bm{i}_{\neg l})}$,  $\tilde{\mathbf{C}}^{(\bm{i}_{\neg l})}$ and $\tilde{\mathbf{e}}^{(\bm{i}_{\neg l})}$ characterize the product $\prod_{\ell \neq l} u_{l\to s}^{(i_{\ell})} \mathcal{N} ( {\mathbf{e}}_{l\to s}^{(i_{\ell})}; \mathbf{H}_{l\to s}^{(i_{\ell})} \mathbf{y}_s, \mathbf{C}_{l\to s}^{(i_{\ell})} )$ of likelihood terms identified by the labels $\bm{i}_{\neg l}$. The resulting components, after multiplication with the prior $\phi_{\rightarrow n}(\mathbf{y}_s)$, lead to an efficient GM approximation of $\theta_k^s$, without
requiring a dedicated separate procedure like Algorithm \ref{alg_prod_gmm_cent} to compute $\theta_k^s$.

\subsection{Target belief via decentralized GM product} \label{sec_tg_belief_gm}	
 Decentralized SCS-MTT algorithms require a distributed evaluation of the target beliefs across the entire network. The computed target belief for a PT $k$ needs to be identical across all the agents, including the agents that do not observe the PT $k$ at time $n$. In this section, we propose an efficient method based on Gibbs sampling and average consensus for GM beliefs. As we shall see, much of the discussion in this section follows Section \ref{sec_ag_belief_gibbs} closely, with the difference that not all the messages are locally available at any single agent. The belief (\ref{eq_tg_belief}) for a PT $k$ can be expressed as
\begin{align}
& b(\mathbf{x}_k,1)=  \sum\nolimits_{j=1}^{J_{\to n, k}} \omega_{\to n, k}^{(j)} \mathcal{N} ( \mathbf{x}_k; \bm{\mu}_{\to n, k}^{(j)}, \bm{\Omega}_{\to n, k}^{(j)} ) \times  \label{eq_tg_gen_gmm_prod1}\\
&   \prod\nolimits_{s=1}^S \Big[ u_{s\to k}^{(0)} + \sum\nolimits_{i_s=1}^{I_{s\to k}} u_{s\to k}^{(i_s)} \mathcal{N} ( {\mathbf{e}}_{s\to k}^{(i_s)}; \mathbf{H}_{s\to k}^{(i_s)} \mathbf{x}_k, \mathbf{C}_{s\to k}^{(i_s)} ) \Big], \nonumber \\
& b(\mathbf{x}_k, 0) = \alpha_{\to n}(\mathbf{x}_k, 0) \prod\nolimits_{s=1}^S \eta_k^s(0) \label{eq_tg_gen_gmm_prod0}
\end{align}
where (\ref{eq_tg_gen_gmm_prod1}) is analogous to (\ref{eq:gen_gauss_prod}) in its generic form. The GM in the first line represents the predicted message $\alpha_{\to n} (\mathbf{x}_k,1)$ in (\ref{eq_msg_alpha}). The likelihood terms in the second line represent the $\gamma_k^s(\cdot,1)$ messages. Note that each agent $s$ only has access to its local message $\gamma_k^s$. We use the generic forms of $\alpha_{\to n}$ and $\gamma_k^s(\cdot,1)$ from Figure \ref{fig_flowchart}. For clarity, the respective superscripts $\alpha, \gamma$ are dropped from the parameters.
Note that if a target is not observed by an agent $s$, i.e., if $k\notin \mathcal{T}_{n,s}$, then $\gamma_{k}^s(\mathbf{x}_k, 1) = \gamma_{k}^s(\mathbf{x}_k, 0) =1$ which, for $r_k = 1$, is represented as $u_{s\to k}^{(0)}=1$ with $I_{s \to k}=0$ and $\log(\eta_k^s(0))=0$. As a consequence of assumption (A10) from Section~\ref{sec:assumptions}, each agent $k$ has access to its local parameters and the predicted message $\alpha_{\to n}(\mathbf{x}_k)$ (Section~\ref{sec_alpha_GM}), the latter being identical across all agents. For compactness, $\forall$ $s\in \mathcal{A}$ and $i_s \neq 0$ we define the parameters of the local $\gamma_k^s(\cdot, 1)$ messages as
\begin{align}
\tilde{\mathbf{e}}_{s\to k}^{(i_s)} &\triangleq \big[\bm{H}_{s\to k}^{(i_s)}\big]^T \big[ \mathbf{C}_{s\to k}^{(i_s)} \big]^{-1} \mathbf{e}_{s\to k}^{(i_s)}, \nonumber\\
\tilde{\mathbf{C}}_{s\to k}^{(i_s)} &\triangleq \big[\mathbf{H}_{s\to k}^{(i_s)} \big]^T\big[ \mathbf{C}_{s\to k}^{(i_s)} \big]^{-1} \mathbf{H}_{s\to k}^{(i_s)}, \label{eq_tg_gamma_param} \\
c_{s\to k}^{(i_s)} &\triangleq \log \left( \mfrac{u_{s\to k}^{(i_s)}}{\sqrt{\det(2 \pi \mathbf{C}_{s\to k}^{(i_s)})}} \right) -\mfrac{1}{2} \big[ \big( \mathbf{e}_{s\to k}^{(i_s)} \big)^{T} \hspace{-1mm} \big( \mathbf{C}_{s\to k}^{(i_s)} \big)^{-1} \hspace{-1mm} \mathbf{e}_{s\to k}^{(i_s)} \big]. \nonumber
\end{align}
Furthermore, for $i_s=0$, let $ \tilde{\mathbf{e}}^{(0)}_{s\to k} = \mathbf{0}_{d_t}$, $\tilde{\mathbf{C}}^{(0)}_{s\to k} = \mathbf{0}_{d_t\times d_t}$ and $c^{(0)}_{s\to k} = \log(u_{s \to k}^{(0)})$. Note that (\ref{eq_tg_gamma_param}) is analogous to (\ref{eq_ag_lkhd_param}). In (\ref{eq_tg_gen_gmm_prod1}), for each likelihood product term denoted by the $S$-length vector $\bm{i}=[i_1, \dots, i_S] \in \bm{I}_S \triangleq \bigtimes_{s=1}^S [0:I_{s\to k}] $, the quantities   
\begin{equation}
\bm{\Xi}^{(\bm{i})} = \sum_{s=1}^S \tilde{\mathbf{C}}_{s\to k}^{(i_s)}, \  \bm{\xi}^{(\bm{i})} = \sum_{s=1}^S \tilde{\mathbf{e}}_{s\to k}^{(i_s)}, \ 
\xi^{(\bm{i})}  = \sum_{s=1}^{S} c_{s\to k}^{(i_s)} \label{tg_belief_cmp}
\end{equation}
require information from across the network and are referred to as global information (this is in contrast to Section \ref{sec_ag_belief_gibbs} where the analogous quantities (\ref{eq_ag_GM_comp}) are locally available). As a result, $b(\mathbf{x}_k,1)$ is $\text{GM} ( \mathbf{x}_k; \{ \omega_k^{(j,\bm{i})}, \bm{\mu}_k^{(j,\bm{i})}, \bm{\Omega}_k^{(j,\bm{i})} \}_{j\in [1:J_{\to n,k}], \bm{i}\in  \bm{I}_S} )$ with parameters
\begin{align}
\bm{\Omega}_{k}^{(j,\bm{i})} &= \Big[ \big(\bm{\Omega}_{\to n,k}^{(j)}\big)^{-1} + \bm{\Xi}^{(\bm{i})} \Big]^{-1} \label{eq_prod_gm_exact_P_tg}\\
\bm{\mu}_{k}^{(j,\bm{i})} &= \bm{\Omega}_{k}^{(j,\bm{i})} \Big[ \big( \bm{\Omega}_{\to n,k}^{(j)} \big)^{-1}\bm{\mu}_{\to n,k}^{(j)} + \bm{\xi}^{(\bm{i})} \Big] \label{eq_prod_gm_exact_m_tg} \\
\omega_k^{(j,\bm{i})} &=  \omega_{\to n,k}^{(j)} \exp \left( \xi^{(\bm{i})} -\mfrac{1}{2} \big(\bm{\mu}_{\to n,k}^{(j)}\big)^T \big( \bm{\Omega}_{\to n,k}^{(j)}\big)^{-1} \bm{\mu}_{s\to k}^{(j)} \right) \nonumber \\
& \hspace*{-7mm} \times\exp \left( \mfrac{1}{2} \big(\bm{\mu}_{k}^{(j, \bm{i})}\big)^T \big( \bm{\Omega}_{k}^{(j, \bm{i})}\big)^{-1} \bm{\mu}_{k}^{(j, \bm{i})}  \right)  \sqrt{\mfrac{\det(\bm{\Omega}_{k}^{(j,\bm{i})})}{\det(\bm{\Omega}_{\to n,k}^{(j)})}}. \label{eq_prod_gm_exact_w_tg}
\end{align}
Again, (\ref{eq_prod_gm_exact_P_tg})-(\ref{eq_prod_gm_exact_w_tg}) are analogous to (\ref{eq_prod_gm_exact_P})-(\ref{eq_prod_gm_exact_w}) in Section \ref{sec_ag_belief_gibbs}. However, directly applying here the sequential Gibbs approach (Algorithm \ref{alg_prod_gmm_cent}) becomes impractical. This is because the evaluation of the parameters of selected Gaussian components (indexed by $\bm{i}$) in (\ref{tg_belief_cmp}) requires the aggregation of parameters from across the entire network. This process happens sequentially for each new label (line \ref{alg_prod_gmm_cent_new} of Algorithm \ref{alg_prod_gmm_cent}). In a decentralized algorithm, this incurs a high communication cost and latency. Thus, a parallel sampling mechanism is favored, where agents sample local labels $i_s \in [0:I_{s\to k}]$ in parallel to form a new label vector $\bm{i}=[i_1 \dots, i_S]^T$. This is followed by synchronization, that is, the computation of the parameters (\ref{tg_belief_cmp}) corresponding to $\bm{i}$ via average consensus.
\begin{table}[t!]	
	\centering
	\begin{tabular}{ c|c } 
		\hline
		Sequential Gibbs & Hogwild! Gibbs \\
		\hline
		$\pi_1({i}_1'\vert \bm{i}_{2:S})$ & \multirow{5}{5cm}{\centering In parallel: \\ $\pi_1({i}_1'\vert \bm{i}_{\neg 1})$, $\cdots$ $\pi_s({i}_s'\vert \bm{i}_{\neg s})$, $\cdots$,  $\pi_S({i}_S'\vert \bm{i}_{\neg S})$  } \\
		$\vdots$ &  \\ 
		$\pi_s({i}'_s\vert \bm{i}'_{1:s-1}, \bm{i}_{s+1:S})$ &  \\
		$\vdots$ &  \\
		$\pi_S({i}'_S\vert \bm{i}'_{1:S-1})$ &  \\ 
		\hline	
	\end{tabular}
	\caption{Conditional sampling of a new label vector $\bm{i}'$ given previous labels $\bm{i}$ in synchronous and Hogwild! Gibbs.}
	\label{table:sync_hog}
\end{table}

Such sampling schemes are referred to as partially synchronous Gibbs sampling~\cite{bibl:terenin_async_gibbs} or Hogwild! Gibbs~\cite{bibl:wilsky_hogwild_gibbs_NIPS2013}. In general, in Hogwild \cite{bibl:wilsky_hogwild_gibbs_NIPS2013} or asynchronous methods, the agents perform sampling/updating as fast as they can, while periodic global synchronization is achieved across the network. The difference between the sequential Gibbs of Algorithm \ref{alg_prod_gmm_cent} and the Hogwild! Gibbs employed here is shown in Table \ref{table:sync_hog}. Starting from a label vector $\bm{i}$, the sequential Gibbs effectively samples new labels $\bm{i}'$ sequentially according to the marginal densities (\ref{eq_Gibbs_cent_cond}). The Hogwild! Gibbs method samples, in parallel at each agent $s \in [1, S]$, a local label ${i}'_s$ conditioned on the previous labels $\bm{i}_{\neg s}$. In contrast to the sequential Gibbs, Hogwild! Gibbs requires the computation of global parameters only after all the agents have locally sampled a new index ${i}'_s$ $\forall$ $s$. 

\subsubsection{Hogwild! Gibbs for GM product} \label{sec_dist_gmm_hogwild}
The proposed Hogwild! Gibbs algorithm produces a set of high-weight Gaussian components. The resulting GM approximates the target belief (\ref{eq_tg_gen_gmm_prod1}) and is presented in Algorithm \ref{alg_prod_gmm_dist}, which is executed synchronously and in parallel at all agents for each PT. The main steps of Algorithm~\ref{alg_prod_gmm_dist} are detailed in the following:
\begin{enumerate}[leftmargin=*]
	\item[a)] \textit{Initialization} (line~\ref{alg3_init}). Each agent $s$ samples an initial label $i_s$ from the local labels $[0:I_s]$ with probability $\text{Pr}({i}_s) \propto \varrho(i_s) $, where
	$\varrho(0) = u_{s\to k}^{(0)} \sum\nolimits_{j=1}^{J_{\to n, k}} \omega_{\to n, k}^{(j)}$, and for $i_s \neq 0$
    \begin{align}
        & \varrho(i_s) = u_{s \to k}^{(i_s)} \sum\nolimits_{j=1}^{J_{\to n, k}} \omega_{\to n, k}^{(j)} \ \times \label{eq_alg_dist_init_label} \\
        & \ \mathcal{N} \big( {\mathbf{e}}_{s\to k}^{(i_s)}; \mathbf{H}_{s\to k}^{(i_s)} \bm{\mu}_{\to n,k}^{(j)}, \mathbf{C}_{s\to k}^{(i_s)} + \mathbf{H}_{s\to k}^{(i_s)} \bm{\Omega}_{\to n,k}^{(j)} \big[ \mathbf{H}_{s\to k}^{(i_s)} \big]^T \big). \nonumber
    \end{align}
    In particular, the weight $\varrho(i_s)$ of the $i_s$-th likelihood component from $\gamma_k^s(\cdot, 1)$, given in (\ref{eq_alg_dist_init_label}), is high if it leads to high-weight Gaussian components after updating the prior $\alpha_{\to n}(\mathbf{x}_k)$. Note that (\ref{eq_alg_dist_init_label}) is analogous to (\ref{eq_alg_cent_init_label}).
	
	\item[b)] \textit{Global parameter evaluation} (line~\ref{alg3_Xi}). Corresponding to the selected labels $\bm{i}$, the global parameters of (\ref{tg_belief_cmp}) are evaluated via average and max consensus \cite{olfati07proc, bibl:rabbat_gossip_proc2010}. 
	The weights we use are Metropolis weights \cite{bibl:boyd_MC_wISPN2005}. Convergence
	is guaranteed as long as the communication graph spanning the agents is connected~\cite{bibl:rabbat_gossip_proc2010}. In practice, we stop 
	after a sufficiently large number $Q$ of consensus iterations.
	An additional max-consensus is carried out to ensure identical values for all agents. The consensus is reached across the entire network, even for agents $s$ that do not observe the target $k$, i.e, for which $k\notin \mathcal{T}_{n,s}$. Note that this is a decentralized implementation of the analogous step (line \ref{alg2_global}) in Algorithm (\ref{alg_prod_gmm_cent}).

	\item[c)] \textit{Computing local Gaussian components} (lines~\ref{alg3_forq}-\ref{alg3_forq_end}). Given the globally computed parameters of the product indexed by $\bm{i}$ (\ref{tg_belief_cmp}), each agent $s$ constructs the Gaussian indexed by $(j,\bm{i}_{*q})$, with parameters given by (\ref{eq_prod_gm_exact_P_tg})-(\ref{eq_prod_gm_exact_w_tg}). This is achieved by first replacing the $i_s$-th component of $\gamma_k^s(\cdot,1)$ with the $q$-th component of $\gamma_k^s(\cdot,1)$. This is done locally, since the previous label $i_s$ and the parameters of all the $q \in [0:I_{s\to k}]$ components of $\gamma_k^s(\cdot,1)$ are available locally. 
	\item[d)] \textit{Computing conditional probabilities and sample} (line \ref{alg3_samp2}). The weights of the Gaussian components indexed by $(j,\bm{i}_{*q})$, $\omega_k^{(j, \bm{i}_{*q})}$, are employed to compute the conditional density $\pi_s(q\vert \bm{i}_{\neg s}) \propto \sum_{j=1}^{J_{\to n, k}} \omega_k^{(j,\bm{i}_{*q})}$ from which a new local label is sampled $i_s \sim \pi_s(q\vert \bm{i}_{\neg s})$. This follows from (\ref{eq_ideal_GM_samp}) and is analogous to line \ref{alg_prod_gmm_cent_weight_q} in Algorithm \ref{alg_prod_gmm_cent}.
	\item[e)] \textit{Repeat for $T$ iterations} the steps b)-d) and return the Gaussian components with distinct labels $(j,\bm{i})$ for $b(\cdot, 1)$. An additional network consensus (line \ref{alg3_belief0}) is required for the non-existence case $r_k=0$ of (\ref{eq_tg_gen_gmm_prod0}), where the scalar value $\log(b_0) \triangleq \sum_{s=1}^S \log(\eta_k^s(0)) $ is evaluated.
    \item[f)]  \textit{Normalization of PT belief} (not shown in Algorithm \ref{alg_prod_gmm_dist}) is necessary in order to obtain an approximate pdf for PT $k$. The pdf of PT $k$ is given as $f_k(\mathbf{x}_k, 1) = \frac{1}{N}\sum_{(j,\bm{i})}  \omega_k^{(j, \bm{i})} \mathcal{N}(\mathbf{x}_k; \bm{\mu}_k^{(j,\bm{i})}, \bm{\Omega}_k^{(j,\bm{i})}) $ and $f_k(\mathbf{x}_k, 0) = \frac{b_0}{N}\alpha_k(\mathbf{x}_k,0)$ where $N = b_0\left[1-\sum_{j=1}^{J_{\to n,k}} \omega_{\to n,k}^{(j)}\right] + \sum_{(j,\bm{i})}  \omega_k^{(j, \bm{i})}$ is the normalization constant. 
\end{enumerate}
\begin{algorithm}[t!]
\caption{Dcent. Gibbs--PT $k$ at agent $s$ (in parallel $\forall\,s$)}
\label{alg_prod_gmm_dist}
\begin{algorithmic}[1]
	\State{\textbf{Input} parameters of (\ref{eq_tg_gen_gmm_prod1})-(\ref{eq_tg_gen_gmm_prod0}).}
	\State{\label{alg3_init}Sample ${i}_s\in [0:I_{s\to k}]$ as described at Section~\ref{sec_dist_gmm_hogwild}-a).}	
	\State{\label{alg3_Xi}\textbf{Network consensus} for $\bm{\Xi}^{(\bm{i})}$, $\bm{\xi}^{(\bm{i})}$ and ${\xi}^{(\bm{i})}$ of (\ref{tg_belief_cmp}).}
	\For{\label{alg3_forq} $q \leftarrow 0$ to $I_{s\to k}$}
	\State{Set $\bm{i}_{*q} \leftarrow [i_1, \cdots, i_{s-1}, q, i_{s+1}, \cdots, i_S]$.}
	\State{\label{alg3_diff} Compute $\bm{\Xi}^{(\bm{i}_{*q})} = \bm{\Xi}^{(\bm{i})} - \big( \mathbf{C}_{s \to k}^{(i_s)} \big)^{-1} + \big( \mathbf{C}_{s \to k}^{(q)} \big)^{-1} $,}
	\Statex{ $\bm{\xi}^{(\bm{i}_{*q})} = \bm{\xi}^{(\bm{i})}-\tilde{\mathbf{e}}_{s \to k}^{(i_s)} + \tilde{\mathbf{e}}_{s \to k}^{(q)}$, $c^{(\bm{i}_{*q})} = {\xi}^{(\bm{i})} -c_{s \to k}^{(i_s)} + c_{s \to k}^{(q)}$.}	
	\For{$j \leftarrow 1$ to $J_{\to n,k}$}
	\State{Compute $\omega_k^{(j,\bm{i}_{*q})}$, $\bm{\mu}_k^{(j,\bm{i}_{*q})}$, $\bm{\Omega}_k^{(j,\bm{i}_{*q})}$ as in (\ref{eq_prod_gm_exact_P_tg})-(\ref{eq_prod_gm_exact_w_tg}).}			
	\EndFor
	\EndFor\label{alg3_forq_end}
	\State{\label{alg3_samp2}Set $\pi_s(q\vert \bm{i}_{\neg s}) \propto \sum_{j=1}^{J_{\to n, k}} \omega_k^{(j, \bm{i}_{*q})} $, sample $\bm{i}_{s} \sim \pi_s(q\vert \bm{i}_{\neg s})$.}
	\State{\label{alg3_repeat}Repeat the steps \ref{alg3_Xi}-\ref{alg3_samp2} for $T$ iterations.}
	\State{\label{alg3_belief0}}\textbf{Network consensus} for $\log(b_0) \triangleq \sum_{s=1}^S \log(\eta_k^s(0)) $.
	\State{\textbf{Return} $ \{\big( w_k^{(j, \bm{i})}, \bm{\mu}_k^{(j,\bm{i})}, \bm{\Omega}_k^{(j,\bm{i})} \big) \}$ $\forall$ distinct $(S+1)$-tuples $(j,\bm{i})$ and $b_0$ to provide a GM approximation for $b(\mathbf{x}_k,r_k)$.}
\end{algorithmic}
\end{algorithm}
	 
During the consensus step in line \ref{alg3_Xi} of Algorithm \ref{alg_prod_gmm_dist}, it is assumed that agent $s$ learns the labels $\bm{i}_{l}$ for all $l\in \mathcal{A} \setminus \{s\}$. This can be achieved by diffusing the scalars $\bm{i}_{l}$ throughout the network and which involves only a mild increase in communication load as compared to the average consensus communication requirements. Note however that only the values taken by the global parameters $\bm{\Xi}^{(\bm{i})}$, $\bm{\xi}^{(\bm{i})}$, ${\xi}^{(\bm{i})}$ are necessary for the computation of local Gaussian components, the conditional probabilities and the ensuing sampling. The label values are only necessary for returning the distinct Gaussian components, i.e., for district $(S+1)$-tuples $(j,\bm{i})$. An alternative decentralized algorithm that avoids the diffusion of the label values ${i}_s$ is possible by modifying Algorithm \ref{alg_prod_gmm_dist} to return only the Gaussian components with distinct weights $\omega_k^{(j, \bm{i})}$, as these form a subset of the set of Gaussian components returned by Algorithm \ref{alg_prod_gmm_dist}.

\subsubsection{Complexity and Convergence} \label{sec_dist_gmm_conv}
The time complexity of Algorithm \ref{alg_prod_gmm_dist} is $O(T J_{\to n, k} I_{s\to k} d_t^3)$. Assuming $Q$ average consensus iterations and denoting with $D_{\mathcal{G}}$ the diameter of the communication graph, the communication load of the consensus step of Algorithm \ref{alg_prod_gmm_dist} is $ (Q+D_{\mathcal{G}}) [T(d_t^2+d_t+1)+1]$ real values and also incurs a latency of $(Q+D_{\mathcal{G}})(T+1)$ communication slots. Note that the Gibbs method of Algorithm \ref{alg_prod_gmm_dist} does not represent a Markov chain as the agents sample in parallel and not sequentially as in Algorithm \ref{alg_prod_gmm_cent}. The existence of a stationary distribution as well as the convergence of the samples drawn with Algorithm \ref{alg_prod_gmm_dist} to such a stationary distribution is not guaranteed outside of special cases \cite{bibl:wilsky_hogwild_gibbs_NIPS2013}. Nonetheless, Hogwild! Gibbs methods have been successfully employed in latent Dirichlet Allocation~\cite{bibl:semola_ldaGibbs2010}. In Section~\ref{sec_sim}, we numerically show the performance of the SCS-MTT filter with Algorithm \ref{alg_prod_gmm_dist} to be close to that of the centralized filter, where a fusion center has access to all the measurements, and carries out all the computations.	 
	 
\subsubsection{Computation of the GM extrinsic information $\delta_k^s (\cdot)$} \label{sec_delta_GMM}
The computation of the $\delta_k^s$ messages in (\ref{eq_msg_delta}) requires again the product of several GM likelihood terms available at different agents in the network. Note that since both $b_k(\mathbf{x}_{k},1)$ (obtained via Algorithm \ref{alg_prod_gmm_dist}) and $\gamma_k^s(\mathbf{x}_{k},1)$ are available as GMs at agent $s$, a GM approximation for the extrinsic information $\delta_k^{s}$ can be constructed in the following manner. First note from line~\ref{alg3_diff} of Algorithm~\ref{alg_prod_gmm_dist}, that the parameters $ \bm{\Xi}^{(\bm{i})} - ( \mathbf{C}_{s \to k}^{(i_s)} )^{-1}$, $ \bm{\xi}^{(\bm{i})}-\tilde{\mathbf{e}}_{s \to k}^{(i_s)}$ and $ {\xi}^{(\bm{i})} -c_{s \to k}^{(i_s)}$ characterize the product $\prod_{\ell \neq s} u_{\ell \to k}^{(i_{\ell})} \mathcal{N}(\mathbf{e}_{\ell\to k}^{(i_{\ell})}; \mathbf{H}_{\ell \to k}^{(i_{\ell})}\mathbf{x}_k, \mathbf{C}_{\ell \to k}^{(i_{\ell})})$ for a label vector $\bm{i}$. Thus, at each iteration of Algorithm~\ref{alg_prod_gmm_dist}, we can construct the following Gaussian components $\{(\omega_{k\to s}^{\delta,(j)}, \bm{\mu}_{k\to s}^{\delta,(j)}, \bm{\Omega}_{k\to s}^{\delta,(j)})\}_{j=1}^{J_{k\to s}}$ of $\delta_k^{s} (\mathbf{x}_{k}, 1)$ (with the same notations as in Figure \ref{fig_flowchart}) as
\begin{align}
& \bm{\Omega}_{k\to s}^{\delta,(j)} = \big[(\bm{\Omega}_{\to n, k}^{(j)})^{-1} + \bm{\Xi}^{(\bm{i})} - ( \mathbf{C}_{s\to k}^{(i_s)})^{-1}\big]^{-1}, \nonumber\\ 
& \bm{\mu}_{k\to s}^{\delta,(j)} = \bm{\Omega}_{k\to s}^{\delta,(j)} \big[  (\bm{\Omega}_{\to n,k }^{(j)})^{-1}\bm{\mu}_{\to n,k}^{(j)} + \bm{\xi}^{(\bm{i})}-\tilde{\mathbf{e}}_{s\to k}^{(i_s)}  \big], \nonumber\\
& \omega_{k\to s}^{\delta,(j)} = \omega_{\to n, k}^{(j)}  \exp \left( {\xi}^{(\bm{i})} -\mfrac{1}{2} \big( \bm{\mu}_{\to n,k}^{(j)} \big)^{T} \big( \bm{\Omega}_{\to n,k }^{(j)} \big)^{-1} \bm{\mu}_{\to n,k}^{(j)} \right) \nonumber\\
& \times \exp \left( \mfrac{1}{2} (\bm{\mu}_{k\to s}^{\delta,(j)})^T (\bm{\Omega}_{k\to s}^{\delta,(j)})^{-1}\bm{\mu}_{k\to s}^{\delta,(j)} - c_{s\to k}^{(i_s)} \right) \sqrt{\mfrac{\det(\bm{\Omega}_{k\to s}^{\delta,(j)})}{\det(\bm{\Omega}_{\to n, k}^{(j)})}}. \nonumber
\end{align}
Note that the expressions above are analogous to the GM parameters for $b(\mathbf{x}_k,1)$ in (\ref{eq_prod_gm_exact_P_tg})-(\ref{eq_prod_gm_exact_w_tg}). The only difference being the absence of the terms (\ref{eq_tg_gamma_param}) corresponding to $\gamma_k^s (\mathbf{x}_k,1)$ After the $T$ iterations of Algorithm \ref{alg_prod_gmm_dist}, the Gaussian components with highest distinct weights $\omega_{k\to s}^{\delta,(j)} $ are retained to form an approximation of $\delta_k^{s} ( \mathbf{x}_{k}, 1)$ while $\delta_k^{s} ( \mathbf{x}_{k},0)$ is obtained as $b( \mathbf{x}_{k},0)/ \eta_k^s(0)$. This procedure allows for the local computation of an approximate GM representation for $\delta_k^{s} ( \mathbf{x}_{k}, r_k)$ without additional network consensus operations.

\section{Decentralized Gaussian SCS-MTT filter}
\label{sec_dist_gauss}
A special case of the GM SCS-MTT filter of the previous section is obtained when all the agent and target densities are represented as single Gaussians. Single Gaussian expressions for the messages exchanged by the SCS-MTT filter can be obtained by specializing the expressions in Section~\ref{sec_gmm}. However, due to the measurement-to-target association uncertainty (see assumption (A8) of Section~\ref{sec:assumptions} and the summation over $m$ in (\ref{eq_msg_Lambda}), (\ref{eq_msg_gamma})), the agent and target beliefs become GMs even if their predicted messages are single Gaussians. The Probabilistic Data Association filter~\cite{bibl:bar_shalom_PDAF2009}, addresses this by performing a single Gaussian approximation of the resulting GM via first and second order moment matching. This is applied straightforwardly to the case of the agent beliefs $b_s(\cdot)$ and their extrinsic information $\theta_k^s(\cdot)$ as their GM computation is done locally as shown in Section~\ref{sec_ag_belief_gibbs}.

The DG-SCS-MTT filter achieves a single Gaussian representation of the target beliefs with a lower communication load than the Hogwild! Gibbs of Algorithm \ref{alg_prod_gmm_dist}. Suppose $b_s(\mathbf{x}_k,1) \triangleq \sqrt[S]{\alpha_{\to n}(\mathbf{x}_k,1)} \gamma_k^s(\mathbf{x}_k,1)$. Observe that (\ref{eq_tg_belief}) becomes $b(\mathbf{x}_k,1) = \prod_{s\in \mathcal{S}} b_s(\mathbf{x}_k,1)$. If $\alpha_{\to n}(\mathbf{x}_k,1) = c_k \mathcal{N}(\mathbf{x}_k; \mathbf{m}, \mathbf{P}) $ then $\sqrt[S]{\alpha_{\to n}(\mathbf{x}_k,1)} = c_k'\mathcal{N}(\mathbf{x}_k; \mathbf{m}, S\mathbf{P})$ is a scaled Gaussian~\cite[eq. 36]{battistelli13_jstsp}, where $c_k' = \sqrt[S]{c_k} \frac{(\det(2\pi S \mathbf{P} ))^{1/2}}{\sqrt[2S]{\det(2\pi \mathbf{P})}}$. Furthermore, let $\gamma_k^s(\mathbf{x}_k,1) $ be a GM of the form (\ref{eq_msg_gamma_gmm}). Then, a locally computed GM $b_s(\mathbf{x}_k, 1) = \sum_{i=1}^{I_s} w_s^{(i)} \mathcal{N}(\mathbf{x}_k; \mathbf{m}_s^{(i)}, \mathbf{P}_s^{(i)}) $ is given as a special case of (\ref{eq_tg_gen_gmm_prod1}) with $J_{\to n, k}=S=1$. The scaled single Gaussian $\hat{b}_s(\mathbf{x}_k,1) = \hat{c}_s \mathcal{N}(\mathbf{x}_k;\hat{\mathbf{m}}_s, \hat{\mathbf{P}}_s)$ that matches the first and second order moments of the GM $b_s(\mathbf{x}_k, 1)$ has parameters~\cite{bibl:orguner_GM2G_2007}
\begin{align}
    \hat{c}_s &= \sum\nolimits_{i=1}^{I_s} w_s^{(i)}, \hspace{1cm}	\hat{\mathbf{m}}_s = \mfrac{1}{\hat{c}_s} \sum\nolimits_{i=1}^{I_s} w_s^{(i)} \mathbf{m}_s^{(i)}, \label{eq:hatm} \\
     \hat{\mathbf{P}}_s &= \mfrac{1}{\hat{c}_s} \sum\nolimits_{i=1}^{I_s} w_s^{(i)} \big[\mathbf{P}_s^{(i)} + \hspace{-1mm} (\mathbf{m}_s^{(i)} \hspace{-1mm} - \hspace{-1mm} \hat{\mathbf{m}}_s)(\mathbf{m}_s^{(i)} \hspace{-1mm} - \hspace{-1mm} \hat{\mathbf{m}}_s)^T \big]. \label{eq:hatP}
\end{align}
Note that (\ref{eq:hatP}) also accounts for the spread of the means of the initial GM. A global $\hat{b}(\mathbf{x}_k,1) = \hat{c} \mathcal{N}(\mathbf{x}_k;\hat{\mathbf{m}}, \hat{\mathbf{P}})$, as a single Gaussian approximation of $b(\mathbf{x}_k,1)$, is obtained via network (average and max) consensus over the weights $ \log(\hat{c}) = \sum_{s=1}^S \log(\hat{c}_s) $, matrices $ \hat{\mathbf{P}}^{-1} = \sum_{s=1}^S \hat{\mathbf{P}}_s^{-1} $ and vectors $\hat{\mathbf{m}} = \hat{\mathbf{P}} \sum_{s=1}^S \hat{\mathbf{P}}_s^{-1} \hat{\mathbf{m}}_s$. The computation of $b(\mathbf{x}_k, 0)$ remains unchanged from Section~\ref{sec_tg_belief_gm}. In contrast to the Hogwild! Gibbs of Algorithm \ref{alg_prod_gmm_dist}, the DG-SCS-MTT filter only performs network consensus once for each PT which involves an exchange of $(Q+D_{\mathcal{G}_k})( d_t^2 + d_t +2)$ real values at each outer-BP loop. Furthermore, we note that computing the local GM belief $b_s(\cdot, 1)$ takes $O ( I_s d_t^3 )$ operations; single Gaussian compression is $O ( I_{s\to k} d_t^2 )$; and the computations required for average consensus are $O ( Q d_t^2)$ (assuming the number of neighbors of an agent is small compared to $Q$). Hence, the overall computational complexity of the DG-SCS-MTT filter is $O ( I_{s\to k} d_t^3 )$ for each PT, at each outer loop iteration. 

The DG-SCS-MTT filter also achieves a single Gaussian approximation for the extrinsic information message $\delta_k^s (\cdot)$ without the need of additional network consensus operations. Based on the parameters of $b(\mathbf{x}_k,1)$, we evaluate $b_{\neg s}(\mathbf{x}_k, 1) \triangleq \hat{c}_{\neg s}\mathcal{N}(\mathbf{x}_k; \hat{\mathbf{m}}_{\neg s},\hat{\mathbf{P}}_{\neg s} ) $ where the parameters $\hat{c}_{\neg s} = \hat{c}_{}/ \hat{c}_{s}$, $\hat{\mathbf{P}}_{\neg s}^{-1} = \hat{\mathbf{P}}_{}^{-1} - \hat{\mathbf{P}}_{ s}^{-1}$,  and $\hat{\mathbf{m}}_{\neg s} = \hat{\mathbf{P}}_{\neg s} \left[ \hat{\mathbf{P}}^{-1}\hat{\mathbf{m}} - \hat{\mathbf{P}}_s^{-1} \hat{\mathbf{m}}_s \right]$ are computed locally. Finally, we evaluate $\delta_k^s (\mathbf{x}_k,1) = b_{\neg s}(\mathbf{x}_k, 1) \sqrt[S]{\alpha(\mathbf{x}_k, 1)}$, which has a scaled single Gaussian form, and $\delta_k^s (\mathbf{x}_k,0) = b(\mathbf{x}_k,0)/\eta_k^s(0)$.

\section{Simulation Results}
\label{sec_sim}

\begin{figure}[!t]
\centering
\begin{subfigure}[a]{.33\textwidth}
	\includegraphics[width=\textwidth]{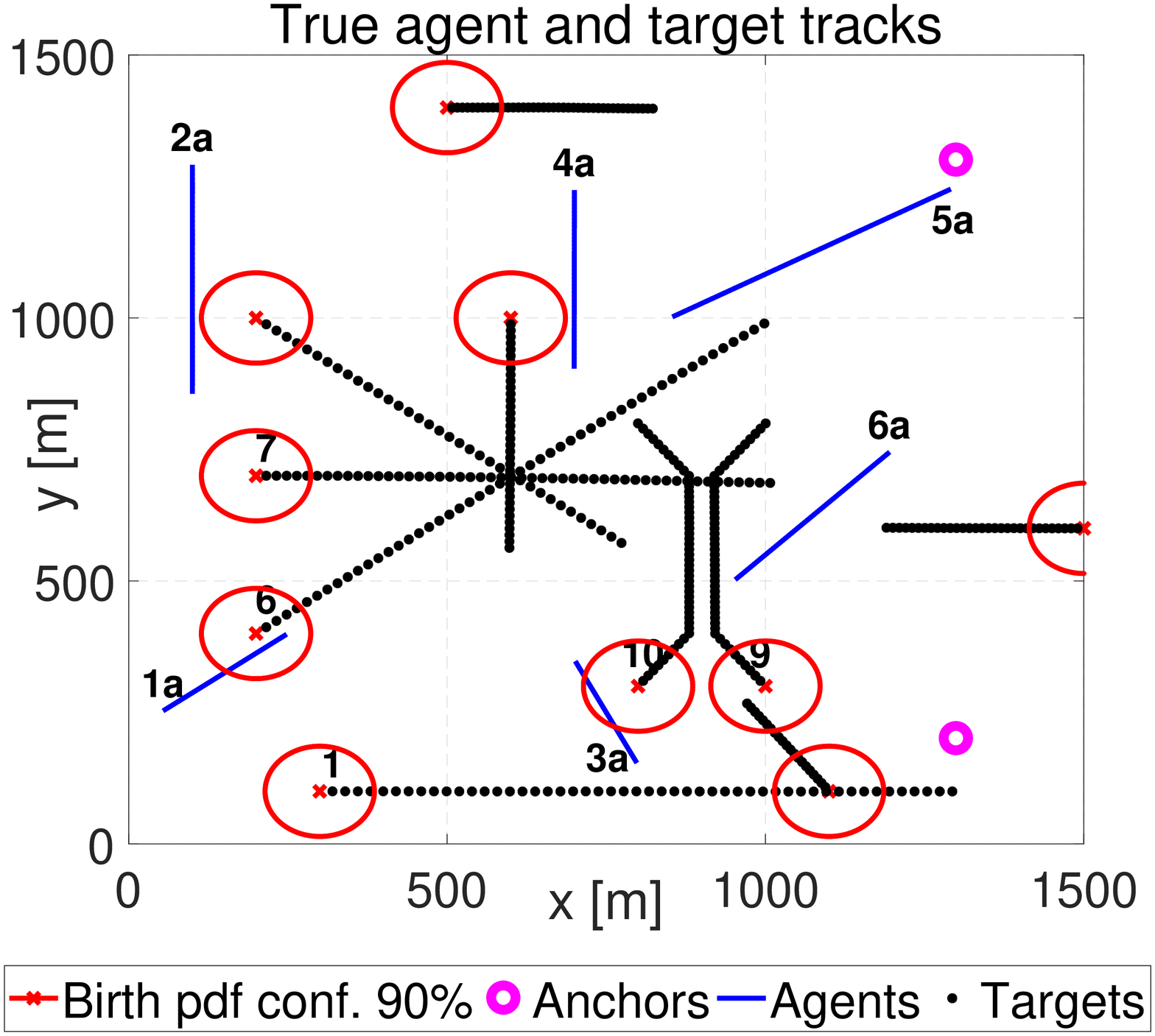}
	\caption*{}
	\label{fig_network1}
\end{subfigure}
\\
\vspace{-8mm}
\begin{subfigure}[a]{.3\textwidth} \includegraphics[width=\textwidth]{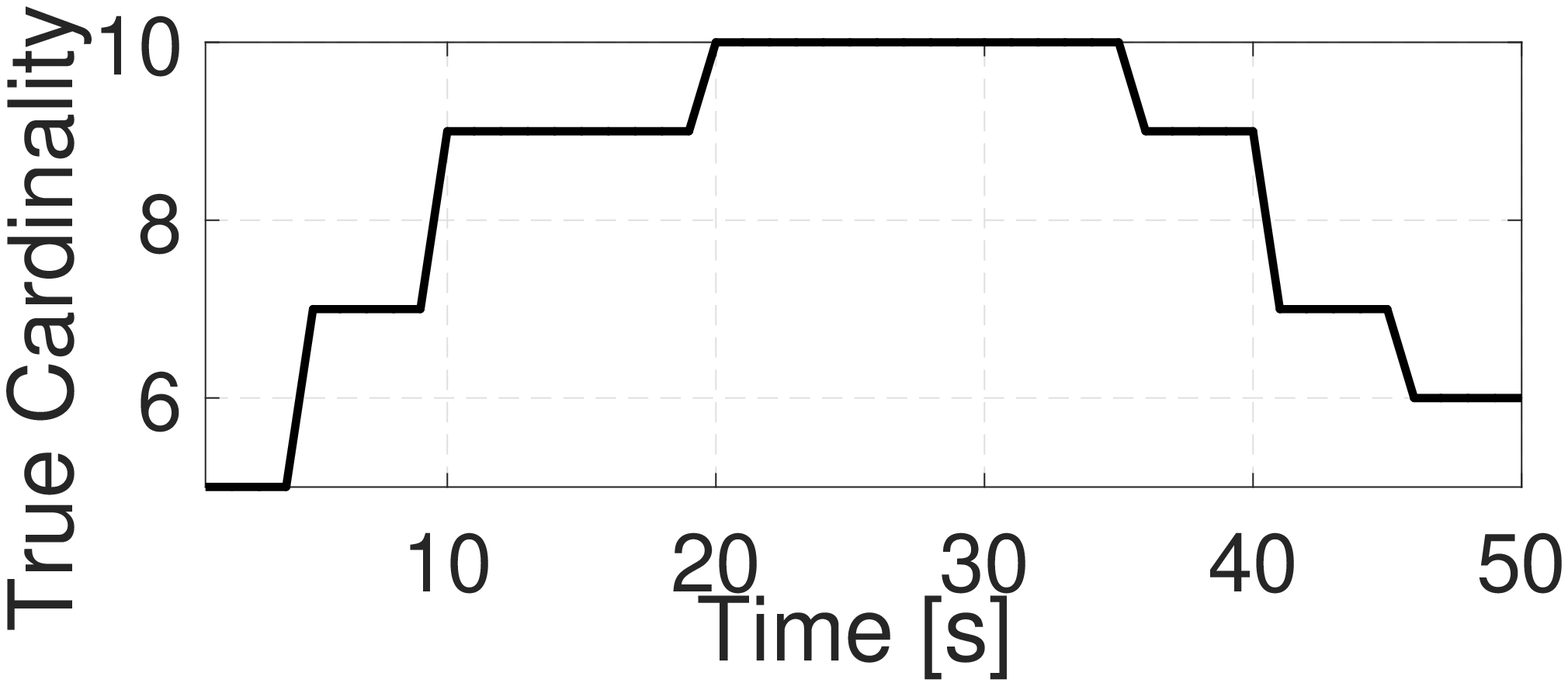}
	\caption*{}
	\label{fig_true_card}
\end{subfigure}
\vspace{-8mm}
\caption{Ground truth trajectories of agents and targets plotted over time (top), along with the true target cardinality over time (bottom). The uncertainty in target births is shown as red ellipses that delineate the area containing $90\%$ of the mass of $f_b(\mathbf{x}_{n,k})$.}
\label{fig_network}
\end{figure}

	
In this section, we numerically evaluate the performance of our proposed GM-SCS-MTT and G-SCS-MTT filters for both decentralized and centralized versions. The centralized GM (CGM-SCS-MTT) version employs the centralized Gibbs Algorithm \ref{alg_prod_gmm_cent} for both agent and PT beliefs while the decentralized GM (DGM-SCS-MTT) filter employs the Hogwild! Gibbs method of Algorithm \ref{alg_prod_gmm_dist} for PT beliefs. We also consider a reference method, named here SPAWN, which consists of the agent self-localization method of~\cite{spawn09} followed by the MTT method of~\cite{meyer17_tsp}. Centralized GM (CGM-SPAWN) and single Gaussian (CG-SPAWN) versions of SPAWN are employed, where the GM product of messages is computed using the centralized Gibbs Algorithm \ref{alg_prod_gmm_cent}. Figure~\ref{fig_network} shows the ground truth tracks of all the agents and targets, over a span of $50$ time steps and the true target cardinality as a function of time. Our network has two stationary agents (called anchors), $6$ mobile agents, and a maximum of $10$ targets over a $\left[ 0,1500 \text{m} \right] \times \left[ 0,1500 \text{m} \right]$ region of interest (ROI). Each mobile agent has a measurement and communication range of $1000$m. The anchors have a communication range of $1000$m and a measurement range of $1500$m. 
	

Agent and target state vectors are constructed as $\mathbf{x} = [p_x,\, p_y, \, \dot{p}_x, \, \dot{p}_y]^T$, where $p_x$ and $p_y$ represent the $x$-$y$ target coordinates and $\dot{p}_x$ and $\dot{p}_y$ are its velocity components along the two axes. All targets have the same dynamical model $f(\mathbf{x}_n \vert \mathbf{x}_{n-1}) = \mathcal{N}(\mathbf{x}_n; \mathbf{B}_{n} \mathbf{x}_{n-1}, \bm{\Sigma}_n)$, where the state transition matrix is
\mbox{$\mathbf{B}_n = \bigl [\begin{smallmatrix}
	\mathbf{I}_2   & T_S\, \mathbf{I}_2           \\
	\mathbf{0}_2  & \mathbf{I}_2          
	\end{smallmatrix} \bigr]$}
with a sampling period of $T_s = 1$s and $\mathbf{0}_n$ and $\mathbf{I}_n$ are the zero and identity matrices of size $n \times n$. The covariance matrix is
\mbox{$\bm{\Sigma}_n = \sigma_q^2\bigl [\begin{smallmatrix}
	0.25{T_S^4}\mathbf{I}_2   & 0.5{T_S^3}\, \mathbf{I}_2           \\
	0.5{T_S^3}\mathbf{I}_2  & T_S^2 \mathbf{I}_2          
	\end{smallmatrix} \bigr]$}, with $\sigma_q = 0.5$. Similarly, all agents have the same linear-Gaussian kinematic model with $\sigma_q = 0.1$. Each agent $s$, with coordinates $(p_x^{s}, p_y^{s})$, observes with probability $P_D^{s}$ a target with state vector $\mathbf{x}$ through a range-bearing model:  
\begin{equation}
\mathbf{z}_{n}^{s} = 
\begin{bmatrix}
\sqrt{(p_x-p_x^{s})^2 + (p_y-p_y^{s})^2} \\
\text{tan}^{-1}(\frac{p_y-p_y^{s}}{p_x-p_x^{s}})        
\end{bmatrix}
+ \mathbf{n}_{n}^{s}
\label{eq:range_bearing}
\end{equation}
where the measurement noise is $\mathbf{n}_{n}^{s} \sim \mathcal{N}(0, \mathbf{R}_n)$. The same range-bearing measurement model (with potentially different parameter values) is employed for inter-agent measurements. Linearization of the non-linear range-bearing observation model is performed before applying the GM or Gaussian SCS-MTT filter. Similar to the extended Kalman filter~\cite[Ch. 2.1]{bibl:beyondKalman}, this is achieved locally at each agent by evaluating the Jacobian of the transformation (\ref{eq:range_bearing}) at the weighted mean of the PT prediction message $\alpha_{\to n}(\cdot)$. Analogously, the inter-agent range-bearing measurement model is linearized with the Jacobian being evaluated at the mean of the agent prediction message $\phi_{\to n}(\cdot)$. Birthed PTs are appended to the existing PTs in the prediction step of the filters. The birth locations are shown in Figure \ref{fig_network}. Unless stated otherwise, the birth probabilities of existence are set to $0.25$, the probability of target survival $P_s=0.99$, the probability of target detection $P_D^s = 0.95$, and the measurement noise covariance $\mathbf{R}_n = \text{diag} (10, 100)$. At each frame, the clutter process for agent $s$ follows a Poisson distribution with rate $\lambda^s=25$, and the clutter points are distributed uniformly over the ROI. Target inference is achieved as indicated in Section \ref{sec:agtg_infer}. The number of outer BP iterations is fixed to $P=1$, to avoid over-confident beliefs \cite{meyer16_tsipn}.

In the GM filters, the initial positions of the agents are modeled using Gaussian mixture densities. More precisely, each agent track is initialized with $4$ equal-weighted GM components, with means at a distance of $R_a=50$m along the $x$ and $y$ directions, from the position shown in Figure \ref{fig_network}. All the $4$ GM components have the same covariance $\text{diag} (1600, 1600, 40, 40)$. In the single Gaussian filters, the mean and covariance of the agent tracks are initialized by the respective values achieved via moment matching \cite{bibl:orguner_GM2G_2007}. Target tracks are initialized with single Gaussian densities in all filters, with means given by the birth locations and covariance matrices $\text{diag} (1600, 1600, 16, 16)$.

\begin{figure*}[t!]
	\centering
	\begin{subfigure}[a]{0.49\textwidth}
		\includegraphics[width=0.49\textwidth]{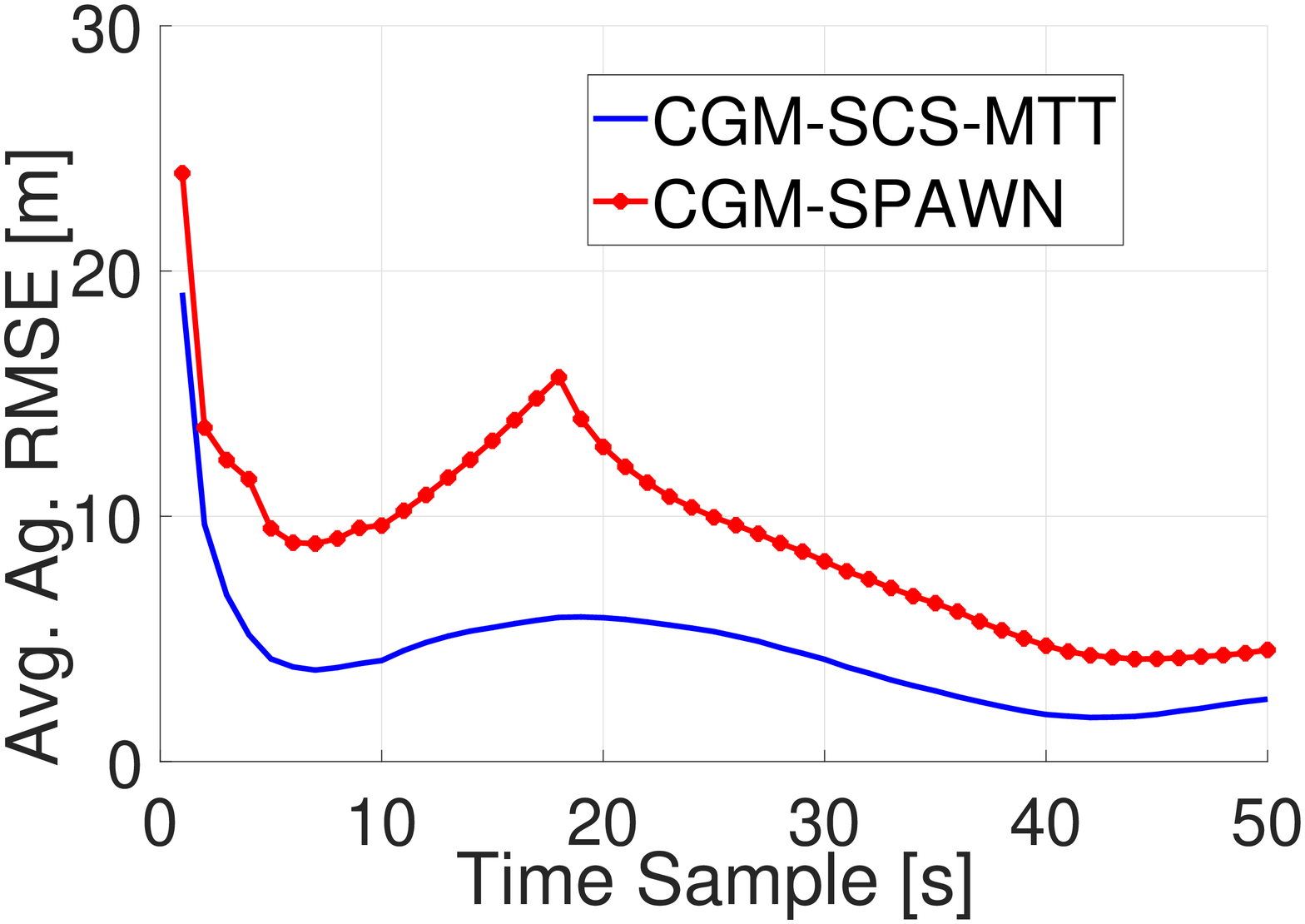}
		\hfill
	    \includegraphics[width=0.49\textwidth]{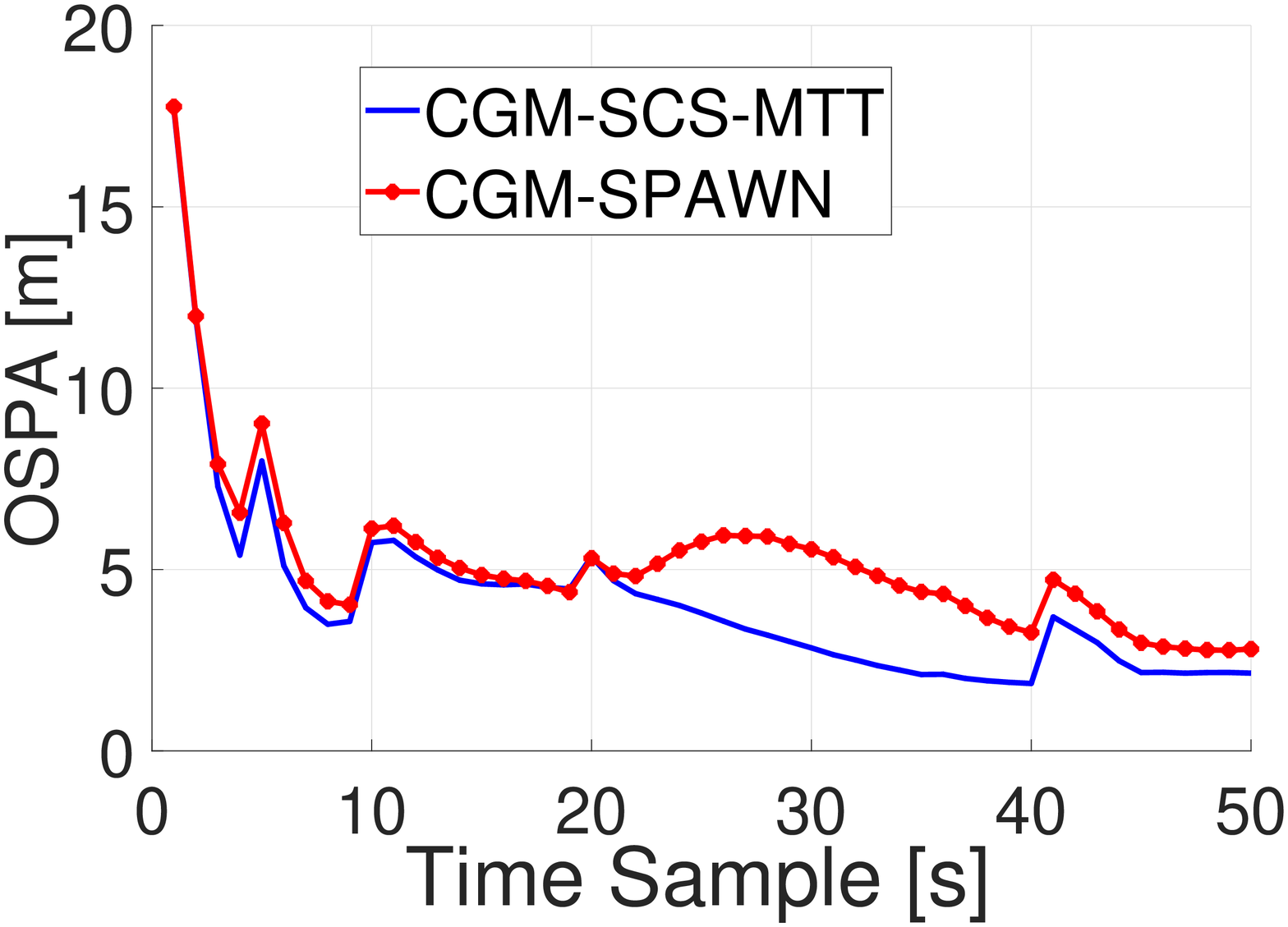}
		\caption{CGM-SCS-MTT vs CGM-SPAWN filters}
		\label{fig_spawn_vs_slt_gauss}
	\end{subfigure}
	\hfill
	\begin{subfigure}[a]{0.49\textwidth}
		\includegraphics[width=0.49\textwidth]{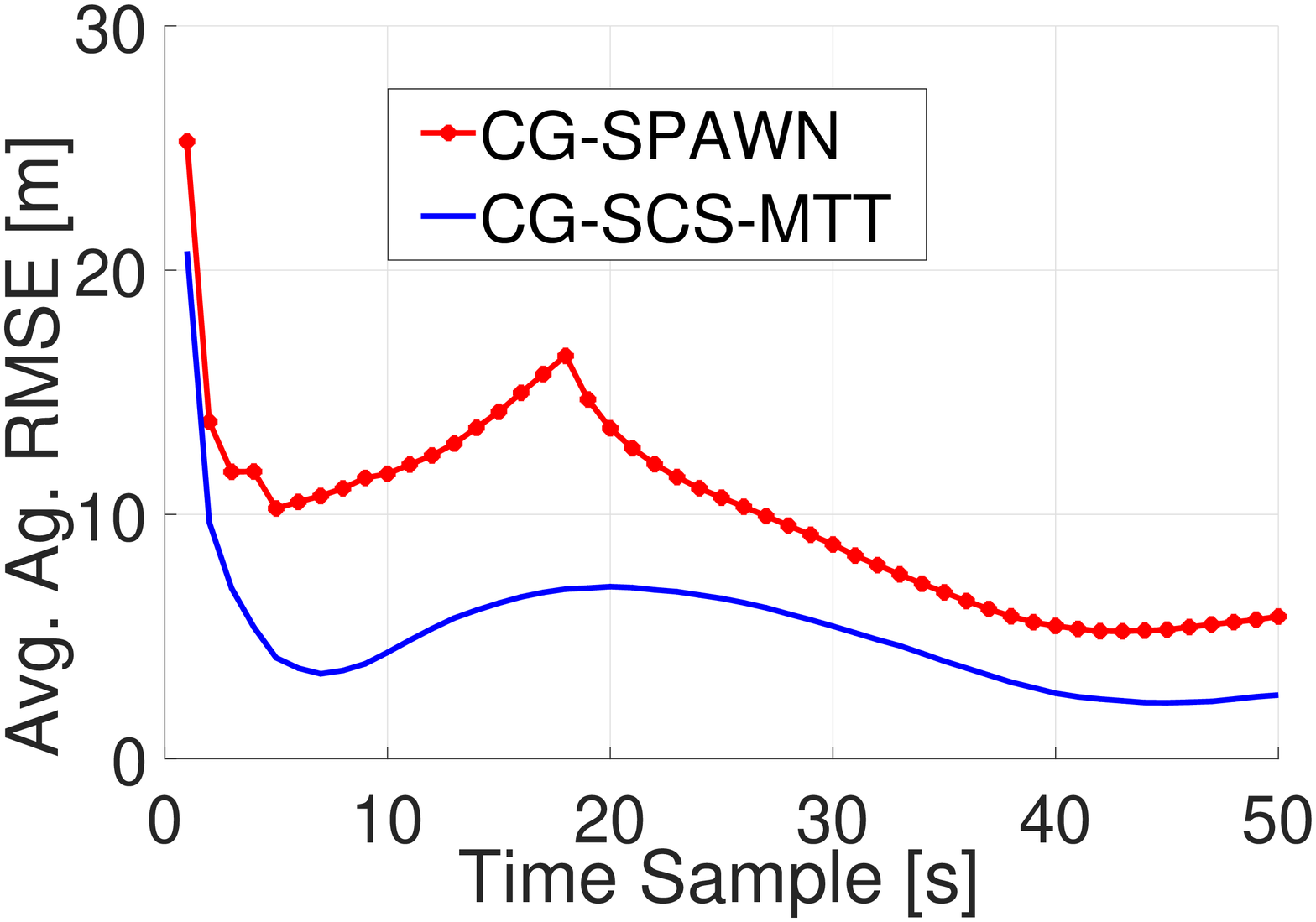}
		\hfill
	    \includegraphics[width=0.49\textwidth]{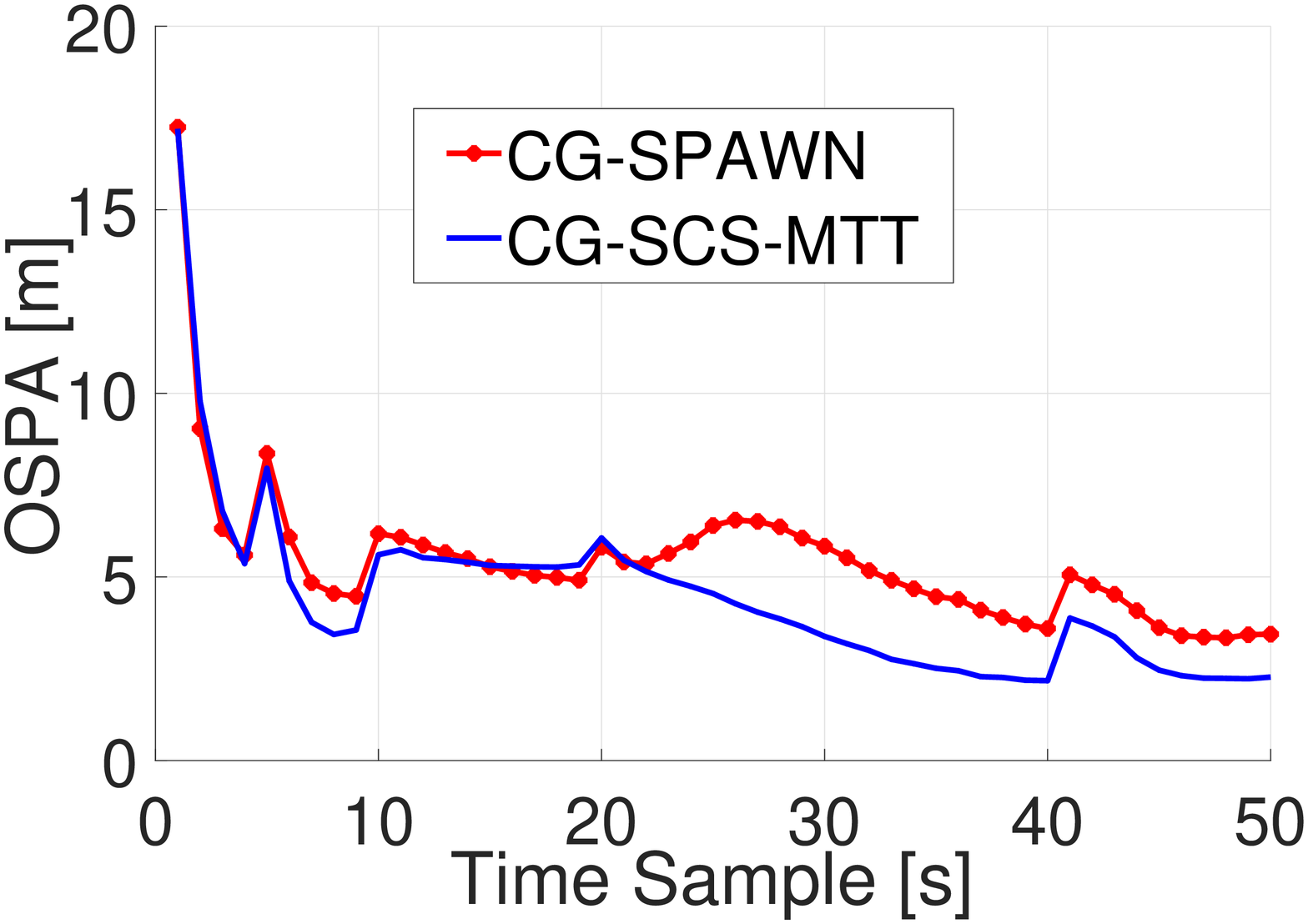}
		\caption{CG-SCS-MTT vs CG-SPAWN filters}
		\label{fig_spawn_vs_slt_gm}
	\end{subfigure}
	\caption{ Comparison of the average agent RMSE and target OSPA error for SPAWN and the proposed SCS-MTT filter.
	}
	\label{fig_spawn_vs_slt}
\end{figure*}

\begin{figure*}
	\centering
	\begin{subfigure}[a]{0.49\textwidth}
		\includegraphics[width=0.49\textwidth]{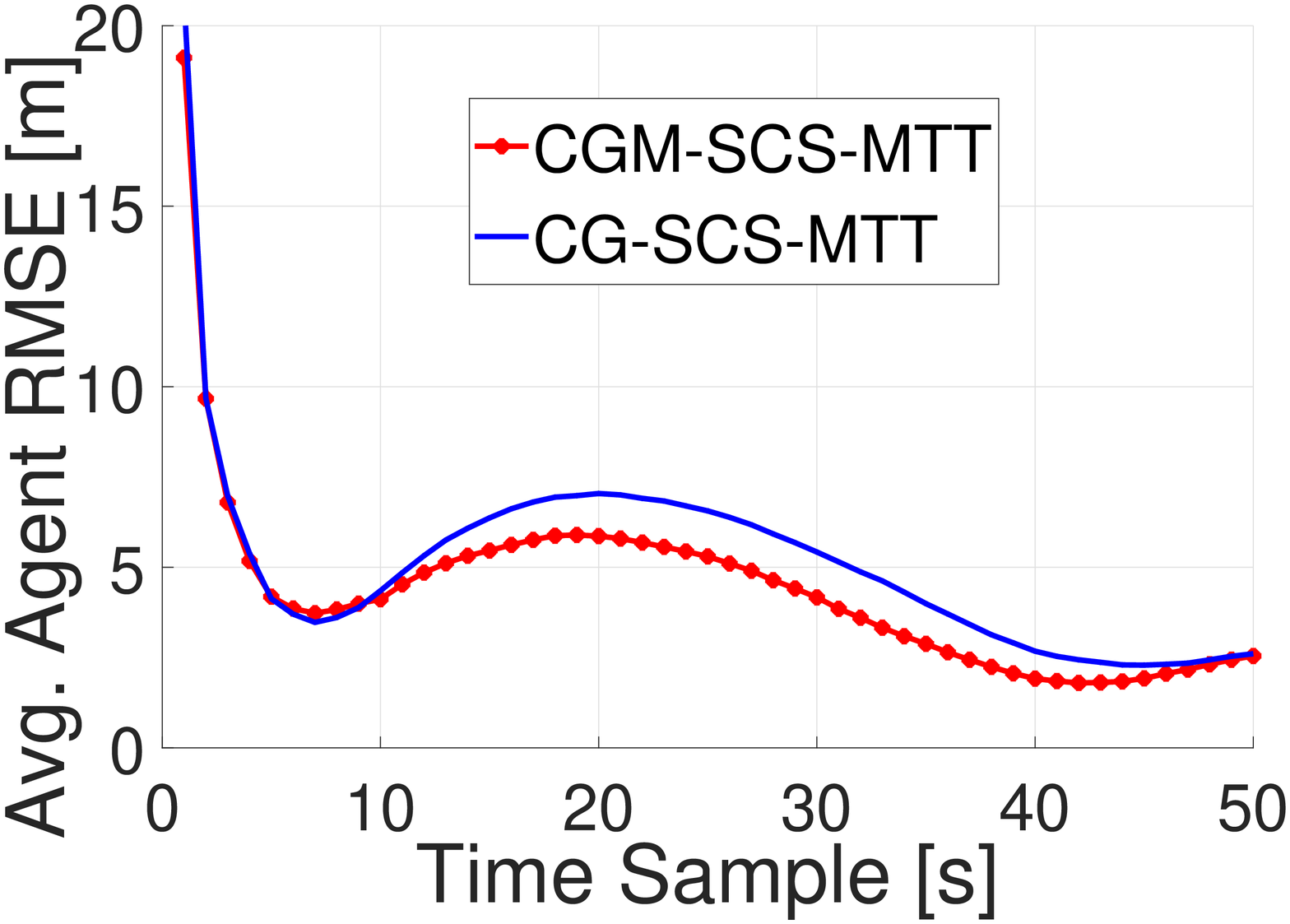}
		\hfill
	    \includegraphics[width=0.49\textwidth]{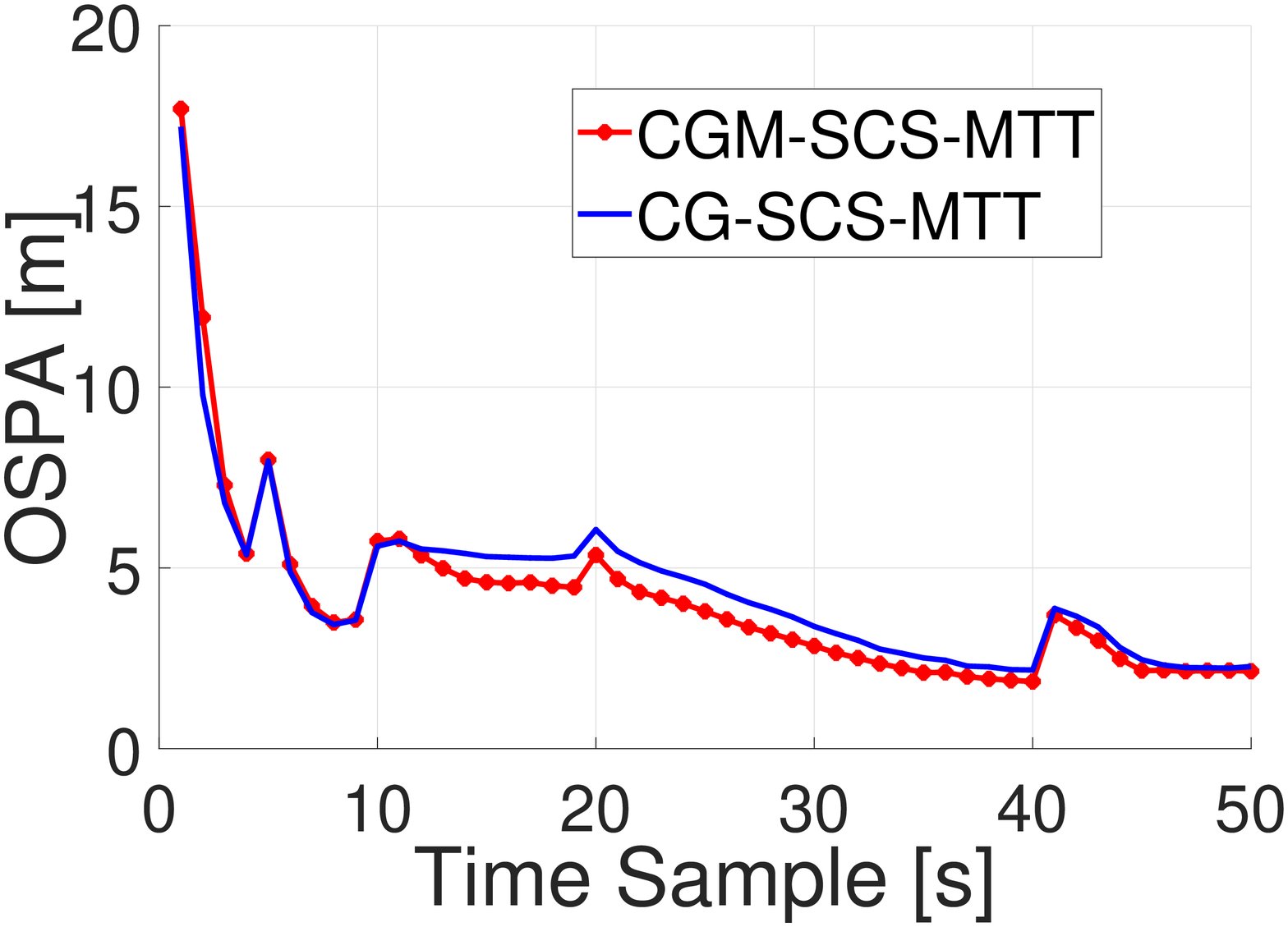}
		\caption{CGM-SCS-MTT vs CG-SCS-MTT (centralized filters)}
		\label{fig_Gauss_vs_gm_cent}
	\end{subfigure}
	\hfill
	\begin{subfigure}[a]{0.49\textwidth}
		\includegraphics[width=0.49\textwidth]{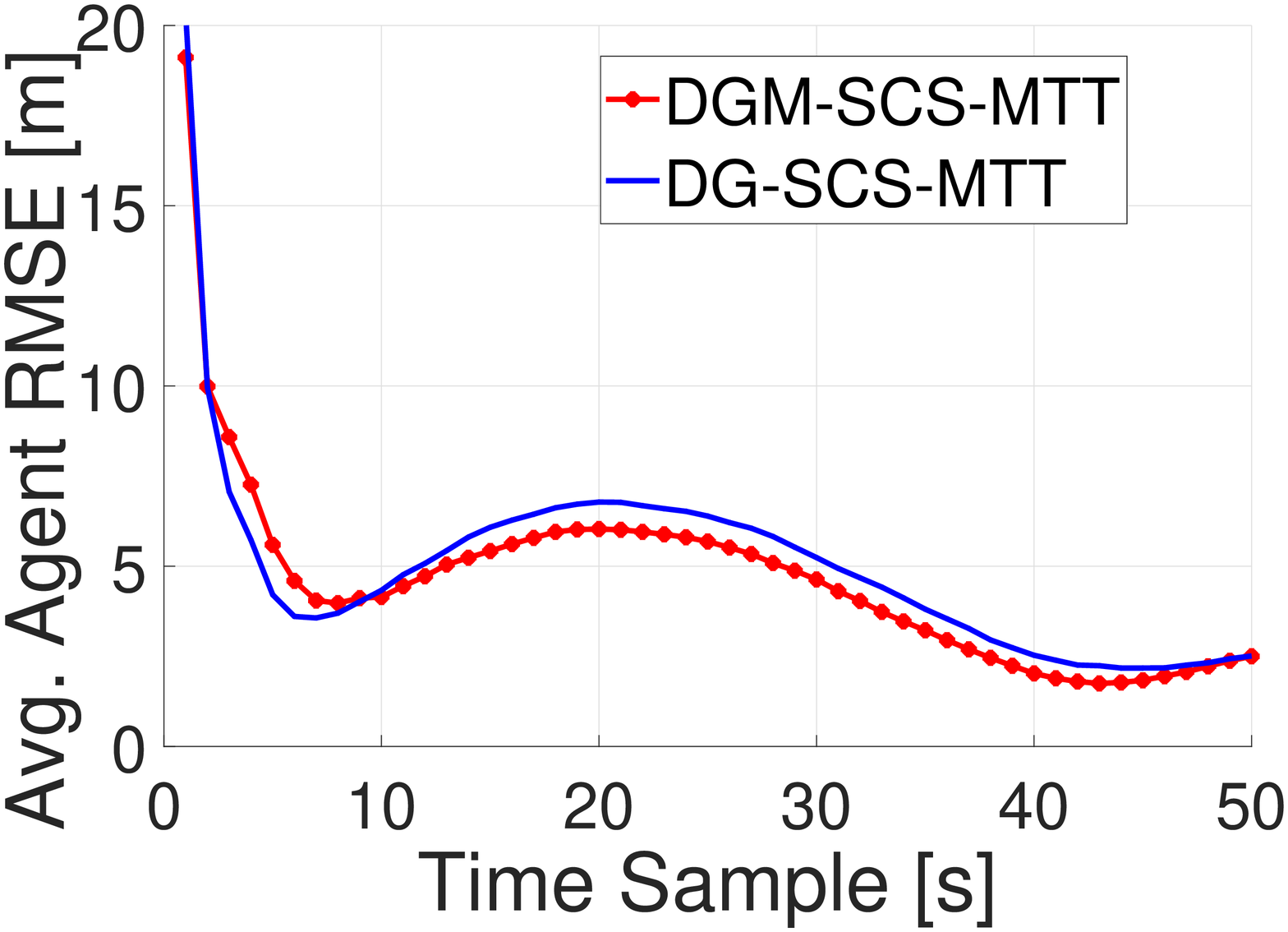}
		\hfill
	    \includegraphics[width=0.49\textwidth]{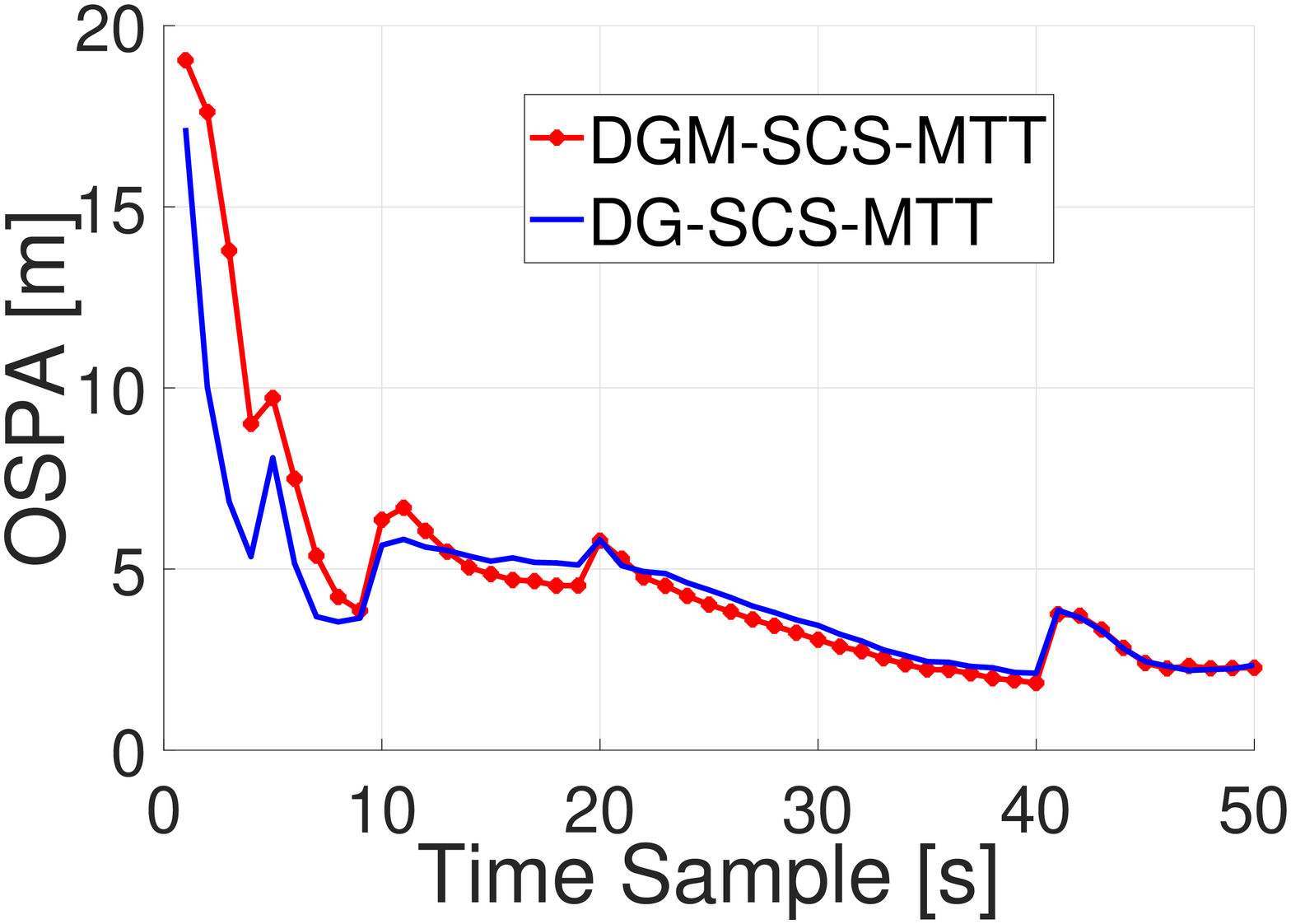}
		\caption{DGM-SCS-MTT vs DG-SCS-MTT (decentralized filters)}
		\label{fig_Gauss_vs_gm_dist}
	\end{subfigure}
	\caption{Comparison of the average agent RMSE and target OSPA error, for the single Gaussian (G) and Gaussian mixture (GM) filters. 
	}
	\label{fig_Gauss_vs_gm}
\end{figure*}

\begin{figure*}
	\centering
	\begin{subfigure}[a]{0.28\textwidth}
		\includegraphics[width=\textwidth]{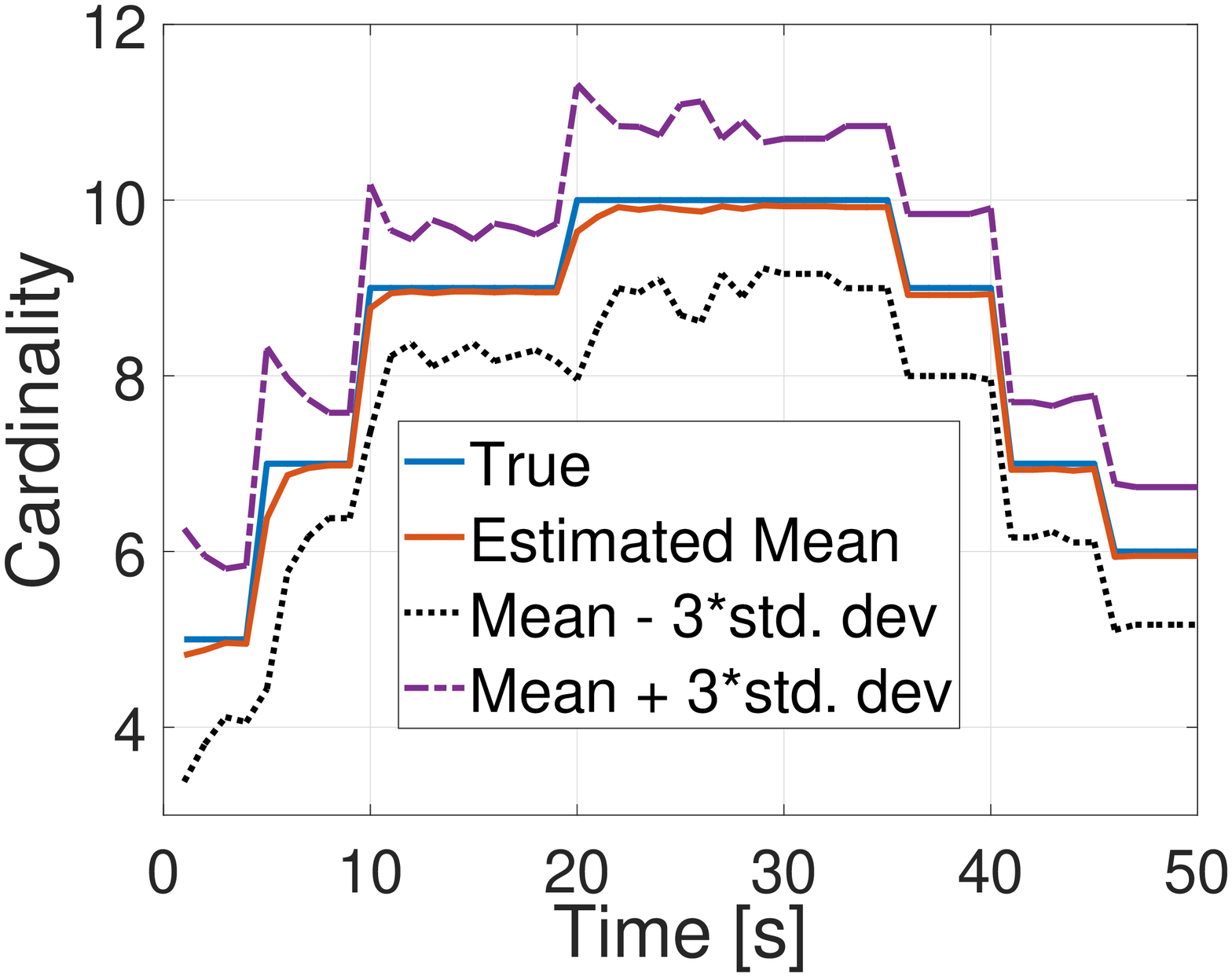}
		\caption{CGM-SPAWN filter}
		\label{fig_card_spawn}
	\end{subfigure}
	~
	\begin{subfigure}[a]{0.28\textwidth}
		\includegraphics[width=\textwidth]{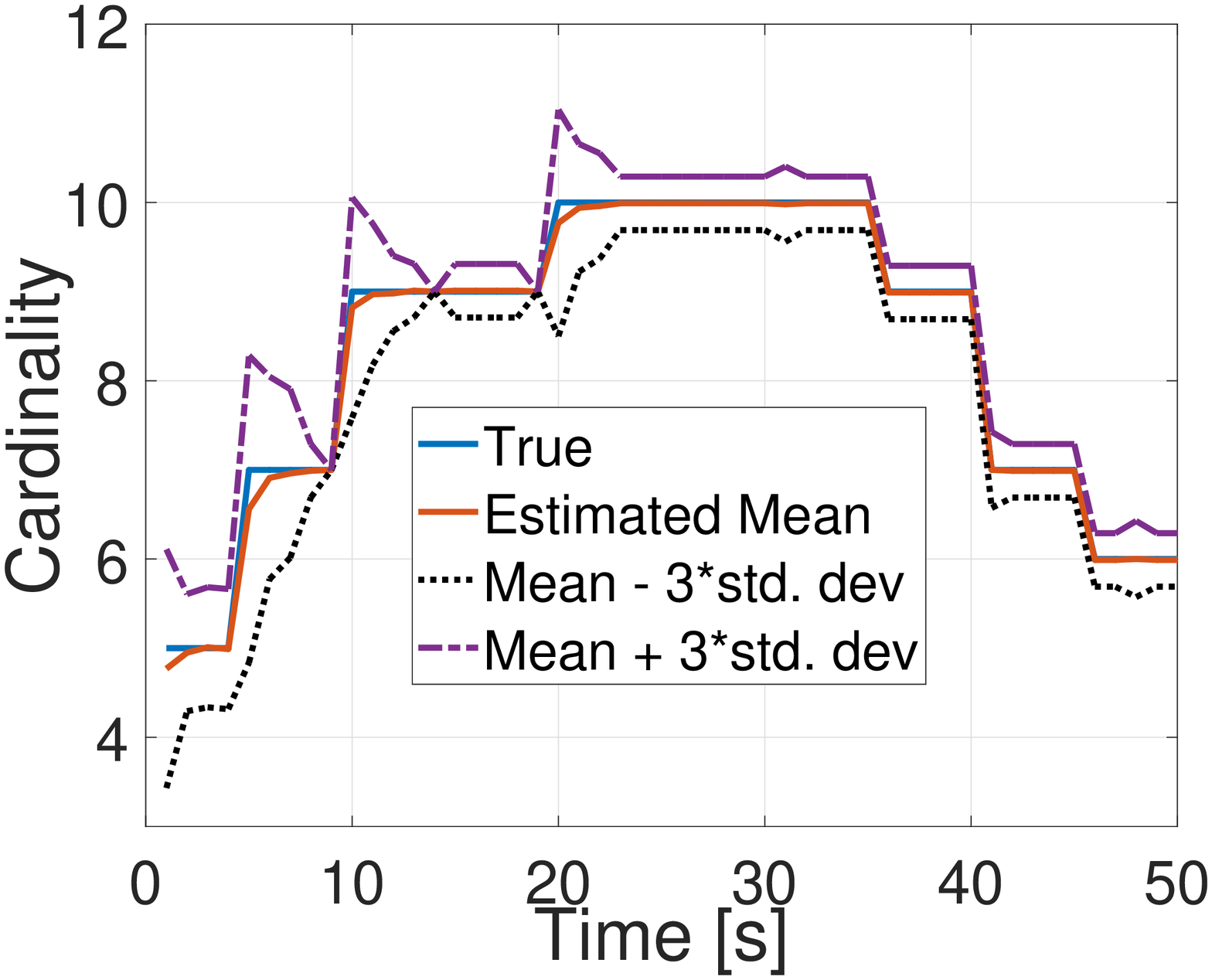}
		\caption{CGM-SCS-MTT filter}
		\label{fig_card_Gibbs}
	\end{subfigure}
	~
	\begin{subfigure}[a]{0.28\textwidth}
		\includegraphics[width=\textwidth]{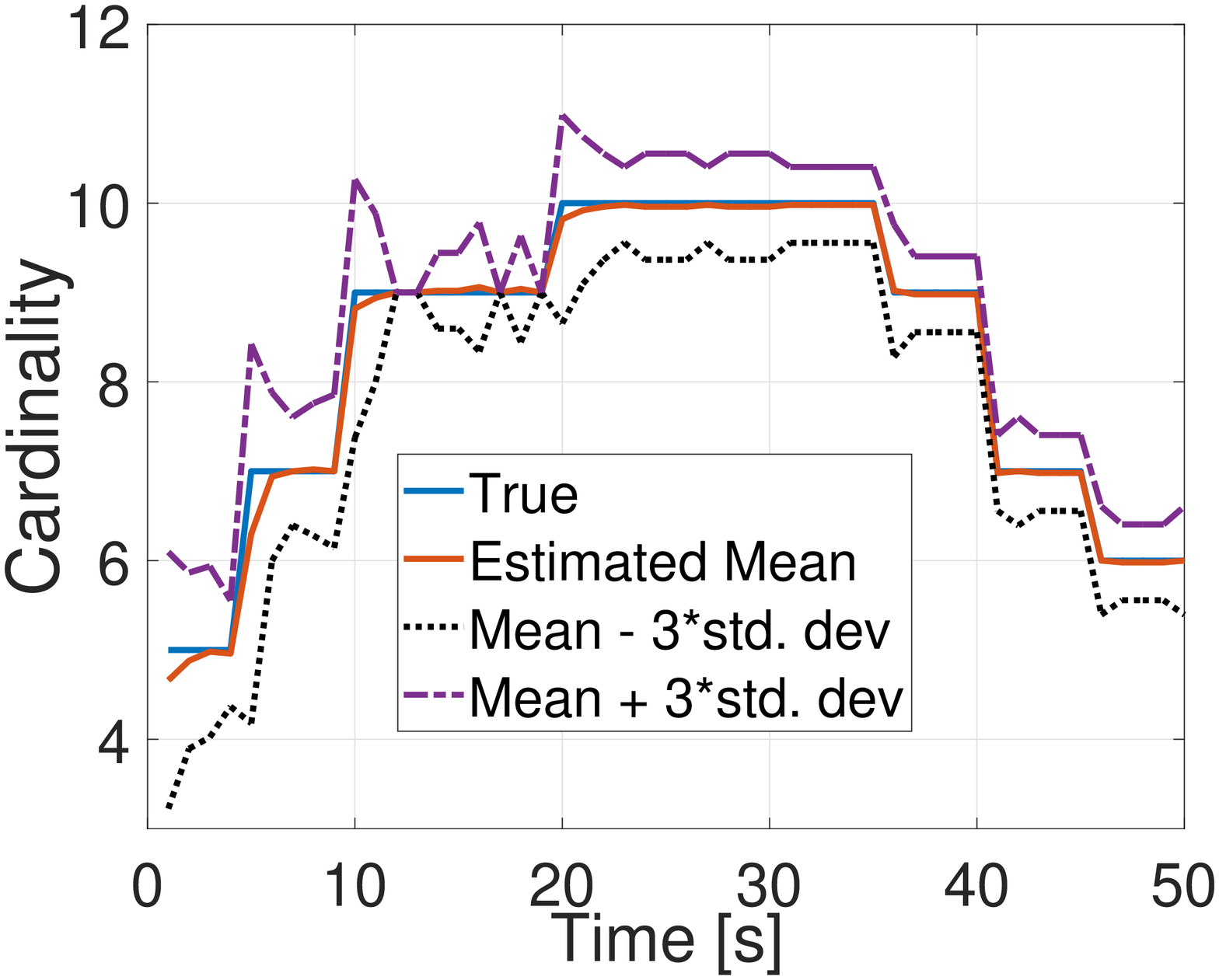}
		\caption{DGM-SCS-MTT filter}
		\label{fig_card_dist}
	\end{subfigure}
	\caption{Comparison of cardinality estimates for the different GM filters. 
	}
	\label{fig_card}
\end{figure*}

\begin{figure}
	\centering
	\begin{subfigure}[a]{.33\textwidth}
		\includegraphics[width=\textwidth]{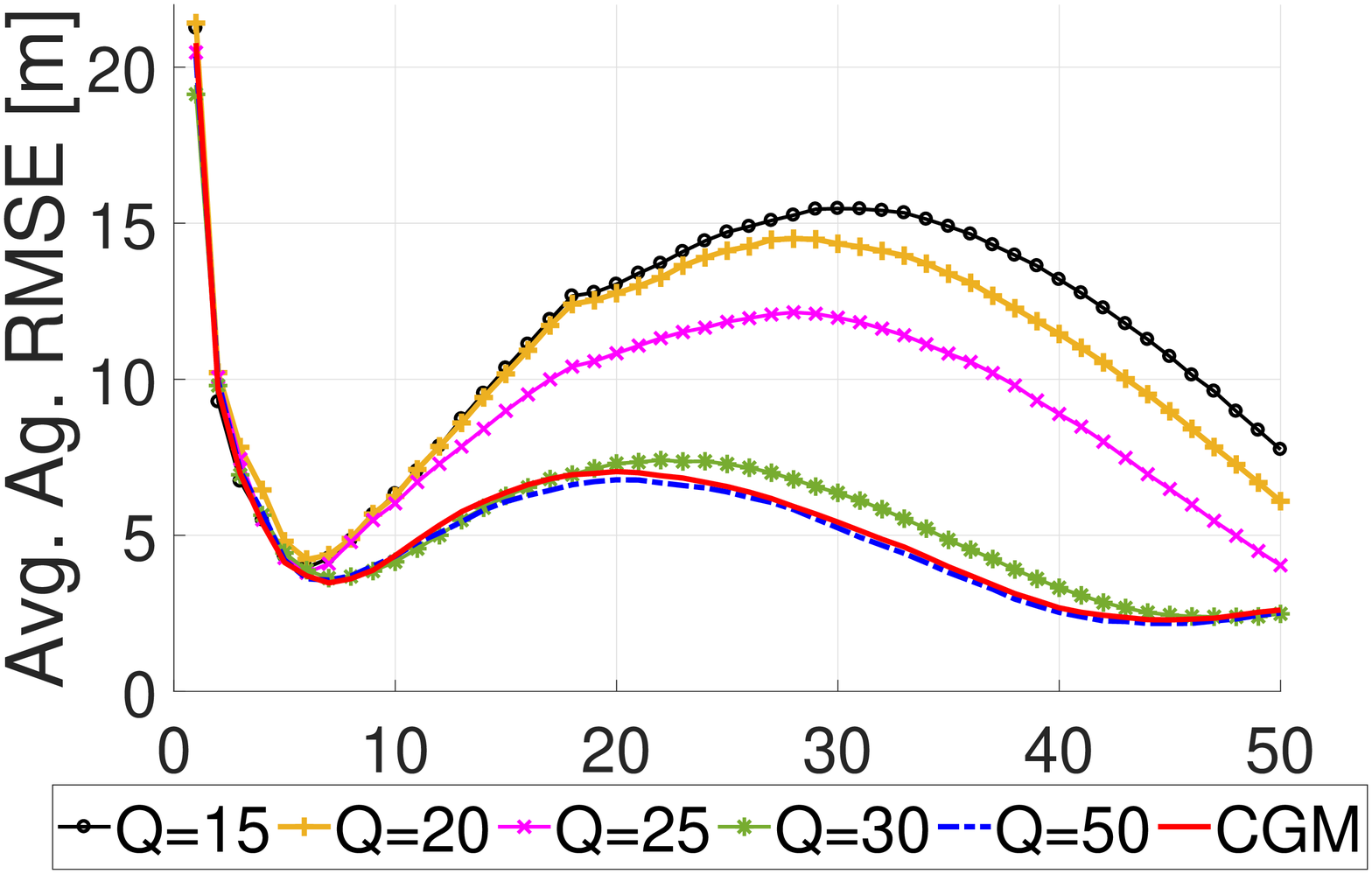}
		\caption*{}
		\label{fig_cent_vs_dist_GM_ag}
	\end{subfigure}
	\vspace{-5mm}
	\begin{subfigure}[a]{.33\textwidth} \includegraphics[width=\textwidth]{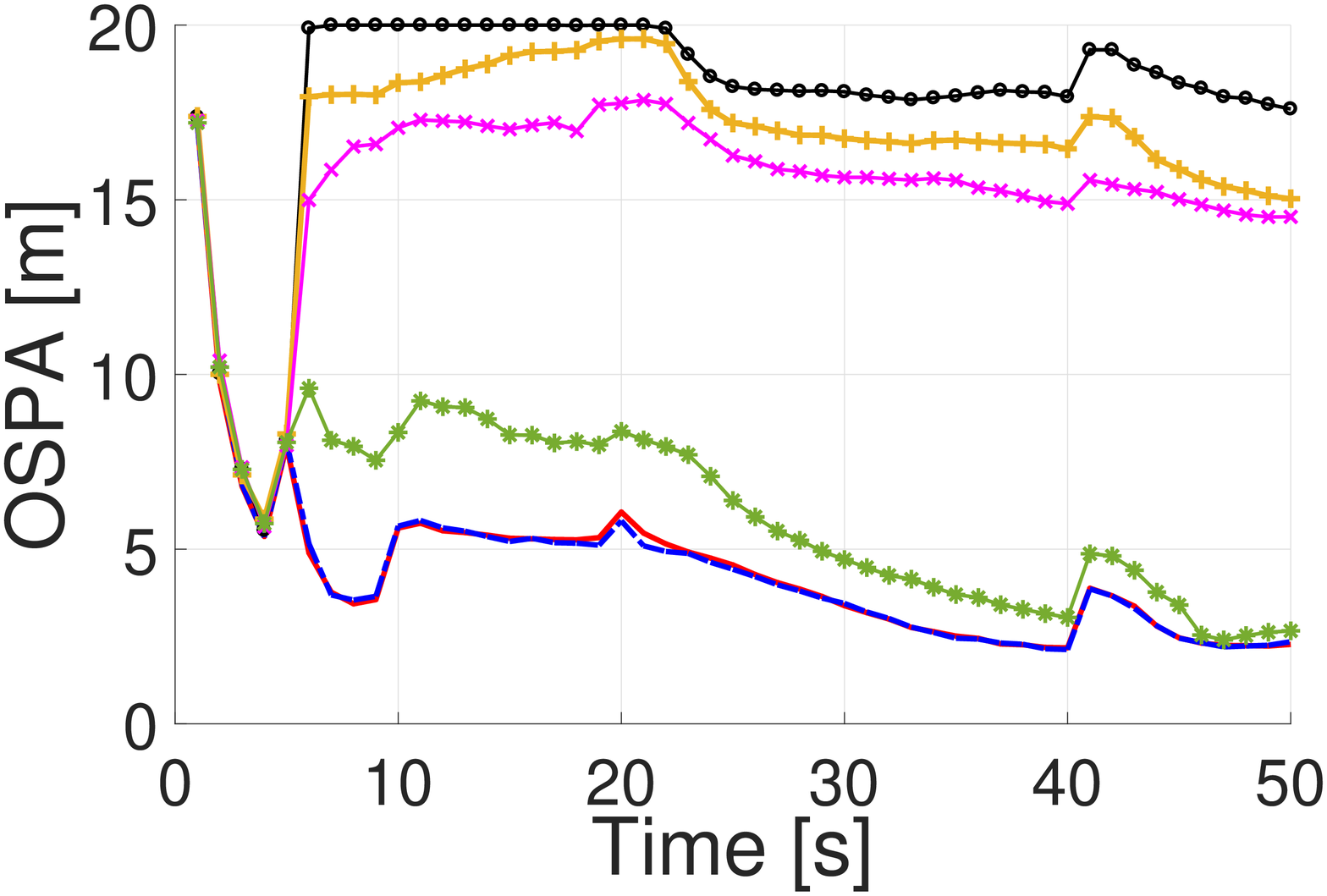}
		\caption*{}
		\label{fig_cent_vs_dist_GM_tg}
	\end{subfigure}
	\caption{Comparison of average agent RMSE and target OSPA error for the DGM-SCS-MTT and CGM-SCS-MTT filters. We have plotted the DGM curves for $Q=15, 20, 25, 30, 50$ consensus iterations. 
	}
	\label{fig_cent_vs_dist_gmm}
\end{figure}

Keeping the agent and target tracks fixed, $100$ independent Monte Carlo (MC) simulation runs are carried out by regenerating the measurement sets. In Figures \ref{fig_spawn_vs_slt}, \ref{fig_Gauss_vs_gm}, \ref{fig_cent_vs_dist_gmm}, we have plotted: (i) the average root mean squared errors (RMSE) in the agent location estimates (averaged across the MC runs and across all the mobile agents); and, (ii) the average target tracking performance via the Optimum Sub-Pattern Assignment (OSPA) error \cite{schuhmacher08_tsp}. The OSPA metric is capable of taking into account errors in estimating both the number of targets (i.e., cardinality) and their tracks. The OSPA employs two parameters: cut-off, set to $20$m and order, set to $1$. In Figure \ref{fig_card}, we have explicitly plotted the average estimated cardinality over time.


\subsection{SCS-MTT vs SPAWN}
\label{subsec_spawn_vs_slt}
In Figure \ref{fig_spawn_vs_slt}, we compare the average agent localization error, and the average OSPA error for targets, of our approach (SCS-MTT) against SPAWN. Figure \ref{fig_spawn_vs_slt_gauss} shows the comparison when the agent and target densities are modeled as Gaussian mixtures. Figure \ref{fig_spawn_vs_slt_gm} showcases the same comparison for single Gaussian densities. Our approach significantly improves the localization performance by taking into account the contribution of the $\Lambda_k^s$ messages from the within-range targets to the agents. This would be especially beneficial for agents which are not in range of the anchor nodes, and have few neighboring agents (e.g., agents $1a, 2a$). Due to the presence of anchors, the agent localization improvements of the SCS-MTT algorithms transpire to a lesser extent into improvements on target tracking performance. The spikes in OSPA error correspond to the time instants of target births ($t=5,10,20$s) and deaths ($t=40$). Note that due to the dynamic nature of the network (frequent target births and deaths), the localization performance cannot be expected to converge over time. This is also evident from the slight increase in the localization error after $t = 40$s, in Figures \ref{fig_spawn_vs_slt} and \ref{fig_Gauss_vs_gm}.

Figure \ref{fig_card} shows the true cardinality, the mean estimated cardinality, and mean $\pm 3 \times$ standard deviation curves for the CGM-SPAWN (Figure \ref{fig_card_spawn}), CGM-SCS-MTT (Figure \ref{fig_card_Gibbs}) and DGM-SCS-MTT (Figure \ref{fig_card_dist}) filters. In all cases, the mean estimated cardinality is close to the true cardinality with the CGM-SPAWN filter having higher cardinality variance. Both the centralized and decentralized GM-SCS-MTT filters have smaller cardinality variance than CGM-SPAWN, while the DGM-SCS-MTT filter has a slightly higher variance than the CGM-SCS-MTT filter. This is attributed to the differences between the sequential Gibbs and the parallel Hogwild! Gibbs samplers and to the network consensus process.

\subsection{Single Gaussian vs Gaussian mixture SCS-MTT filters}
\label{subsec_dist_gauss_vs_gmm}
In Figure \ref{fig_Gauss_vs_gm}, we present the performance of the GM-SCS-MTT, which employs GM representations for both target and agent beliefs, with respect to the single Gaussian G-SCS-MTT filter. Figure \ref{fig_Gauss_vs_gm_cent} shows this comparison for the centralized SCS-MTT filters. Figure \ref{fig_Gauss_vs_gm_dist} showcases the same comparison for the decentralized SCS-MTT filters. 
The GM-SCS-MTT filters exhibit only a slight gain in terms of localization and tracking performance as compared to the G-SCS-MTT filters. This  is  because  the  agent  and  target  dynamic  models,  as discussed in Section \ref{sec_gmm}, are linear with additive Gaussian noise. Additionally, the considered measurement model is only moderately nonlinear. 
Applying our GM-SCS-MTT filter to a highly nonlinear and/or non-Gaussian setting is one of the directions we wish to pursue in a subsequent study.

\subsection{Centralized vs decentralized Gaussian mixture}
In Figure \ref{fig_cent_vs_dist_gmm}, we compare the performance of the centralized (CGM-SCS-MTT) and decentralized (DGM-SCS-MTT) filters. The decentralized method achieves performance similar to the centralized filter, given a sufficient number of consensus iterations $Q$. Compared to the centralized approach, the decentralized approach does not have to rely on a central fusion center and is scalable with the number of sensors. We have plotted the average localization and tracking error for different values of $Q$. For $Q=50$, the DGM filter performance is almost identical to the CGM filter. For small $Q$, since the network is sparsely connected (some agents have only one neighboring agent), the target belief product of (\ref{eq_tg_gen_gmm_prod1})-(\ref{eq_tg_gen_gmm_prod0}) is not accurately evaluated. This leads to poor tracking of targets which subsequently leads to poor localization of agents, due to the interdependence of localization and tracking for SCS-MTT algorithms. 
Similar results and conclusions hold for the single Gaussian case, which is omitted due to space constraints.

	\section{Conclusion}
\label{sec_conc}
In this paper, we proposed a novel decentralized method for simultaneous agent localization and multi-target tracking, for an unknown number of targets, under measurement-origin uncertainty. We proposed decentralized single-Gaussian as well as Gaussian-mixture implementations for our proposed filter. The two cases capture the trade-off between computational and communication efficiency (single Gaussian) and modeling accuracy (Gaussian mixtures). For the Gaussian-mixture case, we proposed a novel decentralized Gibbs method for efficiently computing products of Gaussian mixtures. We have demonstrated the robustness of our approach in a challenging range-bearing measurement model, which showcases the improved performance of the proposed methods with respect to the SPAWN method that performs agent localization and target tracking separately. 

	
\section*{Acknowledgments}
The authors would like to thank Dr. Florian Meyer at LIDS, MIT for his suggestions during the initial phases of this work, and the anonymous reviewers for their insightful comments that helped improve the article significantly.

\appendices 

	\bibliographystyle{IEEEtran}
	\bibliography{abrv,References}

\end{document}